\documentclass[12pt]{article}

\usepackage{times}
\usepackage[hidelinks]{hyperref}
\usepackage{amsmath,amssymb}
\usepackage[skip=2pt,font=footnotesize]{caption}
\usepackage{color}
\usepackage{graphicx}
\usepackage{enumitem}
\usepackage{multirow}
\usepackage{booktabs}

\title{
Machine learning assisted tracking of magnetic objects using quantum diamond magnetometry
}

\author
{
Fernando Meneses,$^{1,2\dag}$ 
Christopher T.-K. Lew,$^{1\dag}$ 
Anand Sivamalai,$^{3}$ 
\\
Andy Sayers,$^{1,3}$ 
Brant C. Gibson,$^{4}$ 
Andrew D. Greentree,$^{4}$ 
\\
Lloyd C. L. Hollenberg,$^{1,2}$ 
David A. Simpson$^{1,5\ast}$ 
\\
\normalsize{$^{1}$School of Physics, University of Melbourne, Parkville, Victoria 3010, Australia}\\
\normalsize{$^{2}$Centre for Quantum Computation and Communication Technology,}
\\
\normalsize{School of Physics, University of Melbourne, Parkville, Victoria 3010, Australia}\\
\normalsize{$^{3}$Phasor Innovation Pty Ltd, 155 Straws Lane, Hesket, Victoria 3442, Australia}\\
\normalsize{$^{4}$School of Science, RMIT University, Melbourne, Victoria 3000, Australia}\\
\normalsize{$^{5}$ARC Centre of Excellence in Quantum Biotechnology,}
\\
\normalsize{School of Physics, University of Melbourne, Parkville, Victoria 3010, Australia}\\
\normalsize{$^\dag$These authors contributed equally to this work.}
\\
\normalsize{$^\ast$To whom correspondence should be addressed; E-mail: simd@unimelb.edu.au.}
}

\date{}

\begin{document}
\baselineskip24pt 
\maketitle 

\begin{abstract} 

Remote magnetic sensing can be used to monitor the position of objects in real-time, enabling ground transport monitoring, underground infrastructure mapping and hazardous detection. However, magnetic signals are typically weak and complex, requiring sophisticated physical models to analyze them and a detailed knowledge of the system under study, factors that are frequently unavailable. In this work, we provide a solution to these limitations by demonstrating a Machine Learning (ML) method that can be trained exclusively on experimental data, without the need of any physical model, to predict the position of a magnetic target in real-time. The target can be any object with a magnetic signal above the floor noise, and in this case we use a quantum diamond magnetometer to track variations of few hundreds of nanoteslas produced by an elevator moving along a single axis. The one-dimensional movement is a simple yet challenging scenario, resembling realistic environments such as high buildings, tunnels or train circuits, and is the first step towards building broader applications. Our ML algorithm can be trained in approximately 40~min, achieving over 80\% accuracy in predicting the target's position at a rate of 10~Hz, for a positional error tolerance of 30~cm, which is a precise distance compared to the 4-meter spacing between parking levels. Our results open up the possibility to apply this ML method more generally for real-time monitoring of magnetic objects, which will broaden the scope of magnetic detection applications.

\end{abstract}

 
\section{Introduction}
\label{Intro}

Real-time object monitoring plays an essential role across many fields, such as traffic monitoring~\cite{Sun2022RSOD:Monitoring, Ke2017Real-TimeVideos}, environmental control~\cite{Kays2015TerrestrialPlanet} and surveillance~\cite{Gariepy2016DetectionView, Ghafoor2019AnArchitecture, Jiang2024Real-TimeSystem}.
Most well-established techniques are based on the global positioning system (GPS), which relies on radio-wave signals and satellite triangulation~\cite{Bock2016Physical}. Although a powerful technique, GPS has important limitations: first, it requires that the object under tracking is equipped with a GPS receiver, then restricting use to cooperative scenarios. Second, there are several GPS-denied environments where radio-wave signals are blocked or distorted, such as underwater, underground, and indoor spaces~\cite{Langley2008PropagationSignals, Hewawasam2019ComparativeEnvironment}.
For all these cases, a different technology is needed, and this is where remote magnetic sensing comes into play, as magnetic signals can propagate in GPS-denied environments and are generated directly by the tracked object, thus not limited to cooperative scenarios.

The major challenge for remote magnetic sensing is interpreting the data, as each tracked object will have its own magnetic signature, convoluted with the environmental noise. The conventional approach for analyzing this kind of information is to build a physical model that describes the target and its environment, and then fit the experimental data to extract the position information. Examples of this method can be found in submarine navigation~\cite{Wang2022quantum, Callmer2010SilentMagnetometers}, 
detection and classification of ground vehicles~\cite{Wahlstrom2014MagnetometerTargets}, and detection of magnetic anomalies~\cite{Chen2022MagneticAlgorithm}. Recent works have introduced ML tools to complement the data analysis and improve the performance compared to classical algorithms, but they still rely on physical models to pre-process magnetic data or synthesize training datasets. For instance, Wu \textit{et al.}\ developed a deep-learning model to monitor the activity of underground trains~\cite{Wu2021VectorModel}; Cárdenas \textit{et al.}\ used Convolutional Neural Networks (CNN) to detect magnetic anomalies and identify the localization, number and properties of magnetic dipoles~\cite{Cardenas2022MagneticExplainability}; Sun \textit{et al.}\ employed Deep Neural Networks (DNN) to localize complex arrays of underground pipelines~\cite{Sun2022MagneticNetworks}; and Vandavasi \textit{et al.}\ improved underwater navigation with shallow learning AI algorithms~\cite{Vandavasi2023MachineHoming}. 

Although magnetic data interpretation has made remarkable progress, most of the real-world scenarios display too complex magnetic signals which would require extensive physical models. As the models grow in complexity, more prior information about the magnetic system is needed in order to determine numerous parameters and fit the experimental data. As a result, remote magnetic sensing becomes unfeasible in most situations where limited information is available. As such, much research has focused on the development of modeling-free alternatives for interpreting the magnetic data, powered by artificial intelligence. The main idea being that experimental measurements can be analyzed by ML algorithms that work as ``black boxes", meaning that they make their own (hidden) model for interpreting data, ultimately taking magnetic measurements as inputs and returning the position of the tracked object. Some examples of these experimentally-trained ML algorithms include the recognition of submarine cables~\cite{Liu2023ARecognition}; robot's indoor magnetic navigation~\cite{Chiang2020MagneticSampling};
and identification of basic maneuvers of ground vehicles~\cite{Li2022ASensors}. As this area of research is relatively new, applications are still limited, focusing on object recognition, autonomous navigation or qualitative behavior, but none of them have the ability to monitor the precise location of a moving object in real-time.

In this work, we develop an experimentally-trained ML algorithm that predicts the position of an elevator in an indoor environment in real-time, using magnetic measurements as the input data, without the need of developing any physical model to interpret the magnetic information. The target (i.e.\ the elevator car) can move along a vertical axis across 8 levels spaced 4-meter apart at an average speed of 1.4~m/s, producing variations of a few hundreds of nanoteslas in the measured magnetic signals. Although this one-dimensional movement may seem simple, the magnetic system is in fact complex and involves multiple elements, such as the elevator car, counterweight, steel cables and electrical motor, each one contributing with their own magnetic signature. In addition, these signals are convoluted with the strong magnetic fields coming from nearby electrical trams, which we identify as magnetic noise for our tracking purposes. As explained above, building a physical magnetic model for this real-world problem is a challenging task that requires a significant level of prior knowledge about the elevator itself and its environment.

In situations where prior knowledge is limited, we need a highly sensitive magnetic field sensor that can operate in real-world scenarios. In this work, we utilize a quantum diamond magnetometer based on a dense ensemble of nitrogen-vacancy (NV) defects as our sensing platform~\cite{Scholten2021WidefieldProspects, Schloss2018SimultaneousSpins, Rondin2014magnetometry, Abrahams2021integrated, Doherty2013TheDiamond, Jelezko2006SingleReview}. Our magnetometer can operate in a wide temperature range without any magnetic shielding, detecting full-vector magnetic fields with sensitivities of around 400~pT/$\sqrt{\text{Hz}}$ per axis, reported within a $1-100$~Hz measurement bandwidth. Furthermore, quantum diamond magnetometers have inherent capabilities promising accurate and stable measurements during long periods of time without any drift or the need for recalibration.

Our results demonstrate that with 6 hours of training with experimental results, the ML algorithm can predict the position of the elevator car, represented by its lowest point, at a 10~Hz rate with an accuracy of 80\% for a positional error tolerance of 30~cm. As this distance is much smaller than the 4-meter spacing between parking levels, and the accuracy implies that on average 8 out of 10 points are correct every second, the trajectory of the elevator and the arrival at each parking level can be reliably monitored in real-time. The flexibility of both our ML algorithm and magnetic platform allows for easy adaptation to various scenarios and monitoring tasks, promising broad applicability in situations where only experimental data is available.

 
\section{Results}
\label{Results}

\subsection{Application overview}
\label{Application_Overview}

\begin{figure}[!h]
    \centering
    \includegraphics[width=1\linewidth]{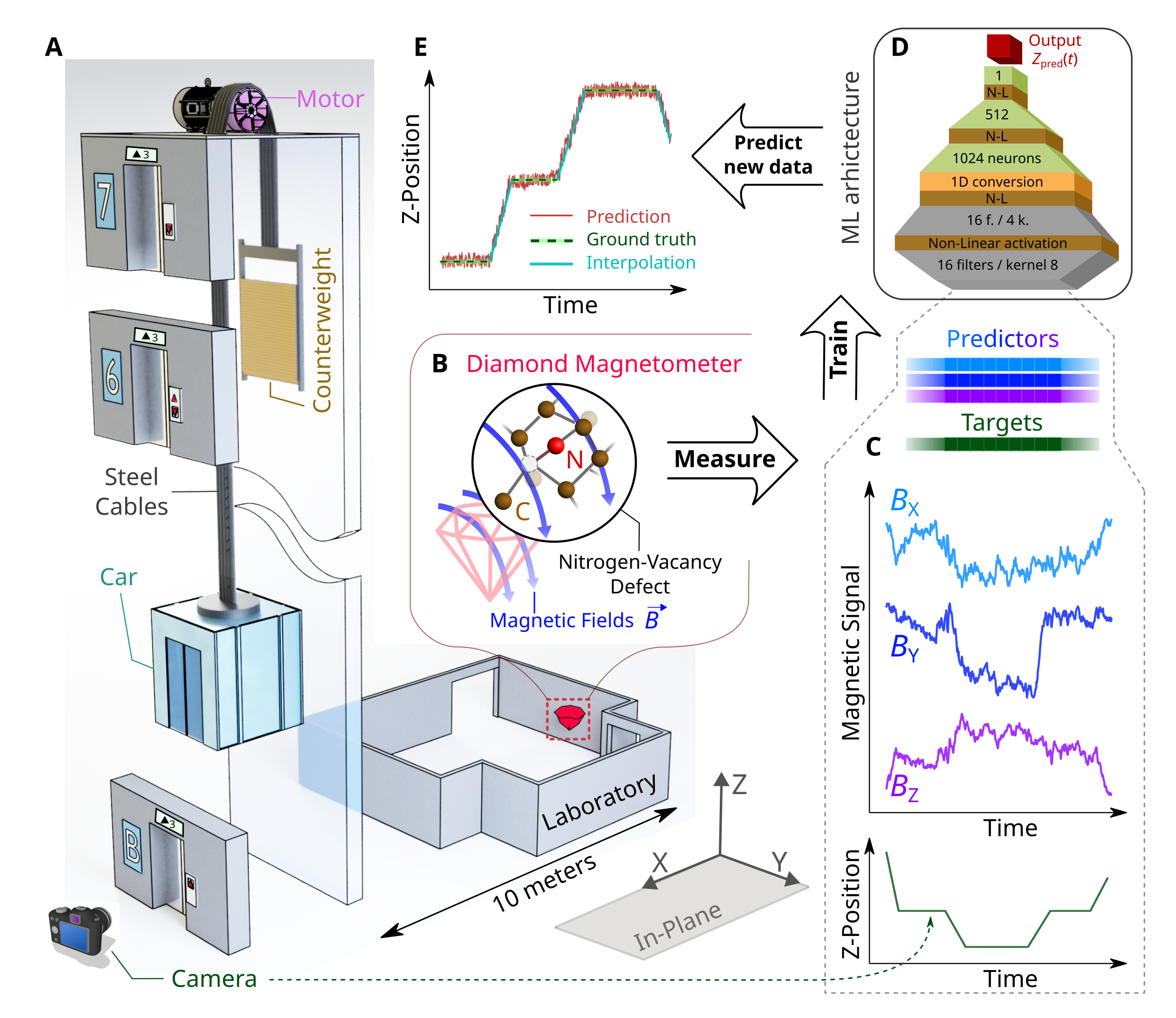}
    \caption{
    \textbf{Real-time object monitoring: application overview}.
    (\textbf{A}) Scheme of the elevator system, where the main magnetic elements are identified: target car, counterweight, steel cables and electrical motor. The sensing platform is located in the lowest level of the building, 10~m away from the elevator's doors. The position of the car is recorded independently by a camera placed in front of the informative display.
    (\textbf{B}) Quantum diamond magnetometer based on a dense ensemble of nitrogen-vacancy defects, capable of sensing external magnetic fields. The inset shows the diamond crystallographic structure containing the NV defect.
    (\textbf{C}) Set of magnetic measurements and Z-positions of the elevator's car as a function of time. They are processed into a one-dimensional timeseries as predictors and targets, respectively.
    (\textbf{D}) Machine Learning architecture, featuring Convolutional neural networks (gray), non-linear activation layers (brown), a one-dimensional conversion layer (orange) and Dense neural networks (green). The inputs are the magnetic predictors and the final outputs are the predicted Z-positions of the car (red).
    (\textbf{E}) Example of the ML predictions (red line) for new, unseen magnetic data. The ground truth measured by the camera indicates the parking levels (dashed green line), while the traveling positions are interpolated by a simple linear approximation (teal line).
    }
    \label{fig1}
\end{figure}

The principle of the real-time object monitoring using the NV quantum diamond magnetometer is illustrated in \textcolor{blue}{Figure~\ref{fig1}}. Our target is an elevator car, with $XYZ$ dimensions $1.8 \times 2.0 \times 2.6$~m$^3$, that can move along 8~levels on the $Z$-axis, with levels spaced 3.7~m apart except for the lowest two, spaced by 4.1~m. The elevator travels at an average speed of 1.4~m/s, meaning that it typically takes almost 3~s to move between adjacent levels.
The quantum diamond magnetometer is located at the basement (lowest level), approximately 10~m from the elevator, and it detects variations of the magnetic field~$\vec{B}$ when the elevator is moving, which are in the order of a few hundreds of~nT.
However, the magnetic signal does not exclusively reflect the magnetic field from the car alone, but also the Earth's magnetic field ($\approx 50 \, \mu$T) and contributions from other magnetic elements in the elevator system, mainly the counterweight (traveling in the opposite direction to the car), the steel cables and the electrical motor located on top of Level~8 (see \textcolor{blue}{Figure~\ref{fig1}A}).
Additionally, there is environmental magnetic noise from nearby electronic equipment and wiring, in the order of few~nT, and a much larger contribution from urban electrical trams operating continuously, located approximately 50~m away from the magnetometer. These vehicles are ground cars with overhead current-carrying cables that generate magnetic fields, providing $1/f$-like random magnetic field fluctuations ranging from 70~nT to 250~nT.

In order to detect full-vector magnetic signals, we use a single-crystal diamond substrate containing a dense ensemble of NV defects (\textcolor{blue}{Figure~\ref{fig1}B}), with sensitivity of $\eta=$~400(50) pT/$\sqrt{\text{Hz}}$ per axis. Considering the $\approx$3~s timescale of the elevator system, we set a 10~Hz sampling rate to ensure a precise time resolution ($dt=0.1$~s) for tracking magnetic field variations. 

As the ML algorithm needs to be trained and evaluated with Z-position reference targets, it is essential to record this ``ground truth" information independently, and we want to use a procedure as simple as possible so our tracking application remains flexible and can be adapted to other scenarios. Following these criteria, we simply place a camera in front of the elevator doors to record its display (\textcolor{blue}{Figure~\ref{fig1}A}) and correlate it in time with the magnetic measurements. As the display only indicates the parking level if the elevator is parked, or the traveling direction and a rough level estimation when the elevator is moving, we can confidently determine only the time intervals in which the elevator is parked. For the complete Z-trajectory, we interpolate the parking positions with a linear approximation during the traveling events, which gives a good estimate of the position. As we will discuss later, the tracking precision can be improved with more sophisticated methods to record the elevator's movement, such as measuring its acceleration as a function of time and using a physical model to reconstruct the Z-position, but this approach is against the original intention of employing minimal information.

Using both sets of magnetic and Z-position measurements, we design a ML algorithm with a deep-learning supervised architecture, in which the full-vector magnetic fields $(B_{\rm X},B_{\rm Y},B_{\rm Z})(t)$ are encoded into 3-channel input predictors and the elevator position $Z(t)$ as target values (\textcolor{blue}{Figure~\ref{fig1}C}). During training, the internal parameters of the ML architecture are tuned to minimize the difference between the predicted positions $Z_{\rm pred}(t)$ and the ground truth (\textcolor{blue}{Figure~\ref{fig1}D}). Once this process is completed, the ML algorithm can predict the elevator position in real-time for new magnetic measurements (\textcolor{blue}{Figure~\ref{fig1}E}). 

To assess the ML predictions' accuracy, we define two different metrics: the parking accuracy $A_{\rm park}$, which takes into account only those intervals in which the elevator is parked (ground truth); and the tracking accuracy $A_{\rm track}$, which uses the entire trajectory, combining both the parking intervals and the linear interpolations. The former metric is the most reliable, as it is based on the ground truth, while the second metric is built on top of a simple approximation for the elevator's movement, which is not strictly precise but gives a good estimation of the real trajectory. For any metric, a single prediction~$P$ at a time~$t$ is considered correct if the absolute difference between the predicted position $Z_{\rm pred}(t)$ and the reference position $Z_{\rm ref}(t)$ (ground truth or interpolated values), is smaller than a fixed position tolerance $Z_{\rm tol}$:

\begin{equation}
    \label{eq_corr_preds}  
    P(t) = \begin{cases}
    1, & 
    \text{if } |Z_{\rm true}(t)-Z_{\rm ref}(t)| \leq Z_{\rm tol}
    \\
    0, &
    \text{if } |Z_{\rm true}(t)-Z_{\rm ref}(t)| > Z_{\rm tol}
    \\
    \end{cases}
\end{equation}

Both metrics $A_{\rm park}$ and $A_{\rm track}$, represented as $A$, are calculated once the ML algorithm is trained, by using a testing dataset with new data and calculating the ratio between the correct and total number~$N$ of predictions:

\begin{equation}
    \label{eq_MLaccuracy}  
    A = \frac{1}{N} \sum^{N}_{i} P(t_{\rm i}) \, .
\end{equation}


\subsection{Magnetic measurements in different frames}
\label{Results_Magnetic_Measurements}

\begin{figure}[!h]
    \centering
    \includegraphics[width=1\linewidth]{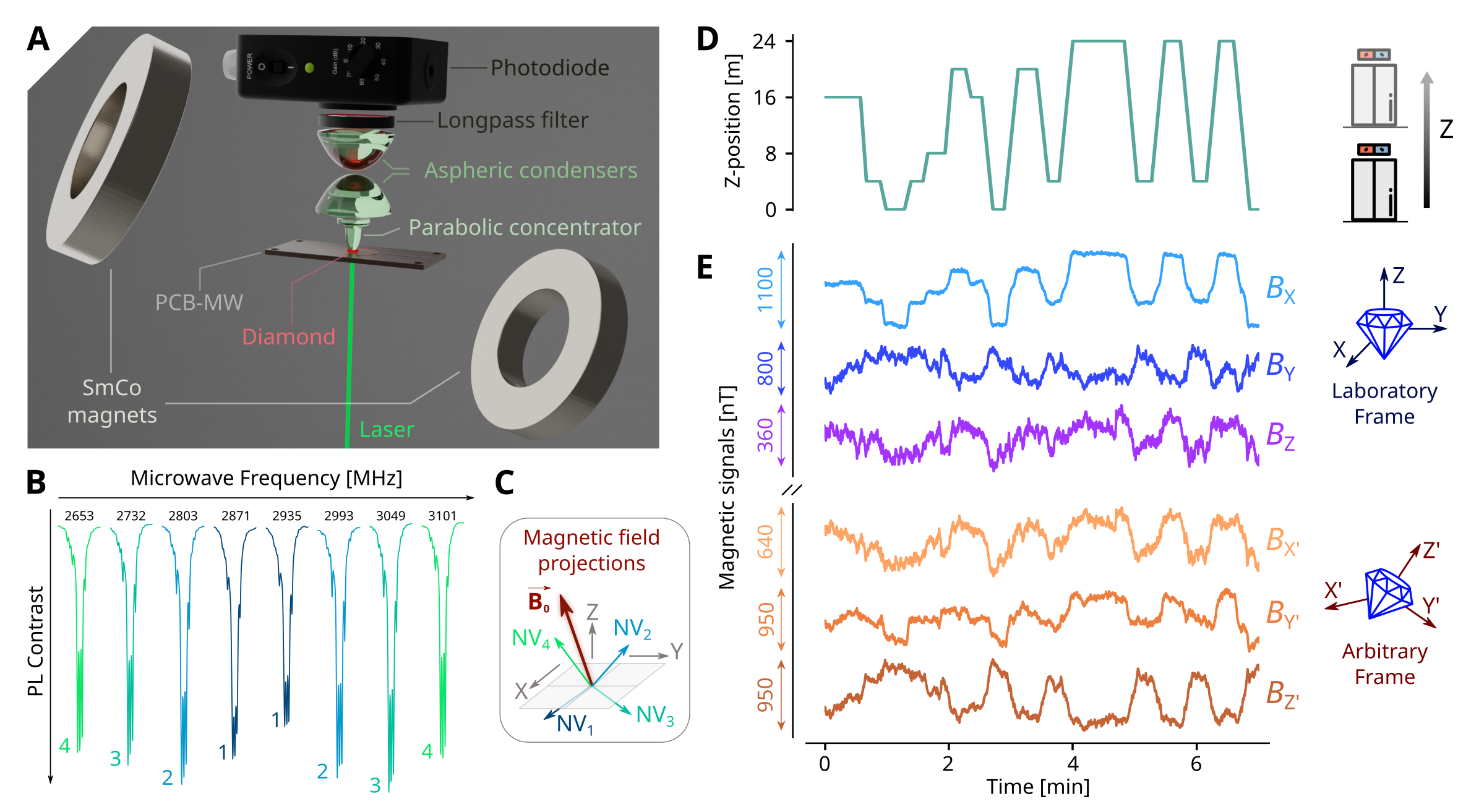}
    \caption{
    \textbf{Magnetic measurements in different frames}.
    (\textbf{A}) Quantum diamond magnetometry platform, with details about the optical and electronic elements, along with the external magnets that provide a constant bias field~$\vec{B}_{0}$.
    (\textbf{B}) Optically Detected Magnetic Resonance spectrum showing the eight resonant frequencies for the ensemble of nitrogen-vacancy (NV) defects, numbered from 1 to 4 according to the NV axes.
    (\textbf{C}) Geometry for the NV axes described in the laboratory frame, along with the $\vec{B}_{0}$ direction.
    (\textbf{D}) Independently recorded positions of the elevator's car by filming the information display. The horizontal segments are the parking intervals (ground truth), connected with each other by linear interpolations. Parking levels are spaced $\approx 4$~m apart.
    (\textbf{E}) Magnetic measurements projected into the laboratory frame (top), with the $Z$-axis aligned with the elevator's axis, and the $XY$ plane squared with the room. The environmental noise affects mainly the $B_{\rm Y}$ and $B_{\rm Z}$ components, leaving a relatively clean $B_{\rm X}$ axis that exhibits a clear correlation with the elevator's position. The same data is shown in a different frame (bottom), where the noise is coupled to all magnetic axes evenly, obscuring the correlation.
    }
    \label{fig2}
\end{figure}

The quantum diamond magnetometer platform utilized in this work is displayed in \textcolor{blue}{Figure~\ref{fig2}A}. Our sensing technique utilizes a multichannel demodulation scheme via four software-defined lock-in amplifiers to simultaneously measure magnetic field variations along the four crystallographic orientations of the NV defects within the diamond relative to a static bias magnetic field. The following outlines the general calibration routine implemented prior to the measurement protocol. 

First, a static bias magnetic field, $\vec{B}_{0}$, is applied with a pair of temperature-compensated samarium cobalt ring magnets. Its projection along each NV axes is shown in \textcolor{blue}{Figure~\ref{fig2}C}, along with the resulting optically detected magnetic resonance (ODMR) spectrum in \textcolor{blue}{Figure~\ref{fig2}B}. The eight resonant frequencies are computed by fitting eight hyperfine-split (i.e.\ $f_{\rm i} \pm 2.158$ MHz) Lorentzians to the ODMR spectrum. A non-linear least square routine based on the Levenberg-Marquardt algorithm is used to minimize the residual error between the eigenvalues solutions to the NV ground-state Hamiltonian and the measured resonant frequencies. From this minimization, the zero-field splitting~$D$, bias magnetic field~$\vec{B}_{0}$ in the laboratory frame, and the longitudinal strain and electric field coupling parameters~$\vec{M}_Z$ to the four  NV orientations are determined. For the ODMR spectrum in \textcolor{blue}{Figure~\ref{fig2}B}, $D=2870.19$~MHz, $\vec{B}_{0}=(4.339, 1.301, 7.915)$~mT, and $\vec{M}_{\rm Z} = (19.602, -4.168, -3.520, -11.914)$~kHz were obtained. 

The NV Hamiltonian is numerically linearized about the measured $D$, $\vec{B}_{0}$, $\vec{M}_{\rm Z}$ (i.e.\ calibration point), resulting in a $4\times4$ Jacobian matrix $\mathbf{A}$ describing the expected frequency shifts due to an external magnetic field and temperature shift relative to where the calibration is performed~\cite{Schloss2018SimultaneousSpins}. The inverse matrix, $\mathbf{A}^{-1}$, is then calculated, which describes the transformation in the opposite direction (i.e., the measured frequency shifts to a vector magnetic field in the laboratory frame and a temperature shift).

In the measurement protocol, the demodulated lock-in signal for each NV-axis is converted to a frequency shift via the measured slope of the lock-in response for each resonance addressed. This slope is determined from a local calibration sweep after the initial ODMR measurement is performed, where a FM-modulated multitone microwave (MW) signal is applied and swept across each resonance addressed sequentially ($\pm20$~MHz span). A linear fit is then used to extract the lock-in signal slope. The frequency shift for each resonance addressed is converted into a magnetic field and temperature shift by multiplying by the $\mathbf{A}^{-1}$ matrix. 

An example of the elevator's $Z$-position for a short time span is presented in \textcolor{blue}{Figure~\ref{fig2}D}, correlated with the magnetic signals $(B_{\rm X},B_{\rm Y},B_{\rm Z})$ projected in the laboratory frame in \textcolor{blue}{Figure~\ref{fig2}E} (top). In this frame, where the $Z$-axis is aligned with the elevator's axis and the $XY$ plane is squared with the room, there is a clear correlation between the $B_{\rm X}$ component and the elevator's movement, while the other projections $(B_{\rm Y},B_{\rm Z})$ are much noisier. This noise directionality originates in the magnetic signals coming from the electrical trams, which exhibit weak fluctuations in the $X$ magnetic axis (up to 70~nT) and strong variations in the $YZ$ plane (up to 250~nT), see \textcolor{cyan}{Figures \ref{fig_SI_Noise_trams_timeseries_histograms}} and \textcolor{cyan}{\ref{fig_SI_Environmental_Noise_Complete}} in Supplementary Text. As we did not have any prior information about the magnetic behavior of our system, our choice of Cartesian projections $XYZ$ for the magnetic fields fortuitously allowed us to easily identify a correlation between magnetic signals and the elevator's position. However, we may have chosen any other frame $X'Y'Z'$ in which the trams' signal is distributed across all magnetic axes evenly, obscuring this correlation. An example of this scenario is shown in \textcolor{blue}{Figure~\ref{fig2}E} (bottom), where the magnetic data is rotated to an arbitrary frame (see Materials and Methods).

From a human perspective, the projection of the magnetic signals plays an important role for data interpretation, and a bad choice such as the last example may hinder the analysis. Notice, however, that the full-vector magnetic information recorded by the magnetometer is equivalent in any frame. This situation is a good example of a real-world scenario, and our challenge is to incorporate it into our object-monitoring application, so the ML learning algorithm can work with equal performance given data projected in any frame. Consequently, we introduce the concept of ``minimum accuracy" $A_{\rm min}$ for performance evaluation, in which a single ML architecture is trained and evaluated in different frames~$F(\theta,\varphi,\alpha)$, defined by the polar~$\theta$ and azimuth~$\varphi$ spherical coordinates of the rotation axis, along with the rotating angle~$\alpha$. The final performance is then defined as the lowest accuracy result:

\begin{equation}
    \label{eq_MLaccuracy_min}  
    A_{\rm min} = \min_{F} A\Big|_F,
\end{equation}

\noindent where $A\Big|_F$ is the accuracy (as defined in \textcolor{blue}{Equation~\ref{eq_MLaccuracy}}) evaluated at a particular frame $F(\theta,\varphi,\alpha)$.

 
\subsection{Machine Learning Architecture}
\label{Results_Architecture}

Considering the particular conditions of our object-monitoring challenge, but also taking into account future adaptations to similar applications, we have studied many ML algorithms that share the following principles: supervised learning, deep learning structure and a regression task. 

In supervised learning, the ML algorithm learns from an experimental dataset containing predictors (magnetic signals) and targets (the elevator position). During training, the algorithm makes predictions based on the input and compares them with the target values. Then, it adjusts its internal parameters to reduce the difference next time, iterating this process over many cycles to improve the accuracy progressively. After training, the ML algorithm is able to predict the elevator position from new magnetic measurements without any target references.

A deep learning structure means that the ML algorithm has multiple hidden layers between the input predictors and the final output, enhancing its ability to analyze complex patterns~\cite{Ahmed2023DeepChallenges, Lecun2015DeepLearning}. This type of architecture works well with our extensive magnetic data: just 1-hour of measurements at 10~Hz generates 36,000 datapoints for each magnetic component. As we will explain later, our ML algorithms perform better when they use the recent magnetic history to predict a single elevator position.

Lastly, a regression task implies that the ML algorithm predicts continuous output values, in this case the $Z$-position of the elevator. As mentioned earlier, our ground truth measurements of the elevator's trajectory are those intervals in which the elevator is parked, while the traveling events are approximated by linear interpolations. The former information suggests a classification task, training the ML algorithm to make categorical predictions such as ``Parked at Level~X" or ``Traveling up/down". However, this approach introduces large uncertainties for the traveling events, as it leaves intermediate positions undefined, which can extend over long durations for the longer flights. Conversely, the continuous predictions provide a good approximation of the elevator's position throughout its entire trajectory, even if the exact location remains uncertain.

\begin{figure}[!h]
    \centering
    \includegraphics[width=1\linewidth]{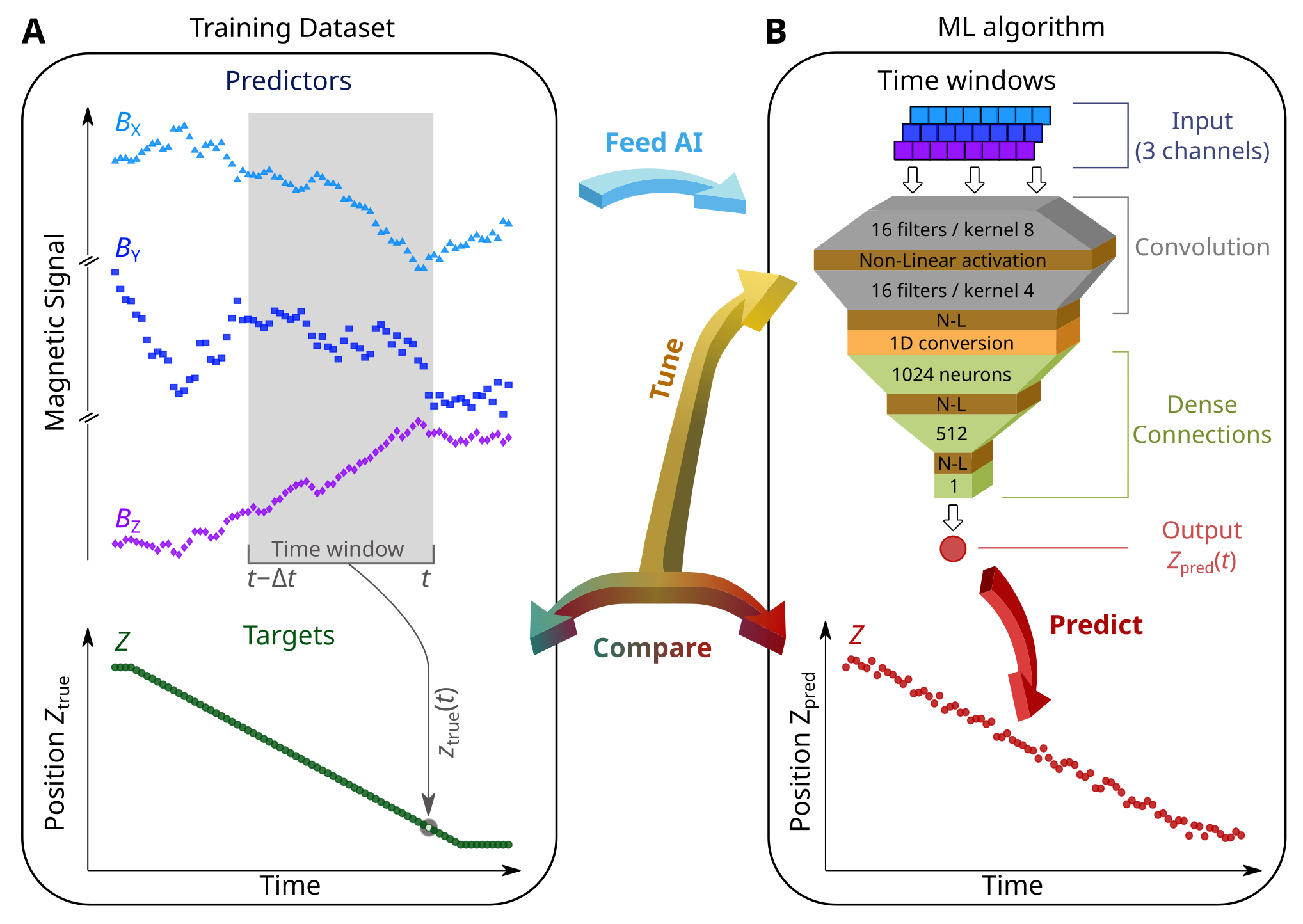}
    \caption{
    \textbf{Training process for the Machine Learning (ML) algorithm.}
    (\textbf{A}) Training dataset, including magnetic signals as predictors and the elevator's positions as targets. At any time~$t$, a predictor is defined as a ``time window" of duration~$\Delta t$, containing the most recent magnetic history. The corresponding target is the ground truth position~$Z_{\rm true}(t)$.
    (\textbf{B}) ML architecture. From top to bottom: the input layer allocates the input time window in a 3-channel matrix, passing this information into a set of Convolutional Neural Networks, characterized by their number of filters and kernel size, along with non-linear activation (N-L) functions. The multi-dimensional output, corresponding to the different filters, is flattened into a one-dimensional (1D) vector by a conversion layer and then input into a series of Dense layers, characterized by the number of neurons, and N-L functions. The final output is a single value representing the predicted position of the elevator~$Z_{\rm pred}(t)$ for the input time window. During a training cycle, the entire set of predictions is compared to the reference targets, computing the ``loss" score as the average value of the absolute difference. The ML internal parameters are then tuned, aiming to minimize the loss for the next cycle. This process is repeated for a fixed number of cycles or until an early stop criterion is met.
    }
    \label{fig3}
\end{figure}

An overview of the ML training process is displayed in \textcolor{blue}{Figure~\ref{fig3}}. In our ML design, at any given time~$t$ a predictor is defined as a ``time window" containing magnetic measurements $(B_{\rm X},B_{\rm Y},B_{\rm Z})$ over a fixed interval~$\Delta t$ (shaded area in \textcolor{blue}{Figure~\ref{fig3}A}). The corresponding target is the elevator position~$Z_{\rm true}(t)$, meaning that predictions are made in real-time by taking into account the most recent magnetic history. Since the ML algorithm is designed to produce continuous outputs, we use all $Z$-positions as ground truth during training, including both parking and traveling intervals. In contrast, during the evaluation process we do distinguish between the ground truth parking positions and the interpolated approximations, by using the $A_{\rm park}$ and $A_{\rm track}$ metrics defined previously.

The architecture of our ML algorithm is illustrated in \textcolor{blue}{Figure~\ref{fig3}B}. From top to bottom, the structure begins with the input layer, which allocates a single predictor: a $3 \times N$ matrix that represents the time window with the $(B_{\rm X},B_{\rm Y},B_{\rm Z})$ components in separate channels (rows), and $N$ points given by the interval's length~$\Delta t$ and the time resolution~$dt=0.1$~s: $N=\Delta t/dt$. This information is processed by a block of one-dimensional CNN and non-linear activation (N-L) functions, which explore long- and short-range correlations within the data and extract the most relevant patterns for the next layers. Our ML algorithm works best with two CNN, each of them having 16 filters (convolutional matrices that analyze the input data), with kernel sizes 8 and 4 for the first and second CNN, respectively. During the Convolutional process, the original data is expanded into multiple-dimensions (as many as the number of filters), which are then flattened into a single vector by a one-dimensional (1D) conversion layer. This format is required by the next processing block: a series of Dense (or Fully Connected) layers and N-L functions. Each Dense layer is characterized by a number of neurons~$n$, which are processing units that receive all input datapoints, multiply them by internal weights and add biases, and finally return an output vector with size~$n$. In our architecture, we progressively downsize the information with Dense layers of sizes $n=$~1024, 512 and~1, until we get a final value as output: the predicted position of the elevator~$Z_{\rm pred}(t)$ for the input time window.

During training, the ML algorithm initializes its internal parameters randomly, and updates them every cycle (or epoch) to improve the future predictions. In each cycle, the complete set of predictors are input into the algorithm and the output predictions $Z_{\rm pred}(t)$ are compared to the ground truth $Z_{\rm true}(t)$, computing a ``loss" score equal to the average value for the absolute difference $|Z_{\rm pred}(t)-Z_{\rm true}(t)|$. The algorithm consequently modifies its internal parameters with a back-propagation method~\cite{Lecun2015DeepLearning}, trying to minimize the loss for the next epoch (see arrows connecting \textcolor{blue}{Figure~\ref{fig3}A-B}). This process is iterated for a fixed number of cycles or until an early stop criterion is met (see Materials and Methods).

In the evaluation process, the ML architecture is fed with a new set of time windows, producing the output predictions $Z_{\rm pred}(t)$ but not updating the internal parameters. The predictions are then compared with the reference values $Z_{\rm true}(t)$, and the accuracies $A_{\rm park}$ and $A_{\rm track}$ are calculated according to \textcolor{blue}{Equation~\ref{eq_MLaccuracy}}. The entire training and evaluation procedures are repeated for different frames, to ensure a robust ML performance, and the final accuracies $A_{\rm park, min}$ and $A_{\rm track, min}$ are computed using \textcolor{blue}{Equation~\ref{eq_MLaccuracy_min}}.


\subsection{Optimization of Hyperparameters}
\label{Results_Optimization}

In any ML algorithm there are two kinds of parameters: internal parameters, which are tuned automatically during the training process; and hyperparameters, which define global structures and require human input. Examples of the later are the type and number of layers within the ML architecture, the type of non-linear activation function to be used, or the number of points in a time window. By optimizing hyperparameters we can significantly boost the ML performance. In this section, we describe a series of optimization stages that we carried out to find our best performing ML algorithm.

As the general condition for each optimization stage, we studied different options for a single hyperparameter, or a reduced combination of them, leaving the other hyperparameters fixed. For each option, we evaluated several possibilities for ML architectures according to the metric $A_{\rm track, min}$ using a tolerance $Z_{\rm tol}=1$~m and several frames for the magnetic data. Among all the results for a single option, we chose the lowest accuracy as the final score. As a reference benchmark, we set a minimum accuracy threshold of 80\%, meaning that, on average, 8 of 10 predictions are correct every second.

\begin{figure}[!h]
    \centering
    \includegraphics[width=0.75\linewidth]{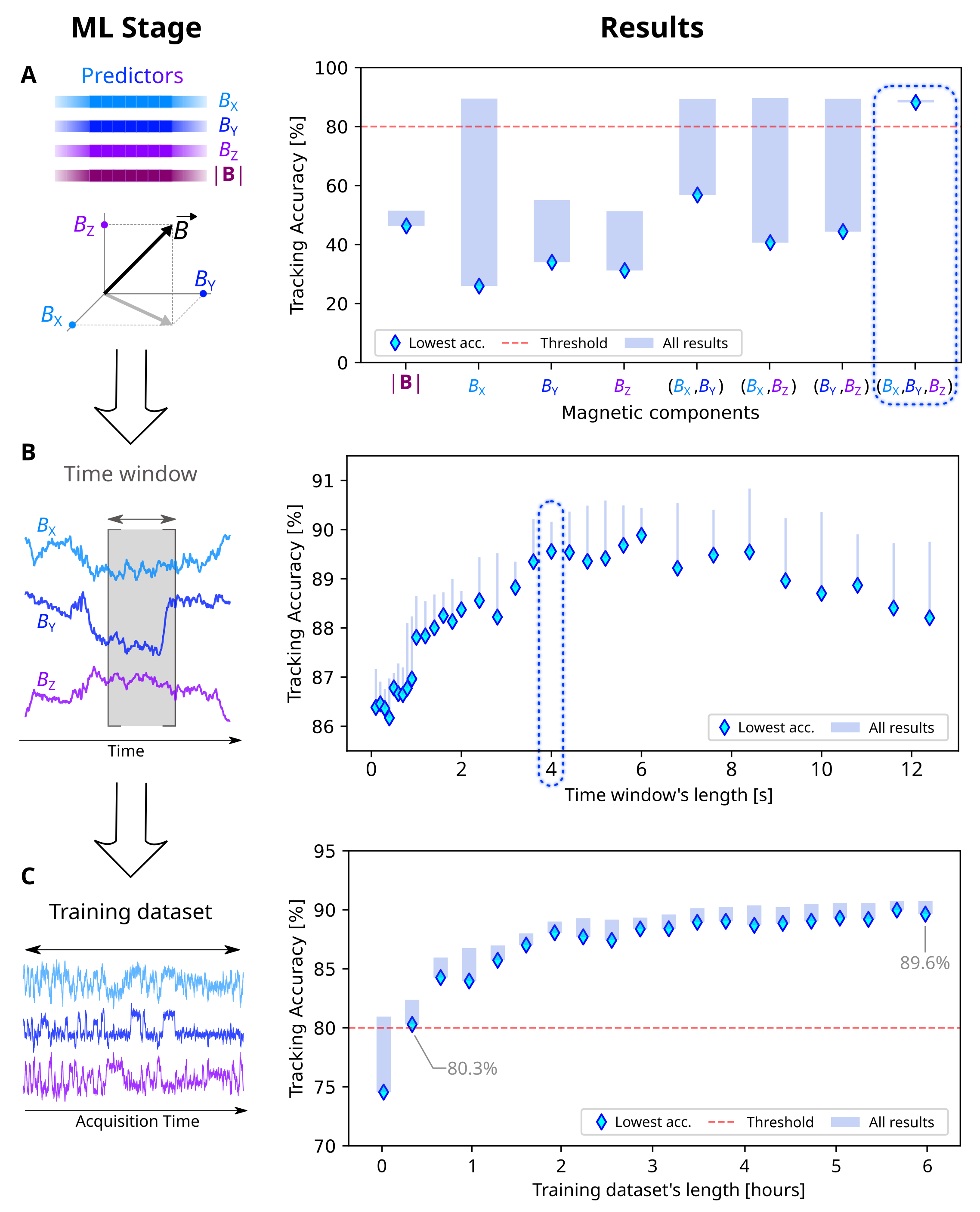}
    \caption{
    \textbf{Optimizing the Machine Learning hyperparameters}.
    In progressive stages, multiple options for a single hyperparameter are studied, training several algorithms and computing the tracking accuracy with a 1-meter tolerance. While the complete range of results are displayed (shaded bars), only the lowest results (blue diamonds) count as the final score. We set a reference benchmark of 80\% (red dashed line).
    (\textbf{A}) First stage: combinations of magnetic components used as input predictors, from which the full-vector option is selected (dashed rectangle).
    (\textbf{B}) Second stage: duration of time windows. There is an optimal range from 4~s to 6~s, from which we select the shorter time (dashed rectangle), favoring a simpler computational structure of the ML algorithm.
    (\textbf{C}) Final stage: impact of the training dataset size. Larger sizes improve the tracking accuracy, up to 89.6\% when using the full dataset (6~hours). If only an 80\% accuracy is required, using just 0.5~hours of measurements is enough.
    }
    \label{fig4}
\end{figure}

The first optimization stage, which addresses the full-vector nature of the measurements, is shown in \textcolor{blue}{Figure~\ref{fig4}A}: do we need all $(B_{\rm X},B_{\rm Y},B_{\rm Z})$ magnetic components for accurate predictions, or can we use just the scalar field $|\vec{B}|$ or a single magnetic projection? This decision has important implications for the complexity of the ML algorithm, as well as for the types of sensing platforms that can be utilized in this application. For each hyperparameter option, the final score is represented by a blue diamond, while the full range of results is shown as a shaded bar. Noticeably, only the full-vector predictors $(B_{\rm X},B_{\rm Y},B_{\rm Z})$ overcome the 80\% accuracy threshold (red dashed line), showing little variation across all results. In contrast, all other magnetic component combinations exhibit low scores (less than 50\%), with great variability in results.

The importance of having the complete magnetic information in the input data becomes clear when analyzing the large spread in accuracy results. For some frames, one or two magnetic projections may capture the essential features of the signal, which explains the higher end for the accuracy results in options such as the single $B_{\rm X}$. However, there are other frames where the reduced information becomes insufficient, hence showing a low performance. The scalar field option~$|\mathbf{B}|$, on the other hand, is invariant to any frame representation at the expense of losing directional information, which negatively impacts the ML performance. Conversely, the full-vector data~$(B_{\rm X},B_{\rm Y},B_{\rm Z})$ always encodes the complete information, irrespectively of the chosen frame, and achieves consistently high accuracy. Consequently, we choose this option for the subsequent stages.

The second optimization stage studies the duration of the time windows: what is the optimal length for accurate predictions? The results are shown in \textcolor{blue}{Figure~\ref{fig4}B}, where we notice that all options score above 86\%, even the single-point time window (left-most result), which is a simple magnetic vector. However, we observe an optimal range approximately from 4~s to 6~s (equivalent to 40 to 60 points, given our 10~Hz measurement rate). We speculate that too short time windows do not unlock the full potential of the Convolutional layers, while too long time windows provide excessive information. As shorter time windows imply less internal parameters for the ML algorithm, which is translated into faster training and less chances to overfit the data, we choose a 4~s duration as the optimal length.

Further optimization stages are described in the Materials and Methods section, resulting in the ML architecture shown in \textcolor{blue}{Figure~\ref{fig3}B}. Fine details of the algorithm include the type of non-linear activation functions, defined as hyperbolic tangent (``tanh") in all cases, the global Adaptive Moment Estimation (``adam") optimizer and the 0.0001 learning rate.

For the best ML architecture, we finally evaluate the effect of the training dataset size (measured in recording hours) on the tracking accuracy performance, as illustrated in \textcolor{blue}{Figure~\ref{fig4}C}. As expected, the larger the training dataset the better the accuracy, reaching an optimal value of 89.6\% with 6~hours of measurements. We also notice that the performance is reaching a limit at this point, suggesting that while the dataset size is critical for training the ML algorithm, it is not the only factor. Other contributions are likely related to the uncertainty in the magnetic measurements, set by the magnetic noise floor, unexpected events and imperfections of the target object. In other words, we do not expect perfect predictions in a real-world scenario. On the other hand, if the monitoring application requires 80\% tracking accuracy, recording just half an hour of measurements is enough to train the ML algorithm, which is a reasonable time if the access to ground truth measurements is limited, or if there is a need for periodical recalibrations. 

As a final comment, we studied the possibility of artificially augmenting the training dataset to reduce the measurements time even further. For this purpose, we reduced the original training dataset to a fraction and then generated similar copies which introduced random variations for the predictors, while the ground truth targets remained unchanged. When the ML algorithms were trained with the augmented predictors, the results showed that data augmentation, by any amount, did not improve the performance, and in most cases made it worse (see \textcolor{cyan}{Figure~\ref{fig_SI_Data_Augmentation}} in Supplementary Text). This result reinforces the advantage of training the ML algorithm exclusively on experimental data: synthetic information cannot be easily used as a substitute of the real measurements. If carefully engineered, artificial data may emulate reality, but that would require prior knowledge about the system, and this is precisely the limitation that we are overcoming with our method.

 
\subsection{Final performance}
\label{Results_Final_Performance}

After selecting the optimal hyperparameters for the ML architecture, we compute the final performance of the algorithm using the testing dataset, which spans 6~hours of measurements, and a collection of 65~randomly generated frames for the magnetic data projection. We compute both the tracking and parking accuracies according to \textcolor{blue}{Equations~\ref{eq_corr_preds}-\ref{eq_MLaccuracy_min}}, using a position tolerance $Z_{\rm tol}=1$~m.

\begin{figure}[!h]
    \centering
    \includegraphics[width=0.85\linewidth]{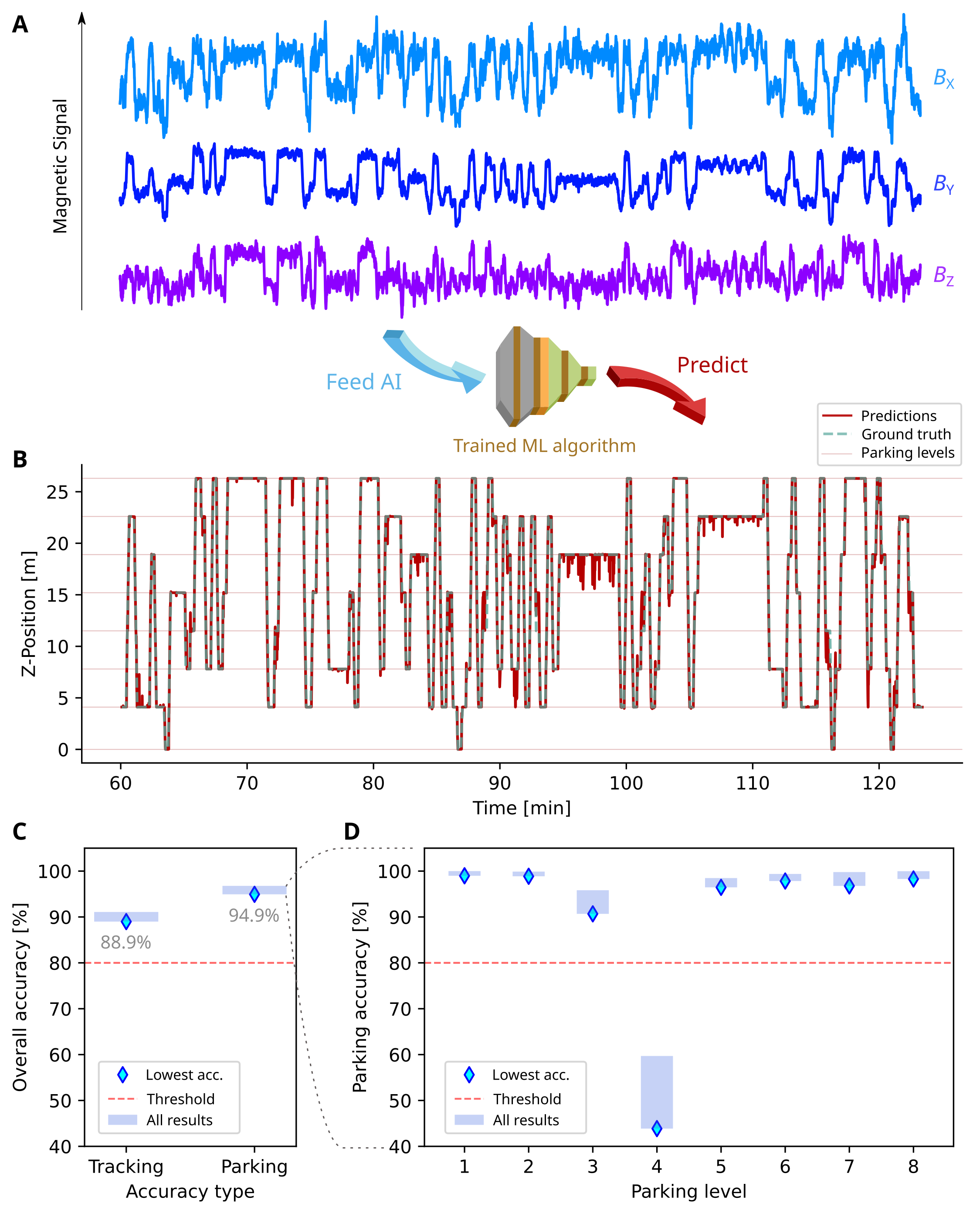}
    \caption{
    \textbf{Final performance for the object-monitoring Machine Learning application}.
    (\textbf{A}) Input magnetic predictors from a segment of the testing dataset, in a particular frame where the environmental noise is distributed across all magnetic components, making the data analysis apparently difficult from a human perspective.
    (\textbf{B}) Predictions for the target's position (red solid line), compared with the ground truth (teal dashed line). Parking levels are indicated with faint horizontal lines.
    (\textbf{C}) Tracking and parking accuracies evaluated in the entire testing dataset (6~hours) across 65~randomly generated frames for the magnetic data projection. The final performance is defined as the lowest result within these frames.
    (\textbf{D}) Parking accuracy examined for each parking level, demonstrating that predictions are more than 90\% accurate except for those related to Level~4.
    }
    \label{fig5}
\end{figure}

The input magnetic fields from a 1-hour segment of the testing dataset are shown in \textcolor{blue}{Figure~\ref{fig5}A}, for a particular frame in which there are not clean magnetic axes, making the data analysis apparently difficult from a human perspective. After this data is fed into the ML algorithm, the predicted Z-positions are compared with the reference ground truth, see \textcolor{blue}{Figure~\ref{fig5}B}. The prediction accuracies, when computed in the entire testing dataset and the 65~frames, are $A_{\rm track, min}=$88.9\% (general tracking) and $A_{\rm park, min}=$94.9\% (only parking positions), as displayed in \textcolor{blue}{Figure~\ref{fig5}C}. Following this result, we demonstrate that our ML method can reliably monitor the position of the elevator in real-time, with an average of 9 out of 10 predictions correct every second. 

In addition to the global results, the parking accuracy is examined in the different parking levels in \textcolor{blue}{Figure~\ref{fig5}D}, meaning that the predictions are separated in groups associated to each (ground truth) parking level. This analysis shows that the parking predictions are very accurate ($A_{\rm park, min}>$~90\%) for all parking levels except for Level~4 ($Z=$11.5~m in \textcolor{blue}{Figure~\ref{fig5}B}). We speculate that at this position the elevator's car and counterweight are close, in a geometric configuration that produces a complex magnetic signal which is easily obscured by the noise and confused by the ML algorithm with the neighboring positions (see \textcolor{cyan}{Figure \ref{fig_SI_Vector_All_measurements}} in Supplementary Text). In fact, if we increase the position tolerance to $Z_{\rm tol}=$~4~m, the prediction's accuracy becomes higher than 99\% for all parking levels (see \textcolor{cyan}{Figure \ref{fig_SI_park_acc_lvls_tol4m}} in Supplementary Text), demonstrating that the incorrect predictions in the previous case were assigned to the adjacent parking levels or in-between, and not anywhere further.

\begin{figure}[!h]
    \centering
    \includegraphics[width=0.95\linewidth]{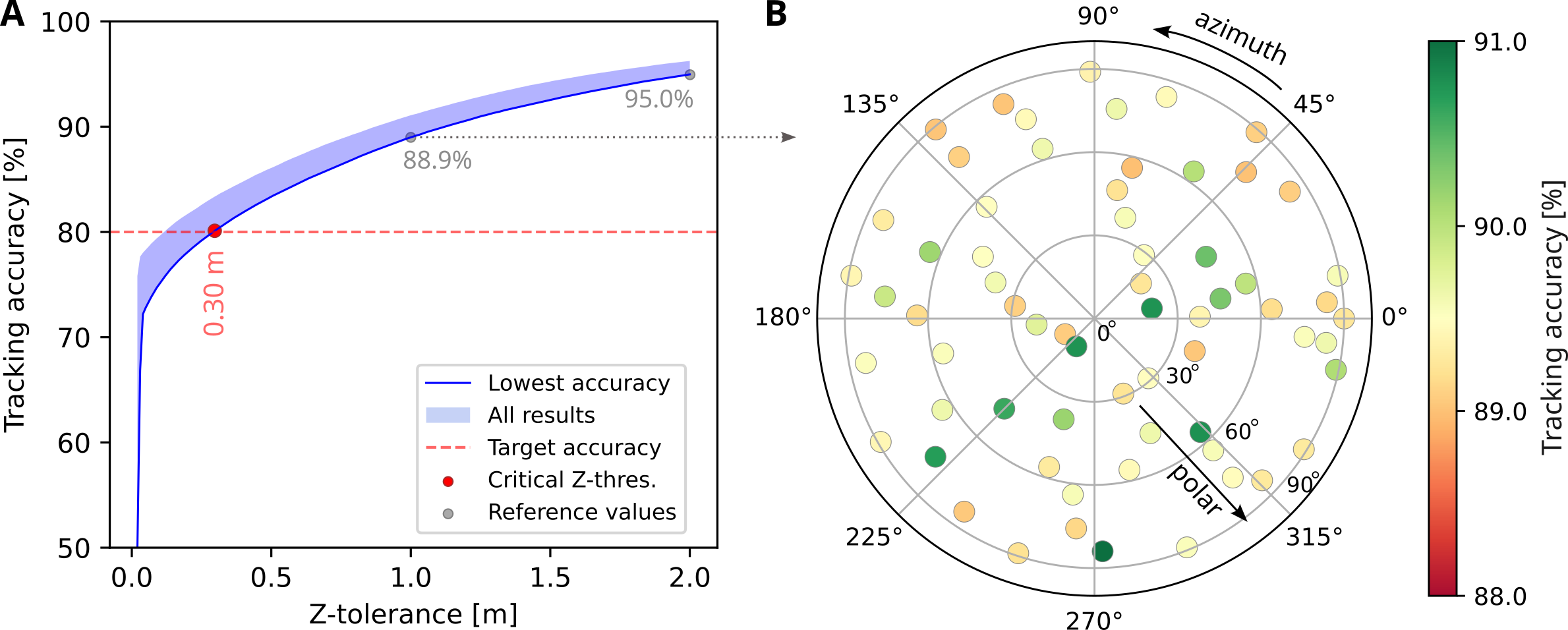}
    \caption{
    \textbf{Tracking accuracy as a function of the Z-position tolerance}.
    (\textbf{A}) The monitoring performance of the Machine Learning algorithm is studied for a continuous range of tolerance values, using a collection of 65~randomly generated frames for the magnetic data projection. The final score is defined as the lowest accuracy across all frames (solid blue line), while the full range of results is displayed as a shaded area. For an accuracy threshold of 80\%, the tolerance can be as low as 30~cm.
    (\textbf{B}) Distribution of tracking accuracy results (color-coded) in the different frames using a 1-meter position tolerance, projected according to the polar and azimuth coordinates that describe each frame.
    }
    \label{fig6}
\end{figure}

Throughout all previous analyses, we have studied the performance of the ML algorithm related to a fixed tolerance value $Z_{\rm tol}=$1~m, which is small compared to the parking levels separation (approximately 4~m), and allows for a certain parking level identification. However, we can extend the analysis and explore how precise our ML method can be for the general tracking task, in terms of position tolerance. By following the same procedure described at the beginning of this section but for a continuous Z-position tolerance, we compute the tracking accuracy~$A_{\rm track, min}$ as a function of $Z_{\rm tol}$, as displayed in \textcolor{blue}{Figure~\ref{fig6}A}. For a minimum accuracy of 80\%, the tolerance can be set as low as 30~cm, demonstrating the precision and robustness of our monitoring application. As previously described in \textcolor{blue}{Figure~\ref{fig5}A}, the tracking accuracy for $Z_{\rm tol}=$1~m is 88.9\%, while the ultimate limit of $Z_{\rm tol}=$~2~m, which still guarantees a reliable identification of the closest parking level, increases the performance up to $A_{\rm track, min}=$95.0\%.

Lastly, we show a complete map of the tracking accuracy~$A_{\rm track}$ for each frame~$F(\theta,\varphi,\alpha)$ in \textcolor{blue}{Figure~\ref{fig6}B}, for a 1-meter position tolerance. The map projects the polar~$\theta$ and azimuth~$\varphi$ coordinates of the different rotating axes (in the laboratory frame), while the rotating angle is fixed to $\alpha=90^{\circ}$ (see Materials and Methods), with each point representing the prediction results, color-coded by the accuracy value. Across all frames, the accuracy variability is less than 3\%, demonstrating the robustness of the ML algorithm. In addition, as there is no clear spatial pattern for the accuracy distribution, we speculate that this small spread is most probably coming from the effect of the random initialization of the ML internal parameters rather than from a geometrical configuration.

 
\section{Discussion} 
\label{Discussion}

In this work, we have developed a Machine Learning algorithm trained exclusively on experimental data, obtained by a quantum diamond magnetometer, to monitor in real-time the activity of a moving magnetic target within a noisy environment. The ML performance was classified into two different accuracies: general tracking~$A_{\rm track, min}$ and parking position~$A_{\rm park, min}$. When using a 1-meter position tolerance to evaluate the correct predictions, the performance is $A_{\rm track, min}=$~88.9\% and $A_{\rm park, min}=$~94.9\%. If only an 80\% accuracy is required for the general tracking task, then the position tolerance can be as low as 30~cm.

While Machine Learning has been applied to similar monitoring challenges, this is the first reported method for real-time magnetic tracking based on pure experimental training, without the need of building intricate physical models or simulations. As a matter of fact, the ML method can be improved with more information, models and simulations, by augmenting the training dataset or providing rules for the predictions. For example, the predictions in \textcolor{blue}{Figure~\ref{fig5}B} show unstable positions at some parking intervals, implying that the elevator is going up and down too fast. If we incorporated physical-informed rules to the deep learning structure, such as penalizing the algorithm for predicting too large ``jumps" in position, the accuracy would increase. Although any improvements are beneficial, the simple ML architecture that we have developed in this work follows a very easy implementation with a demonstrated high performance.

An example of enhancing the ML performance by adding more information is to provide a better interpolation method between parking intervals. Our choice of linear approximation is not entirely realistic as it implies instant acceleration at the beginning and end of each flight, although it provides a good position estimation. Alternatively, we have experimentally measured the elevator's acceleration profile and built a physical model for the position as a function of time, see the Supplementary Text for more details. When the ML algorithm is trained with this type of interpolation, by the end of the final optimization stage the tracking accuracy slightly improves from the previous 89.6\% to 91.1\%. For this particular scenario, a 1.5\% gain probably does not justify the effort of acquiring the additional acceleration measurements and physical model. However, the takeaway message is that the basic ML algorithm can be improved as much as the user wants for their particular application, at the expense of more knowledge and/or computational design, but the baseline performance is already high.

The applicability of this research can be extended to a wide range of situations where magnetic objects move within a relatively controlled (though complex) environment, meaning that the external perturbations are limited. This is the critical factor for the supervised ML algorithm, as it must be trained with experimental data and account for a reasonable number of outcomes. In a continuously changing environment, where significant and distinct magnetic perturbations occur frequently, the training dataset cannot cover all possibilities. Conversely, in a controlled environment where magnetic perturbations remain within some boundaries, like the strong but steady noise from the distant electrical trams in our work, the ML algorithm can learn general trends and make accurate predictions, just from experimental data. For example, we could apply our method to monitor magnetic objects in underground mines or tunnels (where the magnetic environment is relatively free from external factors), such as vehicles or miners carrying magnetic beacons. The more complex the environment and target signals, the larger the required training dataset. Nevertheless, training the ML algorithm is a one-time investment, and it can be a solution for problems where theoretical models or simulations are not available.

Our application integrates a quantum sensing platform with a Machine Learning algorithm to solve a problem that is hardly approachable from a conventional point of view. The adaptability of our approach, where the magnetic sensor and ML design can be customized to the specific requirements of the problem, makes it a versatile and promising solution for a wide range of object monitoring applications based just on experimental measurements.

\noindent\rule{\linewidth}{0.4pt}


\section*{Materials and Methods}

\subsection*{Quantum description of nitrogen-vacancy defects}

The NV defect is a substitutional nitrogen atom adjacent to a vacancy aligned along the $<111>$ crystallographic axes of the diamond lattice, which behaves as a spin-1 system. The ground state spin triplet includes a non-magnetic $m_{\rm S}=0$ state and two energy-degenerate $m_{\rm S} \pm 1$ states which are separated from the $m_{\rm S}=0$ state by a zero field splitting of 2.87~GHz. 
The full Hamiltonian describing the energy levels can be found in the Supplementary Text. When exposed to an external magnetic field, the ground state magnetic levels $m_{\rm S}\pm1$ split due to the Zeeman interaction. 
The energy level splitting is proportional to the magnetic field projection along the N$-$V axis, meaning that each NV defect accounts for specific magnetic information according to its orientation in the crystal lattice. Any external magnetic field vector can be reconstructed by measuring the magnetic field projections along the four possible NV axes within a dense NV ensemble in the diamond.


\subsection*{Experimental setup}

The quantum diamond magnetometry setup is described below in three main sections concerning the optical, electronics and lock-in setups.

\subsubsection*{Optical setup}

The magnetic sensing unit is a $3 \times 3 \times 0.5$ mm$^{3}$ diamond chip (DNV-B14) from Element-6 with a typical NV concentration of 4.5~ppm. A high power laser (Laser Quantum Opus 532, 6~W) is used to illuminate the diamond, at a power of 500~mW during normal operating conditions. The laser output passes through an optical isolator (Thorlabs IO-5-532-HP), half waveplate (Thorlabs WPH10M-532), beam sampler (thorlabs BSF10-A), two steering mirrors (Edmund $\#$38-893) and a $f=$~500~mm focusing lens (Thorlabs AC254-500-A-ML). A final mirror is then used to direct the focused laser beam through a 1~mm diameter hole in the printed circuit board (PCB) with the embedded microwave (MW) antenna design \cite{sasaki2016broadband}. With approximately 320~mW of green excitation measured just before illuminating the diamond, 10.6~mW of NV photoluminscence (PL) is measured. 

PL from the diamond is collected using a compound parabolic concentrator (Edmund $\#$17-709), followed by two back-to-back aspheric lenses condenser (Thorlabs ACL25416U-B), a 635~nm longpass filter (Semrock BLP01-635R-25) and a large area photodetector (Thorlabs PDA36A2). A small portion of the laser output picked off using the beam sampler is directed toward a second photodetector (Thorlabs PDA36A2) for common-mode rejection of laser intensity noise after passing through a continuously variable neutral density filter (Thorlabs NDC-50C-2M) and a 532~nm bandpass filter (Semrock FF01-524/24-25). The common-mode rejected photodetector signal is then simultaneously digitized by eight AC-coupled analog-to-digital converters (ADCs) at a sampling rate of 202.8 kSa/s ($4 \times$National Instruments (NI) PXI-4461 modules housed within a NI PXIe-1062Q chassis) and averaged in software to reduce quantization noise. 

\subsubsection*{Electronics setup}

Four MW synthesizers ($2\times$Agilent E4432B, $2\times$Agilent E4433B) are used to generate four frequency-modulated (FM) multitone signals to address four of the eight NV resonances and their three hyperfine spin transitions spaced $\pm2.158$~MHz apart. The FM square waveforms are generated by four digital-to-analog converters (DACs) at a sampling rate of 202.8~kSa/s from four analog outputs on two of the NI PXI-4461 modules. A 10~MHz oven-controlled crystal oscillation (OCXO) clock source is used to synchronize all MW synthesizers and modules housed within the NI chassis by connecting in a daisy-chain manner.

The MW signal from each synthesizer passes through a DC block (Mini-circuits BLK-89-S+) and an isolator (Microwave Associates MZG-3000) before being combined with each other using three power combiners (Mini-circuits ZX10Q-2-34-S+). An amplifier (Mini-circuits ZHL-5W-422+) is then used to amplify the combined MW signal before sending the signal to the embedded PCB MW antenna.

\subsubsection*{Lock-in setup}

As described in the main text, the lock-in demodulation of the common-mode rejected, averaged, and frequency modulated PL response is performed in software. First, a 10$^{\mathrm{th}}-$order Butterworth highpass filter with a cutoff frequency at 1500~Hz is applied. The filtered signal is then mixed with the four square reference waveforms individually. Each demodulated trace is then lowpass filtered (10$^{\mathrm{th}}-$order Butterworth, 53~Hz), removing the 2$f_{\rm i}$ second harmonic component. Lastly, the demodulated and filtered traces are decimated and averaged by a factor of 20280, downsampling from 202.8~kSa/s to 10~Sa/s.

 
\subsection*{Training and testing datasets}

The total set of measurements, including magnetic signals and the elevator car's position, comprises 12~hours acquired in 5 different segments. We only include daytime measurements, when the elevator is actively used. All timeseries for magnetic and position measurements can be found in the Supplementary Text, along with a histogram for the $B_{\rm X}$ component in the laboratory frame where 8~peaks can be resolved, corresponding to the 8~parking levels within the building (\textcolor{cyan}{Figures \ref{fig_SI_Bx_All_measurements}, \ref{fig_SI_Vector_All_measurements}}). The complete dataset spans 426,496 points for each magnetic component and target position, which is split equally into the training dataset (211,756 points) and testing dataset (214,740 points).

 
\subsection*{Frames for magnetic data}

The original magnetic data, projected into the laboratory frame $XYZ$, can be converted to any other frame~$X'Y'Z'$ by a rotational transformation around an axis~$\hat{n}$ by an angle~$\alpha$. We identify a rotated frame~$F(\theta,\varphi,\alpha)$ by the polar~$\theta$ and azimuth~$\varphi$ spherical angles (expressed in the laboratory coordinates) of the rotating axis~$\hat{n}$, and a rotating angle~$\alpha$. The mathematical description of the rotating transformations can be found in the Supplementary Text. 

In the main sections~\textcolor{cyan}{\ref{Results_Magnetic_Measurements}} and~\textcolor{cyan}{\ref{Results_Architecture}}, the ML algorithms are trained and evaluated in the following frames: laboratory (\textcolor{blue}{Figure~\ref{fig2}E} top), equivalent to~$F(0^{\circ},0^{\circ},0^{\circ})$; $F(67^{\circ},59^{\circ},120^{\circ})$ where the noise is partially distributed; 
and $F(58^{\circ},61^{\circ},90^{\circ})$ where the noise is evenly distributed across all magnetic components (\textcolor{blue}{Figure~\ref{fig2}E} bottom). In the main section~\textcolor{cyan}{\ref{Results_Final_Performance}}, multiple rotational frames are generated by randomly selecting the spherical coordinates within ranges $\theta \in (0^{\circ},90^{\circ}]$ and $\varphi \in (0^{\circ},360^{\circ}]$, and fixing the rotating angle to $\alpha=90^{\circ}$.

 
\subsection*{Data normalization and time windowing}

As Machine Learning algorithms perform better with normalized input data~\cite{cabello2023impact, singh2020investigating}, we apply a normalization process to our magnetic datasets. This process converts the values of each magnetic component into a dimensionless scale, centered at zero and with unitary amplitude. First, we account for the Earth's background magnetic field, which remains effectively constant in our measurement timescale (see Supplementary Text) by subtracting the mean value in each magnetic component. Next, we identify the maximum scalar value~$B_{\rm max}$ within the training dataset and define the normalization factor as $B_{\rm norm}=B_{\rm max}/\sqrt{3}$. Finally, all the data (both training and testing datasets) is divided by $B_{\rm norm}$, resulting in dimensionless datasets with zero mean value and variance smaller than~1. In the main text, the magnetic data shown in figures is already normalized, with the exception of \textcolor{blue}{Figure~\ref{fig2}E}. There, we show the original magnetic variations for each component, after subtracting the Earth's magnetic field.

Following data normalization, the magnetic information is processed into time windows with a fixed duration~$\Delta t$, which are the input units for the Machine Learning algorithm. For each time~$t$ associated with the elevator's car position~$Z(t)$, the corresponding time window contains the magnetic information within the time interval~$[t-\Delta t,t]$. This data is processed as a $N_{\rm channels} \times N_{\rm points}$ matrix, where $N_{\rm channels}$ is the number of magnetic components that are used as input data, e.g.\ $N_{\rm channels}=$~3 for the full-vector option $(B_{\rm X},B_{\rm Y},B_{\rm Z})$, and $N_{\rm points}$ is the number of measurements within the interval $[t-\Delta t,t]$.

 
\subsection*{General architecture for all ML algorithms}

The ML architecture shown in \textcolor{blue}{Figure~\ref{fig3}B} is structured in a series of layers which are representative of all the algorithms we have explored. Although the hyperparameters of each layer may change in a particular design, we always use the following sequence: (1)~input layer (predictors); (2)~set of one-dimensional convolutional neural networks (1D-CNN) layers, non-linear activation functions (N-L) and optional Pooling layers; (3)~one-dimensional conversion layer; (4)~set of Dense layers, N-L and optional Dropout layers, always including a final Dense layer with a single neuron; and (5)~output layer, a single value representing the position of the elevator's car (target). Notice that the architecture illustrated in the main text does not include any Pooling or Dropout layers, which are optional. A detailed explanation for each of the architecture elements can be found in the Supplementary Text.

In addition to the sequence of layers, all the ML algorithms are characterized by the following global hyperparameters: Mean Absolute Error as the Loss Function; validation fraction of 25\% during training; early stop criterion of $N_{\rm early}=$~15 training cycles without any improvement in the validation loss score; and maximum number of training cycles equal to $N_{\rm max}=$~200. For the final training stage, in which we use the optimized hyperparameters, we extend the early stop criterion to $N_{\rm early}=$~30 and the maximum number of training cycles to $N_{\rm max}=$~400.

 
\subsection*{Training stages for hyperparameters optimization}

Our best ML architecture is built after a sequence of 6~optimization stages, focused on the following hyperparameters: (1)~input magnetic components; (2)~time window's length; (3)~main architecture (variations within the layers' structure); (4)~global hyperparameters; (5)~fine architecture (local hyperparameters); and (6)~training dataset size. Stages 1, 2 and 6 are described in the main text (\textcolor{blue}{Figure~\ref{fig4}}), while the complete description is available in the Supplementary Text. After the final optimization stage, we keep the best ML architecture described in the main text, which has 1,019,297 internal parameters.

In all the optimization stages, the performance of every hyperparameter option (i.e.\ the lowest tracking accuracy) includes the accuracy results of several ML architectures. Each architecture design was trained and evaluated in three frames for the magnetic data: $F(0^{\circ},0^{\circ},0^{\circ})$, $F(67^{\circ},59^{\circ},120^{\circ})$ and $F(58^{\circ},61^{\circ},90^{\circ})$; as well as three different random seeds for the initialization of internal parameters. Therefore, each shaded bar containing all accuracy scores in \textcolor{blue}{Figure~\ref{fig4}} includes 9~results per ML architecture, multiplied by the number of evaluated architectures (which varies according to the optimization stage, see Supplementary Text). 

 
\subsection*{ML libraries and Computing resources}

The programming code developed in this research, along with the experimental dataset, is available in the GitHub repository \href{https://github.com/Fertmeneses/ML_QDM_Meneses_et_al}{Fertmeneses/ML\_QDM\_Meneses\_et\_al}, written in \textsc{Python 3.10} and using \textsc{Tensorflow 2.11.0}. All ML training processes were carried out on a personal laptop: Lenovo ThinkPad X1 Extreme Gen 5, 12th Gen Intel® Core™ i7-12800H, equipped with a NVIDIA RTX GeForce 3050 Ti GPU. Considering the architecture of our best model, the training time for 400~epochs is approximately 40~minutes. Once the model is trained, the prediction time for a single time window is approximately 0.6~ms (much faster than our 10~Hz measurement rate).


\section*{Acknowledgments}

We thank A. Silvester and L. T. Hall for fruitful discussions. D.A.S. and L.H.L. thank the ARC Centre of Excellence in Quantum Biotechnology [QUBIC] for its support.

\textbf{Funding:}
This work was financially supported by the Australian Research Council (ARC) Centre of Excellence Scheme (Grant no. CE170100012 and CE230100021). D.A.S. and L.T.H. acknowledge support from the National Intelligence and Security
Discovery Research Grants Program (NS220100071). This work was partially supported by Defence through the Next Generation Technologies Fund (NGTF) and managed under the Advanced Strategic Capabilities Accelerator. Additional ARC funding support was provided through (IM240100073, DP190101506, DE210101093 and DE200101785) and by The University of Melbourne’s Research Computing Services and the Petascale Campus Initiative.

\textbf{Author contributions:}
Conceptualization: D.A.S., A.Sa., B.A.G., A.G., L.C.L.H. $|$ 
Machine learning methodology: F.M. $|$ 
Experimental methodology: C.T.K.L. $|$ 
Analysis: F.M., C.T.K.L., D.A.S., A.Si., A.Sa., B.A.G., A.G., L.C.L.H. $|$ 
Visualization: F.M. and C.T.K.L. $|$ 
Funding acquisition: D.A.S., L.C.L.H, B.A.G., A.G.. $|$ 
Supervision: D.A.S. and L.C.L.H. $|$
Writing—original draft: F.M., C.T.K.L., D.A.S. $|$ 
Writing—review and editing: F.M., C.T.K.L., D.A.S., A.Si., A.Sa., B.A.G., A.G., L.C.L.H..

\textbf{Competing interests:} 
The authors declare that they have no competing interests.

\textbf{Data and materials availability:} All data needed to evaluate the conclusions in the paper are present in the paper and/or the Supplementary Text.

\include{02_SI}

\bibliographystyle{science}
\bibliography{references}

\begin{thebibliography}{10}

\bibitem{Sun2022RSOD:Monitoring}
W.~Sun, L.~Dai, X.~Zhang, P.~Chang, X.~He, {\it Applied Intelligence\/} {\bf 52}, 8448 (2022).

\bibitem{Ke2017Real-TimeVideos}
R.~Ke, {\it et~al.\/}, {\it IEEE Transactions on Intelligent Transportation Systems\/} {\bf 18}, 890 (2017).

\bibitem{Kays2015TerrestrialPlanet}
R.~Kays, M.~C. Crofoot, W.~Jetz, M.~Wikelski, {\it Science\/} {\bf 348}, aaa2478 (2015).

\bibitem{Gariepy2016DetectionView}
G.~Gariepy, F.~Tonolini, R.~Henderson, J.~Leach, D.~Faccio, {\it Nature Photonics\/} {\bf 10}, 23 (2016).

\bibitem{Ghafoor2019AnArchitecture}
H.~Ghafoor, Y.~Noh, {\it IEEE Access\/} {\bf 7}, 98841 (2019).

\bibitem{Jiang2024Real-TimeSystem}
Y.~Jiang, W.~Chen, X.~Zhang, X.~Zhang, G.~Yang, {\it Sensors\/} {\bf 24} (2024).

\bibitem{Bock2016Physical}
Y.~Bock, D.~Melgar, {\it Reports on Progress in Physics\/} {\bf 79}, 106801 (2016).

\bibitem{Langley2008PropagationSignals}
R.~B. Langley, {\it GPS for Geodesy\/} {\bf 2}, 103 (2008).

\bibitem{Hewawasam2019ComparativeEnvironment}
H.~S. Hewawasam, M.~Y. Ibrahim, G.~Kahandawa, T.~A. Choudhury, {\it Proceedings of the IEEE International Conference on Industrial Technology\/} pp. 19--26 (2019).

\bibitem{Wang2022quantum}
X.~Wang, {\it et~al.\/}, {\it Gravity, Positioning and Reference Frames\/} p.~87 (2022).

\bibitem{Callmer2010SilentMagnetometers}
J.~Callmer, M.~Skoglund, F.~Gustafsson, {\it Eurasip Journal on Advances in Signal Processing\/} {\bf 2010} (2010).

\bibitem{Wahlstrom2014MagnetometerTargets}
N.~Wahlstr{\"{o}}m, F.~Gustafsson, {\it IEEE Transactions on Signal Processing\/} {\bf 62}, 545 (2014).

\bibitem{Chen2022MagneticAlgorithm}
Z.~Chen, {\it et~al.\/}, {\it IEEE Sensors Journal\/} {\bf 22}, 3686 (2022).

\bibitem{Wu2021VectorModel}
X.~Wu, S.~Huang, M.~Li, Y.~Deng, {\it Applied Sciences (Switzerland)\/} {\bf 11} (2021).

\bibitem{Cardenas2022MagneticExplainability}
J.~C{\'{a}}rdenas, C.~Denis, H.~Mousannif, C.~Camerlynck, N.~Florsch, {\it Computers and Geosciences\/} {\bf 169} (2022).

\bibitem{Sun2022MagneticNetworks}
T.~Sun, {\it et~al.\/}, {\it Computers and Geosciences\/} {\bf 159}, 104987 (2022).

\bibitem{Vandavasi2023MachineHoming}
B.~N.~J. Vandavasi, H.~Venkataraman, A.~R. Gidugu, {\it Ocean Engineering\/} {\bf 280}, 114692 (2023).

\bibitem{Liu2023ARecognition}
Y.~Liu, {\it et~al.\/}, {\it Journal of Marine Science and Engineering\/} {\bf 11} (2023).

\bibitem{Chiang2020MagneticSampling}
T.~H. Chiang, Z.~H. Sun, H.~R. Shiu, K.~C.~J. Lin, Y.~C. Tseng, {\it IEEE Sensors Journal\/} {\bf 20}, 13110 (2020).

\bibitem{Li2022ASensors}
P.~Li, M.~Abdel-Aty, Q.~Cai, Z.~Islam, {\it IEEE Transactions on Intelligent Transportation Systems\/} {\bf 23}, 3148 (2022).

\bibitem{Scholten2021WidefieldProspects}
S.~C. Scholten, {\it et~al.\/}, {\it Journal of Applied Physics\/} {\bf 130} (2021).

\bibitem{Schloss2018SimultaneousSpins}
J.~M. Schloss, J.~F. Barry, M.~J. Turner, R.~L. Walsworth, {\it Physical Review Applied\/} {\bf 10}, 034044 (2018).

\bibitem{Rondin2014magnetometry}
L.~Rondin, {\it et~al.\/}, {\it Reports on progress in physics\/} {\bf 77}, 056503 (2014).

\bibitem{Abrahams2021integrated}
G.~Abrahams, {\it et~al.\/}, {\it Applied Physics Letters\/} {\bf 119} (2021).

\bibitem{Doherty2013TheDiamond}
M.~W. Doherty, {\it et~al.\/}, {\it Physics Reports\/} {\bf 528}, 1 (2013).

\bibitem{Jelezko2006SingleReview}
F.~Jelezko, J.~Wrachtrup, {\it Physica Status Solidi (A) Applications and Materials Science\/} {\bf 203}, 3207 (2006).

\bibitem{Ahmed2023DeepChallenges}
S.~F. Ahmed, {\it et~al.\/}, {\it {Deep learning modelling techniques: current progress, applications, advantages, and challenges}\/}, vol.~56 (Springer Netherlands, 2023).

\bibitem{Lecun2015DeepLearning}
Y.~Lecun, Y.~Bengio, G.~Hinton, {\it Nature\/} {\bf 521}, 436 (2015).

\bibitem{sasaki2016broadband}
K.~Sasaki, {\it et~al.\/}, {\it Review of Scientific Instruments\/} {\bf 87} (2016).

\bibitem{cabello2023impact}
K.~Cabello-Solorzano, I.~Ortigosa~de Araujo, M.~Pe{\~n}a, L.~Correia, A.~J.~Tall{\'o}n-Ballesteros, {\it International Conference on Soft Computing Models in Industrial and Environmental Applications\/} (Springer, 2023), pp. 344--353.

\bibitem{singh2020investigating}
D.~Singh, B.~Singh, {\it Applied Soft Computing\/} {\bf 97}, 105524 (2020).

\bibitem{AustralianGeomagnetism}
{Australian Government - Geoscience Australia}, {Australian Government's Geomagnetism}, \url{https://geomagnetism.ga.gov.au/main} (2024). Retrieved November 21, 2024.

\bibitem{Keras2024Layers}
{Keras 3 API documentation / Layers API}, \url{https://keras.io/api/layers/} (2024). Retrieved November 21, 2024.

\bibitem{Keras2024Input}
{Keras 3 API documentation / Layers API / Core Layers / Input Object}, \url{https://keras.io/api/layers/core_layers/input/} (2024). Retrieved November 21, 2024.

\bibitem{Keras2024Conv1Dlayer}
{Keras 3 API documentation / Callbacks API / Convolutional layers / Conv1D layer}, \url{https://keras.io/api/layers/convolution_layers/convolution1d/} (2024). Retrieved November 21, 2024.

\bibitem{Li2022AProspects}
Z.~Li, F.~Liu, W.~Yang, S.~Peng, J.~Zhou, {\it IEEE Transactions on Neural Networks and Learning Systems\/} {\bf 33}, 6999 (2022).

\bibitem{Keras2024LayerActivations}
{Keras 3 API documentation / Layers API / Layer activations}, \url{https://keras.io/api/layers/activations/} (2024). Retrieved November 21, 2024.

\bibitem{Keras2024PoolingLayers}
{Keras 3 API documentation / Layers API / Pooling layers}, \url{https://keras.io/api/layers/pooling_layers/} (2024). Retrieved November 21, 2024.

\bibitem{Keras2024ReshapingLayers}
{Keras 3 API documentation / Layers API / Reshaping layers}, \url{https://keras.io/api/layers/reshaping_layers/flatten/} (2024). Retrieved November 21, 2024.

\bibitem{Keras2024CoreLayers}
{Keras 3 API documentation / Layers API / Core layers / Dense layer}, \url{https://keras.io/api/layers/core_layers/dense/} (2024). Retrieved November 21, 2024.

\bibitem{Keras2024DropoutLayer}
{Keras 3 API documentation / Layers API / Regularization layers / Dropout layer}, \url{https://keras.io/api/layers/regularization_layers/dropout/} (2024). Retrieved November 21, 2024.

\bibitem{Keras2024Losses}
{Keras 3 API documentation / Losses}, \url{https://keras.io/api/losses/} (2024). Retrieved November 21, 2024.

\bibitem{Keras2024Optimizers}
{Keras 3 API documentation / Optimizers}, \url{https://keras.io/api/optimizers/} (2024). Retrieved November 21, 2024.

\bibitem{Keras2024Initializers}
{Keras 3 API documentation / Layers API / Layer weight initializers}, \url{https://keras.io/api/layers/initializers/} (2024). Retrieved November 21, 2024.

\bibitem{Keras2024EarlyStopping}
{Keras 3 API documentation / Callbacks API / EarlyStopping}, \url{https://keras.io/api/callbacks/early_stopping/} (2024). Retrieved November 21, 2024.

\end{thebibliography}
\include{bibliography} 


\pagebreak
\begin{center}
\textbf{\large Supplementary Text for:\\
Machine learning assisted tracking of magnetic objects using quantum diamond magnetometry}
\end{center}

\setcounter{equation}{0}
\setcounter{figure}{0}
\setcounter{table}{0}
\setcounter{page}{1}
\setcounter{section}{0}
\makeatletter
\renewcommand{\theequation}{S\arabic{equation}}
\renewcommand{\thefigure}{S\arabic{figure}}
\renewcommand{\thetable}{S\arabic{table}}
\renewcommand{\thesection}{S\arabic{section}}


This PDF includes:

\begin{itemize}[topsep=0pt,itemsep=-1ex,partopsep=1ex,parsep=0ex]
    \item Supplementary Text, sections \ref{SI_Quantum_NV} to \ref{SI_Accel_model}.
    \item Figure \ref{fig_SI_Magnetometer_Sensitivity}: 
    Magnetic sensitivity of the quantum diamond magnetometer.
    \item Figure \ref{fig_SI_Earth_Dynamics}:
    Earth's magnetic fields.
    \item Figure \ref{fig_SI_Bx_All_measurements}:
    Experimental magnetic signals for the $B_{\rm X}$ component.
    \item Figure \ref{fig_SI_Vector_All_measurements}:
    Complete experimental magnetic measurements.
    \item Figure \ref{fig_SI_Elevator_Position_Speed}:
    Car's position timeseries and speed histogram.
    \item Figure \ref{fig_SI_Noise_trams_timeseries_histograms}:
    Timeseries and histograms for environmental noise.
    \item Figure \ref{fig_SI_Environmental_Noise_Complete}:
    Environmental noise during electrical trams' active and inactive periods.
    \item Figure \ref{fig_SI_Rot_Frames_Example}:
    Position and magnetic signals in different frames.
    \item Figure \ref{fig_SI_Time_Windows}:
    Time windowing process.
    \item Figure \ref{fig_SI_Training_Example}:
    Training example: Early stopping.
    \item Figure \ref{fig_SI_Optim_S3_first}:
    Complete results for optimization stage 3: layers' structure.
    \item Figure \ref{fig_SI_Optim_S3_second}:
    Selected results for optimization stage 3: layers' structure.
    \item Figure \ref{fig_SI_Optim_S4_first}:
    Complete results for optimization stage 4: global hyperparameters.
    \item Figure \ref{fig_SI_Optim_S4_second}:
    Selected results for optimization stage 4: global hyperparameters.
    \item Figure \ref{fig_SI_Optim_S5_first}:
    Complete results for optimization stage 5: fine architecture.
    \item Figure \ref{fig_SI_Optim_S5_second}:
    Selected results for optimization stage 5: fine architecture.
    \item Figure \ref{fig_SI_Example_ML_Architecture}:
    Architectural description of our best performing Machine Learning algorithm.    
    \item Figure \ref{fig_SI_Data_Augmentation}:
    Data augmentation effect on Machine Learning performance.
    \item Figure \ref{fig_SI_park_acc_lvls_tol4m}:
    ML performance for extended position tolerance.
    \item Figure \ref{fig_SI_Acceleration_records}:
    Experimental measurements of the Z-acceleration profile of the elevator.
    \item Figure \ref{fig_SI_Acceleration_Motion_1lvl}:
    Kinematics for the elevator's 1-level flights according to the physical model.
    \item Figure \ref{fig_SI_Optimization_Phys_Model}:
    Optimizing the Machine Learning hyperparameters for different interpolation approaches.
    \item Table \ref{SI_tab_ML_arch}:
    Tuning hyperparameters for each layer within the Machine Learning architecture.
    \item Table \ref{SI_tab_Training_Protocol}:
    Main characteristics of Machine Learning training protocol.
    \item Table \ref{SI_tab_Training_Stages}:
    Training stages for Machine Learning algorithms.
\end{itemize}


\section{Quantum description of the Nitrogen$-$Vacancy defect}
\label{SI_Quantum_NV}

The NV ground state spin Hamiltonian describes how the resonant frequencies determined in an optically detected magnetic resonance (ODMR) spectrum are related to the temperature, external magnetic field, and strain. For each NV orientation, the Hamiltonian can be expressed as:

\begin{equation}
    \nonumber
    \mathcal{H}^{i} = (D+\mathcal{M}_{\rm Z}^{i})(S_{\rm Z}^{i})^{2} + \gamma_{\rm NV} (\vec{B}\cdot \vec{S}^{i})\, ,
\end{equation}

\noindent{with} $i$ denoting each of the four NV orientations. $D\sim2870$ MHz is the zero field splitting and varies with temperature by $dD/dT \sim -74$ kHz/K, and $\mathcal{M}_{\rm Z}$ is the strain and electric field coupling parameter along the Z-axis. Here, the unit vectors are defined in the NV body frame with $\hat{Z}$ parallel to each NV axis. $\gamma_{NV}\sim28.04$ MHz/mT is the NV gyromagnetic ratio, $\vec{B}$ is the external magnetic field, and $\vec{S}$ is the spin-1 operator. It is assumed that the transverse strain and electric field coupling parameters are heavily suppressed by the on-axis component of the external magnetic field and can thus be neglected.


\section*{Magnetic sensitivity}

\begin{figure}[!h]
    \centering
    \includegraphics[width=0.5\linewidth]{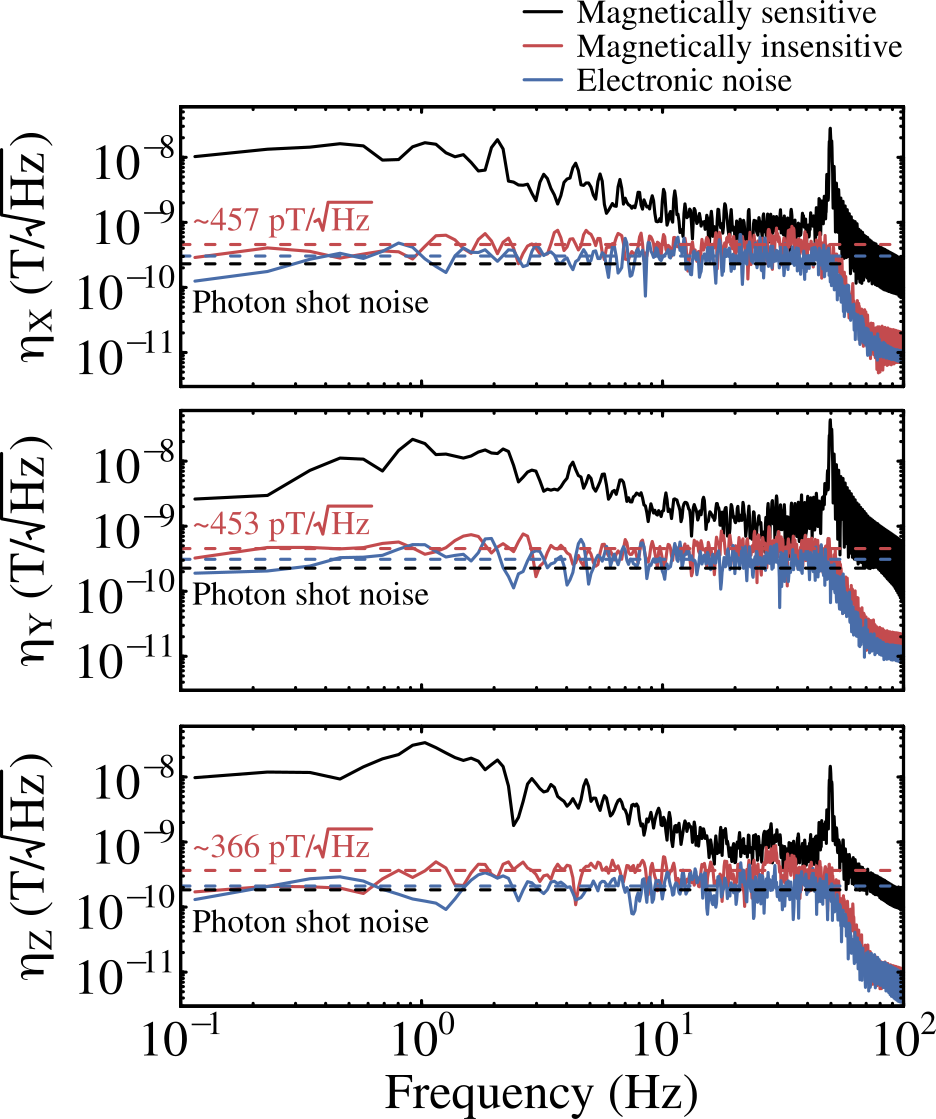}
    \caption{\textbf{Magnetic sensitivity of the quantum diamond magnetometer.}
    Noise Spectral Density for 10~s time traces at the full sampling rate of 202.8 kSa/s, for the magnetic components $(B_{\rm X},B_{\rm Y},B_{\rm Z})$ (from top to bottom, respectively). The magnetically sensitive and insensitive spectra (black and red solid lines, respectively) are associated with the magnetic noise in the laboratory and the inherent noise floor of the quantum diamond magnetometer, respectively. The electronic noise spectrum (blue solid line) accounts for the noise associated with the photodetector, analog-to-digital converter, and lock-in amplifier signal path. The photon shot noise (black dotted horizontal line) indicates the fundamental sensitivity limit of the quantum diamond magnetometer in its current configuration.
    }
    \label{fig_SI_Magnetometer_Sensitivity}
\end{figure}

The magnetic sensitivity, $\eta$, of our quantum diamond magnetometer was evaluated by recording 10~s time traces at the full sampling rate of 202.8 kSa/s without decimation and averaging. An estimate of the noise spectral density (NSD) using the modified Daniell kernel with a weighting coefficient of $m = 2$ was computed for each magnetic field component $(B_{\rm X},B_{\rm Y},B_{\rm Z})$. The resulting NSD spectra are presented in \textcolor{blue}{Figure~\ref{fig_SI_Magnetometer_Sensitivity}}. 

The magnetically sensitive data (black solid line) was measured with the four microwave (MW) tones tuned on resonance and the resulting spectrum represents the magnetic noise in the laboratory. For the magnetically insensitive spectrum (red solid line), the MW tones are tuned off resonance and the spectrum characterizes the inherent noise floor of the quantum diamond magnetometer. Lastly, the electronic noise spectrum (blue solid line) is measured with the MW and laser turned off, accounting for the noise associated with the photodetector, analog-to-digital converter, and lock-in amplifier signal path. The photon shot noise (black dotted horizontal line), calculated separately from the total amount of photocurrent collected based on the method described in \cite{Schloss2018SimultaneousSpins}, indicates the fundamental sensitivity limit of the quantum diamond magnetometer in its current configuration.

By taking the mean of the magnetically insensitive NSD from 0.1~Hz up to 40~Hz before the lowpass filter rolloff imposed by the lock-in amplifier cutoff frequency at 53~Hz, a sensitivity of approximately 450 pT/$\sqrt{\bf{Hz}}$ for the transverse magnetic field direction and 370 pT/$\sqrt{\bf{Hz}}$ for the longitudinal magnetic field direction were computed. These sensitivities are approximately 2~times above the shot noise limit, indicating strong suppression and cancellation of MW and laser-intensity noise.  

 
\section{Earth's magnetic field}
\label{SI_Earth}

In our experimental measurements, the magnetic field as a function of time~$\vec{B}(t)$ sensed by the quantum diamond magnetometer can be described as a superposition of the elevator signal~$\vec{B}_{\rm elevator}(t)$ (with all its magnetic components), the Earth's magnetic field~$\vec{B}_{\rm Earth}(t)$ and the magnetic noise~$\vec{B}_{\rm noise}(t)$ coming from the electrical trams and laboratory environment:
\begin{equation}
\nonumber
    \vec{B}(t) = \vec{B}_{\rm Earth}(t) + \vec{B}_{\rm elevator}(t) + \vec{B}_{\rm noise}(t) \, .
\end{equation}

\begin{figure}[h!]
    \centering
    \includegraphics[width=0.6\linewidth]{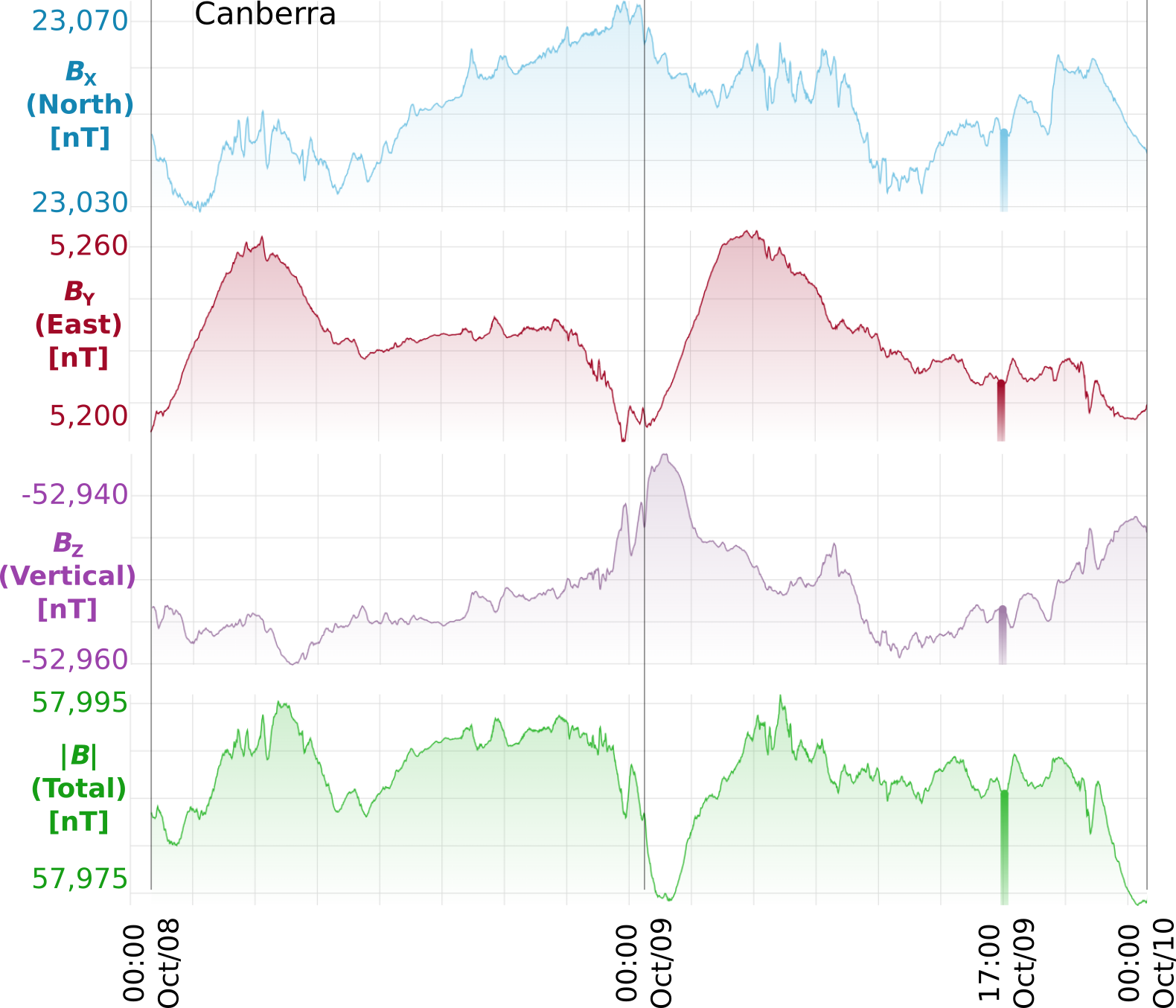}
    \caption{\textbf{Earth's magnetic fields.}
    Magnetic measurements along 48 hours, made by the Australian Government's Geomagnetism department in Canberra~\cite{AustralianGeomagnetism}, the closest station to our laboratory in Melbourne. The scalar field (bottom row) shows an average value very close to 58~$\mu$T, with fluctuations not greater than 20~nT. The other components indicate that magnetic variations can be up to 60~nT along several hours.
    }
    \label{fig_SI_Earth_Dynamics}
\end{figure}

The intensity of these contributions varies significantly: $\vec{B}_{\rm Earth}$ is around 60~$\mu$T, $\vec{B}_{\rm elevator}$ variations are within $\approx 1 \, \mu$T, and $\vec{B}_{\rm noise}$ is in the order of 100$-$200~nT.
Furthermore, their timescales are different: compared to the elevator's traveling times in the order of seconds, the noise fluctuations are much faster, while changes in the Earth's magnetic field are very slow: less than 60~nT over several hours. A magnetic recording from the Australian Government's Geomagnetism department~\cite{AustralianGeomagnetism} in Canberra (the closest station to our laboratory in Melbourne) is shown in \textcolor{blue}{Figure~\ref{fig_SI_Earth_Dynamics}}. The bottom row describes the scalar component of the Earth's magnetic field, with an average value of almost 58~$\mu$T and total fluctuations not greater than 20~nT. For the magnetic projections, instead, variations can be up to 60~nT along several hours.

Since our object monitoring application relies on the magnetic variations of the elevator~$\vec{B}_{\rm elevator}(t)$, our experimental measurements $\vec{B}_{\rm meas}$ are always computed after subtracting a constant value corresponding to the average $\vec{B}_{\rm Earth}$. Then, the experimental signal can be simplified to:
\begin{equation}
\nonumber  
    \vec{B}_{\rm meas}(t) = \vec{B}(t) - [\vec{B}_{\rm Earth}]_{\rm average} 
    \approx \vec{B}_{\rm elevator}(t) + \vec{B}_{\rm noise}(t) \, ,
\end{equation}
\noindent with the small variations from the Earth's magnetic field included in $\vec{B}_{\rm noise}(t)$.


\section{Experimental measurements}
\label{SI_Measurements}

\begin{figure}[!h]
    \centering
    \includegraphics[width=0.75\linewidth]{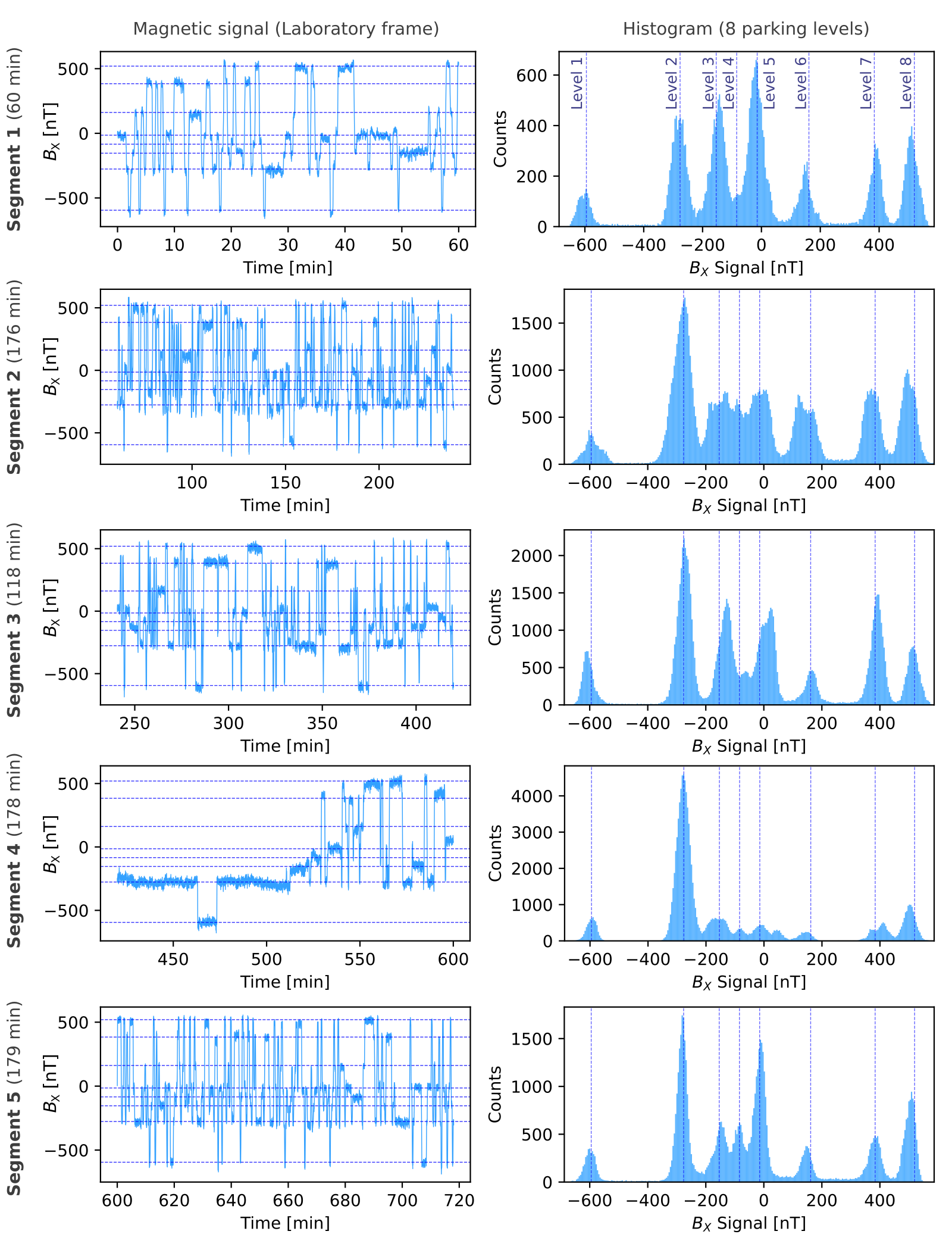}
    \caption{\textbf{Experimental magnetic signals for the $B_{\rm X}$ component.}
    (Left) Magnetic projections along the $X$~axis in the laboratory frame, recorded in 5~consecutive segments. The Earth's magnetic field has already been subtracted. The 8~dashed lines indicate those fields correlated with the parking levels of the elevator's car, in ascending order from bottom to top.
    (Right) Histograms showing that magnetic signals are grouped around the elevator's parking levels (vertical dashed lines).
    }
    \label{fig_SI_Bx_All_measurements}
\end{figure}

The full experimental magnetic dataset recorded by the quantum diamond magnetometer is comprised by 5 consecutive segments, spanning a total acquisition time of 12~hours. The $B_{\rm X}$ projection in the laboratory frame, after subtracting the Earth's constant magnetic field, is presented in \textcolor{blue}{Figure~\ref{fig_SI_Bx_All_measurements}} for all segments, showing a clear correlation with the car's position. The left column depicts the magnetic timeseries, along with the magnetic levels (dashed lines) associated with the elevator's parking positions. The lowest parking position (Level~1) matches the most negative magnetic signal (close to $-600$~nT), while increasing levels are related to increasing magnetic signals.
The histograms for magnetic signals are illustrated on the right column, grouping the signals around 8~peaks correlated with the parking levels.

\begin{figure}[!h]
    \centering
    \includegraphics[width=0.75\linewidth]{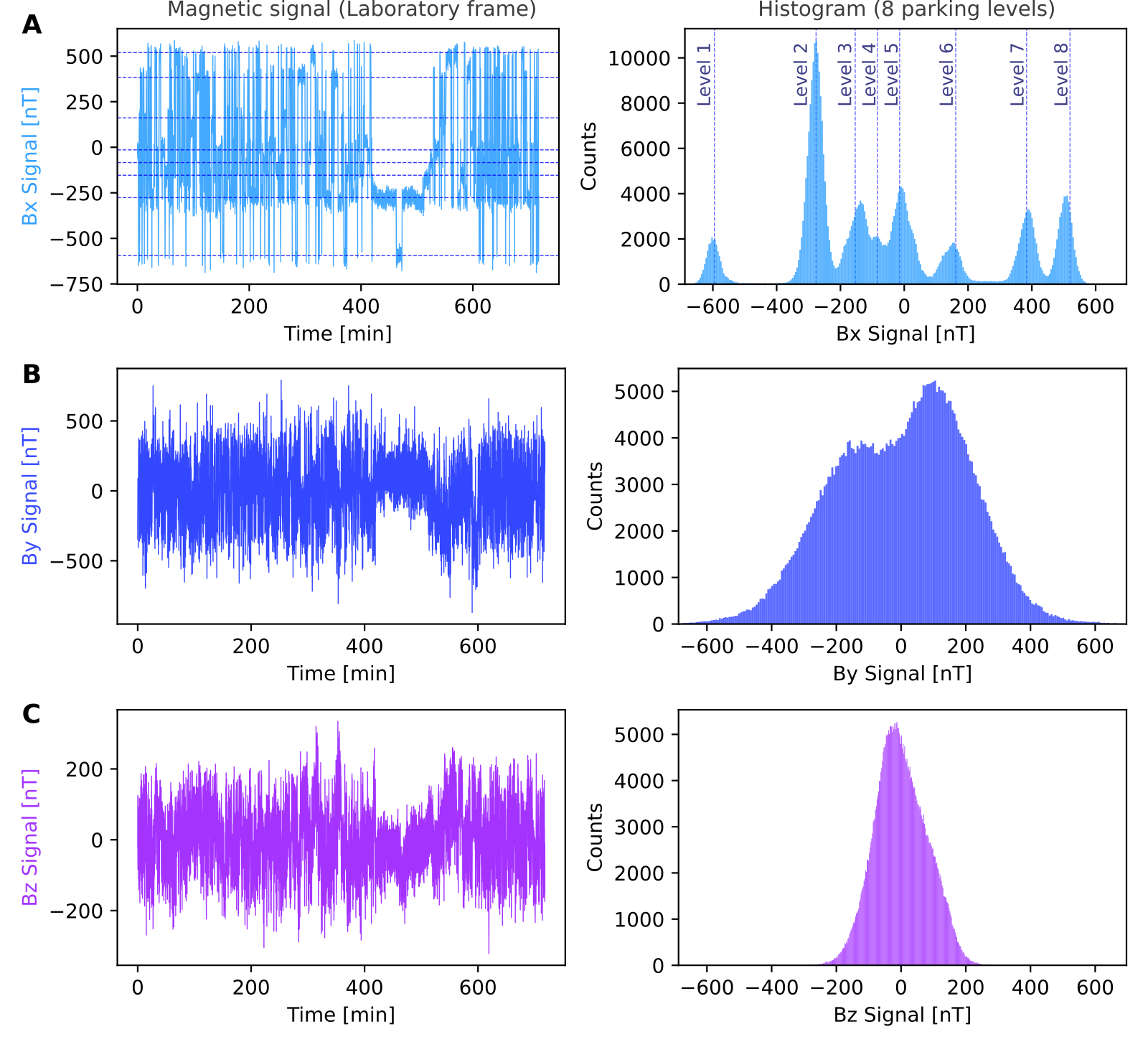}
    \caption{\textbf{Complete experimental magnetic measurements.}
    (Left) Full magnetic timeseries in the laboratory frame, including all recorded segments, for the \textbf{(A)} $B_{\rm X}$, \textbf{(B)} $B_{\rm Y}$ and \textbf{(C)} $B_{\rm Z}$ components. The Earth's magnetic field has already been subtracted. The 8~dashed lines in the $B_{\rm X}$ plot indicate those fields correlated with the parking levels of the elevator's car, in ascending order from bottom to top.
    (Right) Histograms showing a clear correlation between the $B_{\rm X}$ projection and the elevator's parking positions (vertical dashed lines). The other magnetic components are strongly coupled to the environmental noise and no clear associations with the elevator can be made.
    }
    \label{fig_SI_Vector_All_measurements}
\end{figure}

Because of the strong environmental noise, which mainly affects the $YZ$ plane in the laboratory frame, no clear correlations can be found between the $(B_{\rm Y}(t),B_{\rm Z}(t))$ components and the car's position~$Z(t)$. The full timeseries (combining all 5 segments) and histograms for all magnetic projections are shown in \textcolor{blue}{Figure~\ref{fig_SI_Vector_All_measurements}}.

\begin{figure}[h!]
    \centering
    \includegraphics[width=0.85\linewidth]{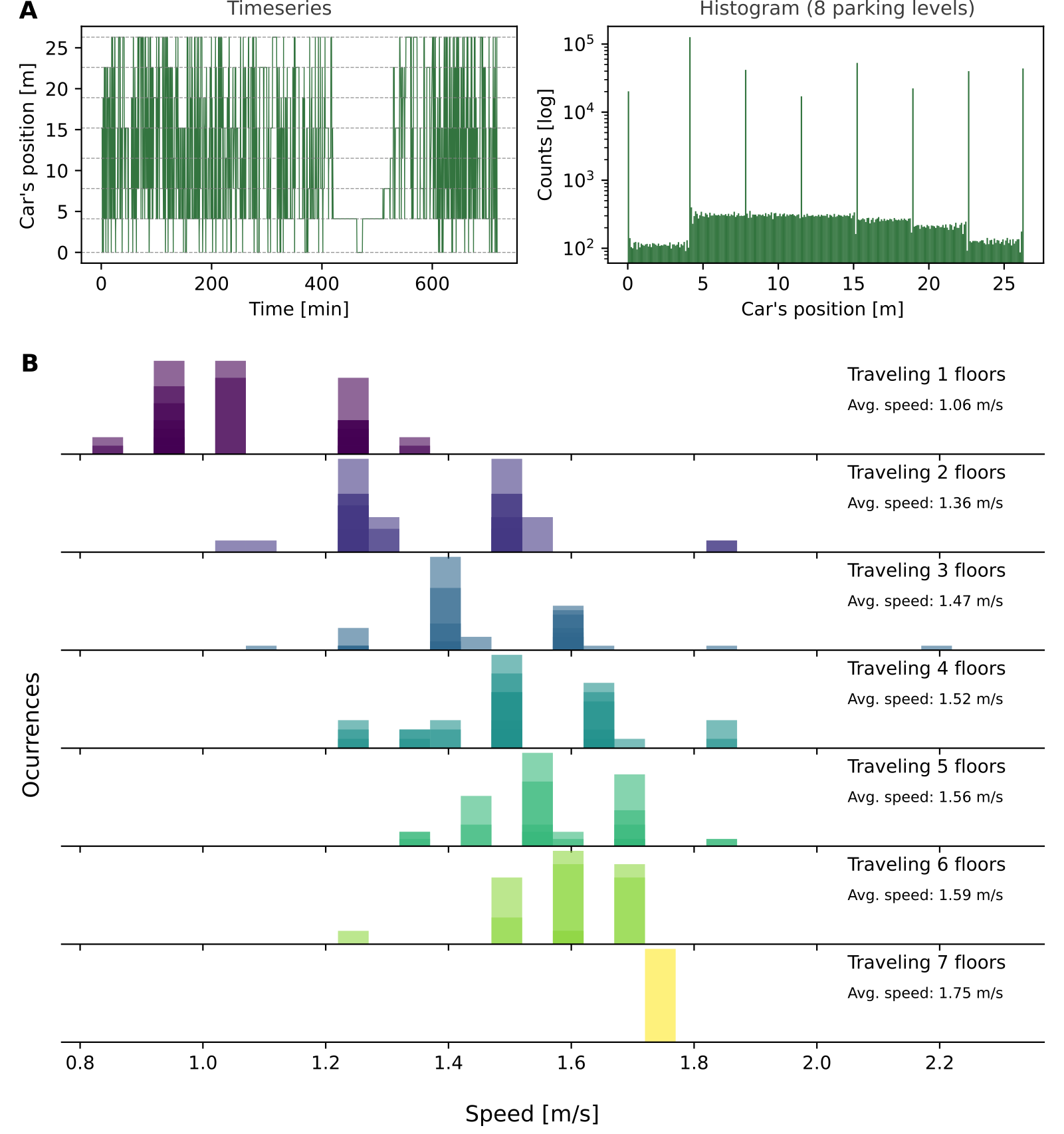}
    \caption{\textbf{Car's position timeseries and speed histogram.}
    \textbf{(A)} (Left) Complete position measurements merged into a single timeseries, dashed lines indicate the parking levels. (Right) Position histogram in logarithmic scale, showing that the elevator spends most of the time at the parking levels.
    \textbf{(B)} Speed histogram grouped by the number of traveling floors, according to the linear interpolation approximation. Within each group, the color gradient is associated to the different initial-final destinations.
    }
    \label{fig_SI_Elevator_Position_Speed}
\end{figure}

The experimental dataset for the car's position is recorded with a camera pointing at the display of the elevator's door (outside), which allows us to certainly determine the parking intervals. The traveling events are interpolated using linear approximations, which are not exact but give a good estimation about the trajectory (we will discuss later a more sophisticated approach). The entire dataset, correlated with the 5 magnetic segments, is shown as a single timeseries in \textcolor{blue}{Figure~\ref{fig_SI_Elevator_Position_Speed}A}, along with the position histogram (logarithmic scale). The similarity with the $B_{\rm X}$ timeseries (\textcolor{blue}{Figure~\ref{fig_SI_Vector_All_measurements}A} left) in the laboratory frame is notorious. However, the position histogram is very sharp compared to the $B_{\rm X}$ histogram (affected by noise fluctuations), and it informs that the elevator spends most of the time at the parking levels.

On the other hand, the car's average traveling speed (according to the linear interpolation approximation) can be calculated from its timeseries, as shown in \textcolor{blue}{Figure~\ref{fig_SI_Elevator_Position_Speed}B}. The plot groups the speeds according to the number of floors covered in each trip, and the color gradient within each category is associated with the initial-final destinations (e.g.\ traveling from Level~1 to Level~2 has a slightly different color than traveling from Level~4 to Level~3). The wide spread in speeds is mainly due to the time uncertainty when identifying the traveling events from the videos. However, we can identify an average speed of about 1.4~m/s, meaning that it takes approximately 3~seconds to travel between two consecutive levels.


\section{Environmental magnetic noise}
\label{SI_Environmental_Noise}

\begin{figure}[!h]
    \centering
    \includegraphics[width=0.75\linewidth]{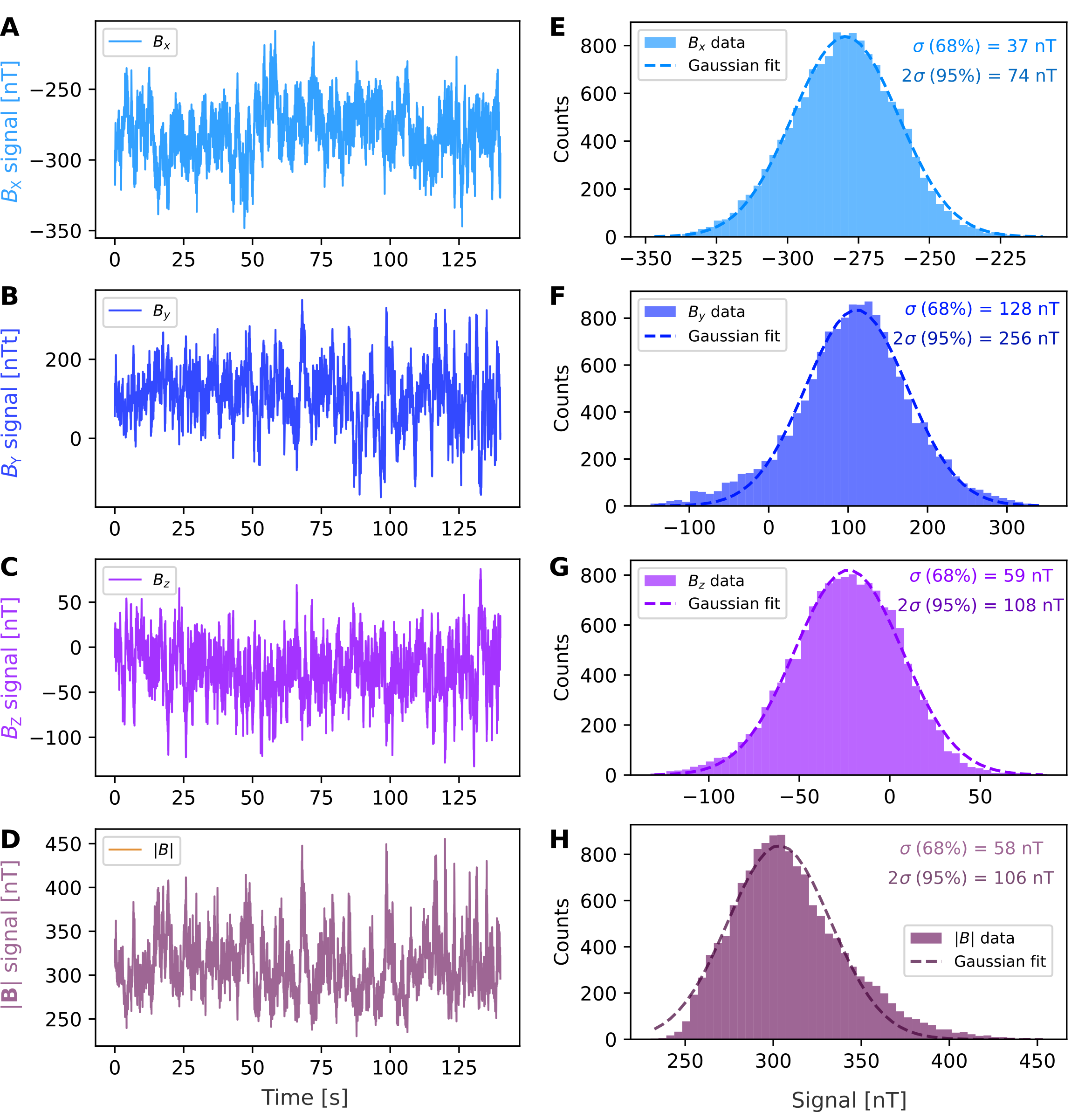}
    \caption{\textbf{Timeseries and histograms for environmental noise.}
    These examples are taken while the electrical trams are active and the elevator is parked at Level~2. 
    \textbf{(A-D)} Timeseries for $(X,Y,Z)$ magnetic projections in the laboratory frame and the scalar magnitude.
    \textbf{(E-H)} Respective histograms including Gaussian fittings and describing one- and two-sigma standard deviations, which include 68\% and 95\% of the values, respectively.
    }
    \label{fig_SI_Noise_trams_timeseries_histograms}
\end{figure}

The environmental magnetic noise in our monitoring application is partly affected by nearby electronic equipment and wiring, but the main contribution comes from distant electrical trams. As our laboratory is close to the tram tracks ($\approx$50~m), which are frequently used by many urban lines, the magnetic noise exhibits an unpredictable pattern with variations in the order of 60~nT for the scalar magnitude, although each magnetic component behaves differently. \textcolor{blue}{Figure~\ref{fig_SI_Noise_trams_timeseries_histograms}(A-D)} shows the noise pattern for a short timeseries while the elevator is parked at Level~2, for each magnetic Cartesian projection in the laboratory frame and the scalar magnitude. The histograms for those patterns are illustrated in \textcolor{blue}{Figure~\ref{fig_SI_Noise_trams_timeseries_histograms}(E-H)}, along with Gaussian fittings and one- and two-sigma standard deviations, which include 68\% and 95\% of the values, respectively.

\begin{figure}[!h]
    \centering
    \includegraphics[width=0.85\linewidth]{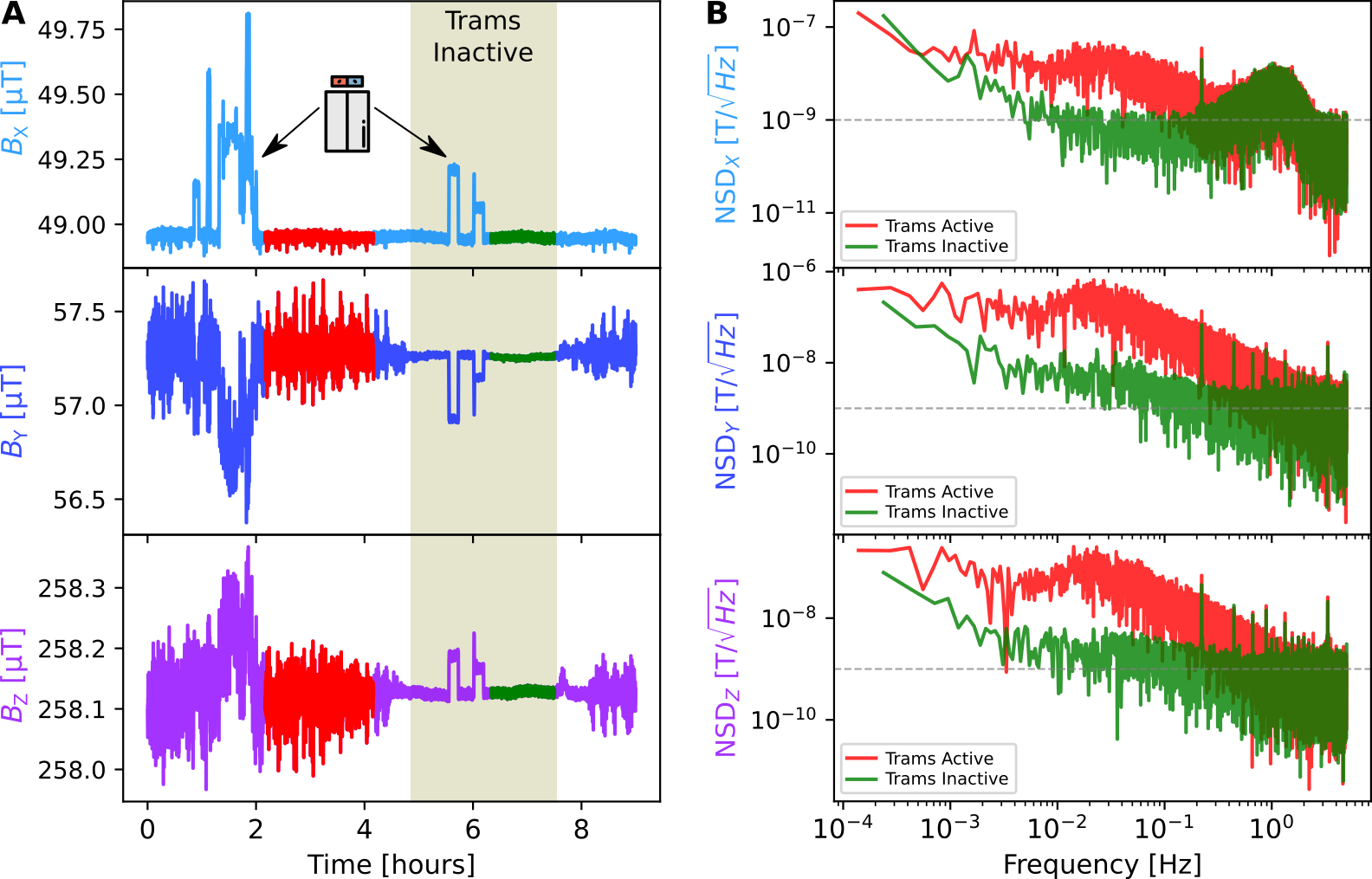}
    \caption{\textbf{Environmental noise during electrical trams' active and inactive periods.}
    \textbf{(A)} Magnetic measurements including active and inactive periods of tram activity, the latter happening overnight for about 3 hours (shaded area). In this dataset, the elevator is only operating during two intervals (depicted by arrows). Some data has been color-coded in red (active period) and green (inactive period), in intervals where the elevator is steady.
    \textbf{(B)} Noise Spectral Density (NSD) for the color-coded intervals. In all magnetic components, the noise roughly follows a $1/f$ behavior. 
    }
    \label{fig_SI_Environmental_Noise_Complete}
\end{figure}

On the other hand, clear evidence of the noise source can be observed in the magnetic measurements along both active and inactive periods of trams activity, which only cease to operate overnight for about 3~hours, as shown in \textcolor{blue}{Figure~\ref{fig_SI_Environmental_Noise_Complete}A} (the Earth's magnetic field has not been subtracted in these measurements). Furthermore, the trams' noise is strongly directional, primarily affecting the $YZ$ plane: the $B_{\rm X}$ component remains relatively clean at all times, while the $(B_{\rm Y},B_{\rm Z})$ projections are only as clean as $B_{\rm X}$ during the inactive period, but much noisier at other times. 
The Noise Spectral Density (NDS) for each magnetic component is analyzed in \textcolor{blue}{Figure~\ref{fig_SI_Environmental_Noise_Complete}B}, for both active and inactive trams periods. In all cases, the noise follows a general $1/f$ behavior (inverse of frequency).

\section{Frames for magnetic data}
\label{SI_RF}

In our experimental measurements, we record magnetic projections along the four NV axes and transform them into Cartesian components in the laboratory frame $(B_{\rm X},B_{\rm Y},B_{\rm Z})$, in which the $Z$-axis is parallel to the elevator's axis and the $XY$ plane is squared with the room. Any other frame, with Cartesian axes $\Tilde{X},\Tilde{Y},\Tilde{Z}$, can be obtained using a unitary transformation matrix~$\mathcal{R}$:

\begin{equation}
\nonumber
    [\mathcal{R}(\vec{n},\alpha)] \, (B_{\rm X},B_{\rm Y},B_{\rm Z}) =
    (B_{\rm \Tilde{X}},B_{\rm \Tilde{Y}},B_{\rm \Tilde{Z}})\,,
\end{equation}

\noindent where the rotated variables are represented by~$(B_{\rm \Tilde{X}},B_{\rm \Tilde{Y}},B_{\rm \Tilde{Z}})$. The rotation operation~$\mathcal{R}(\vec{n},\alpha)$ is defined by a unitary rotating axis~$\vec{n}=(n_{\rm X},n_{\rm Y},n_{\rm Z})$ described in the laboratory frame and a rotating angle~$\alpha$:
\begin{equation}
\nonumber   
    \mathcal{R}(\vec{n},\alpha) = 
    \begin{bmatrix}
    u+n_{\rm X}^2 (1-u) &
    n_{\rm X} n_{\rm Y} (1-u)-n_{\rm Z} v &
    n_{\rm X} n_{\rm Z} (1-u)+n_{\rm Y} v \\
    n_{\rm Y} n_{\rm X} (1-u)+n_{\rm Z} v &
    u+n_{\rm Y}^2 (1-u) &
    n_{\rm Y} n_{\rm Z} (1-u)-n_{\rm X} v \\
    n_{\rm Z} n_{\rm X} (1-u)-n_{\rm Y} v &
    n_{\rm X} n_{\rm Y} (1-u)+n_{\rm X}  &
    u+n_{\rm Z}^2 (1-u) \\ 
    \end{bmatrix} \, , 
\end{equation}
\begin{equation}
\nonumber   
    u = \cos\alpha \; , \; v = \sin\alpha \, .
\end{equation}

When going from the laboratory frame to any other by using the rotational operation, we identify the rotated frame~$\Tilde{X}\Tilde{Y}\Tilde{Z}$ as~$F(\theta,\varphi,\alpha)$, representing the polar~$\theta$ and azimuth~$\varphi$ coordinates of the $\Tilde{Z}$ axis, seen from the laboratory frame system, along with the rotating angle~$\alpha$.

\begin{figure}[h!]
    \centering
    \includegraphics[width=0.9\linewidth]{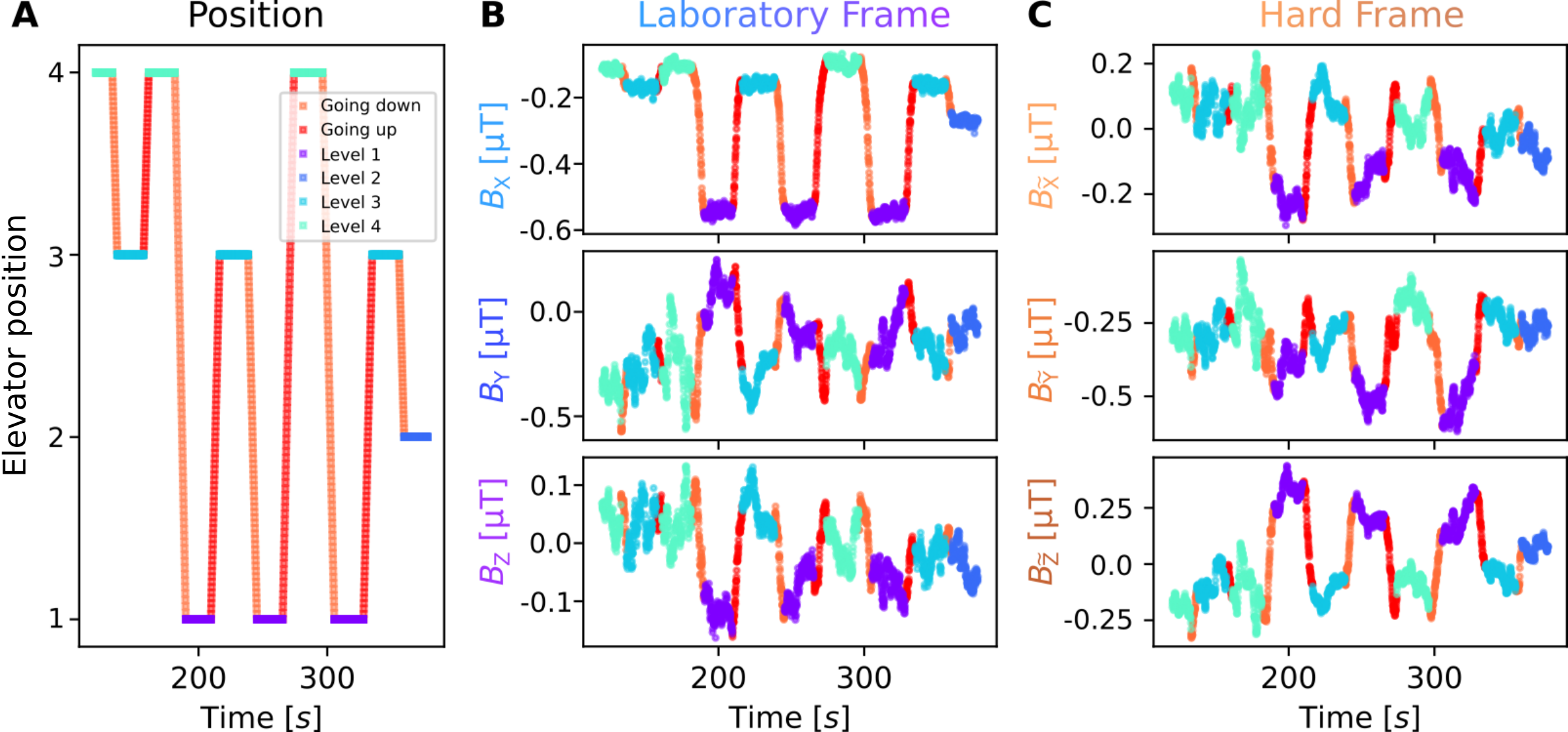}
    \caption{\textbf{Elevator's position and magnetic signals in different frames.}
    \textbf{(A)} Short segment showing the elevator's position as a function of time, color-coded according to the parking level or traveling direction.
    \textbf{(B)} Correlated magnetic measurements in the laboratory frame $XYZ$. 
    \textbf{(C)} Same magnetic measurements, but transformed to the ``hard'' frame $\Tilde{X}\Tilde{Y}\Tilde{Z}$, in which the environmental noise is evenly distributed across all magnetic components.
    }
    \label{fig_SI_Rot_Frames_Example}
\end{figure}

According to the noise distribution in the magnetic axes, we define three distinct frames that represent qualitatively different scenarios:
(1)~the laboratory frame, equivalent to~$F(0^{\circ},0^{\circ},0^{\circ})$, where the noise is strongly coupled to $YZ$ and very weak in $X$;
(2)~an ``intermediate'' frame $F(67^{\circ},59^{\circ},120^{\circ})$, where noise is strongly coupled to $X$ and weak in $YZ$; and
(3)~a ``hard'' frame $F(58^{\circ},61^{\circ},90^{\circ})$, where the noise is evenly distributed across all magnetic projections $XYZ$.
A comparison between the laboratory and hard frames, correlated with the car's parking levels and traveling direction, is illustrated in \textcolor{blue}{Figure~\ref{fig_SI_Rot_Frames_Example}}.

\section{Time windows}
\label{SI_Time_Windows}

\begin{figure}[h!]
    \centering
    \includegraphics[width=0.75\linewidth]{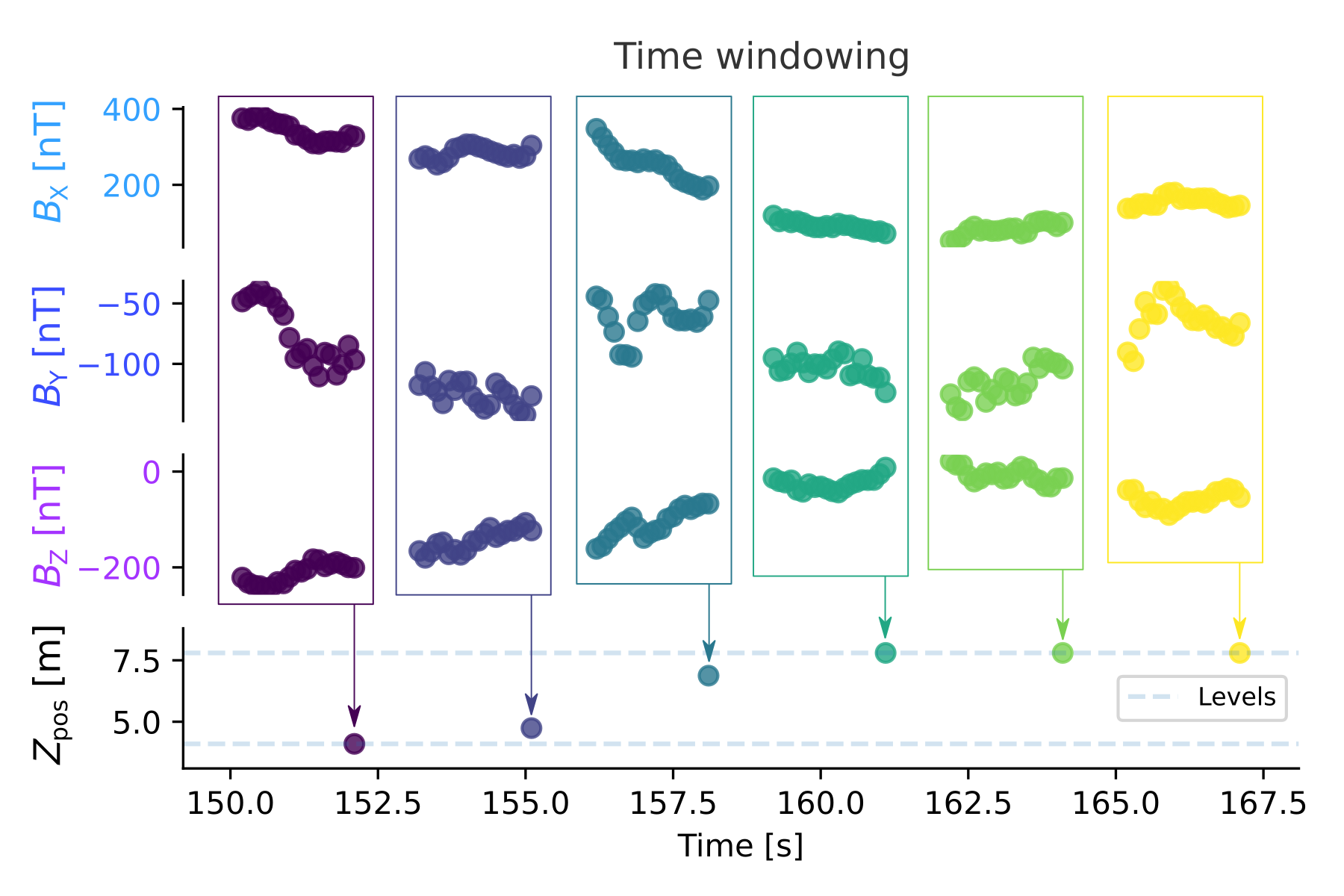}
    \caption{\textbf{Time windowing process.}
    Example of 3-second time windows (boxes), applied to magnetic data and associated to the elevator's $Z$-position (bottom row). Each time window is a matrix with three rows representing the channels $(B_{\rm X},B_{\rm Y},B_{\rm Z})$, and 30~columns with the magnetic field values (time resolution is 0.1~s).
    }
    \label{fig_SI_Time_Windows}
\end{figure}

The time windowing process described in the main text is illustrated in \textcolor{blue}{Figure~\ref{fig_SI_Time_Windows}}, for magnetic data in the hard frame. For this example, we have chosen 3-second intervals~$\Delta t$, meaning that a time window stores 30 points (time resolution is 0.1~s) in each magnetic component, in a $3 \times 30$ matrix format. Since time windows are predictors in the supervised Machine Learning (ML) algorithm, they are associated to the target elevator's position~$Z_{\rm pos}$ during the training process. This is represented by the arrows that connect the time windows' boxes with the single values $Z_{\rm pos}$.

As an important remark, notice that each time window is an independent unit which carries magnetic information, but it does not include the explicit time vector. The time correlations are implicitly given by the order of the points in each magnetic projection, and the only requirement for our object monitoring application is that the time resolution is kept constant along all experiments.

 
\section{Training and Testing datasets}
\label{SI_Datasets}

For ML training and evaluation purposes, the complete measurements dataset (12~hours, see \textcolor{blue}{Figure~\ref{fig_SI_Vector_All_measurements}}) is equally split, using segments $[1, 4, 5]$ for training (5.90~hours) and $[2, 3]$ for testing (5.95~hours). During the training process, the training dataset is further split in two subsets: 75\% for effective training and 25\% for validation. This splitting occurs after the time windowing procedure, using a random selection: since each pair predictor-target is an independent unit, time correlation among different units is irrelevant.

 
\section{Machine Learning architecture}
\label{SI_MLarch}

\subsection{General description}

The architecture for all ML algorithms trained in this work is organized as a series of layers~\cite{Keras2024Layers}, which can be single instances or sets, in which many instances and variations of the same type of layers are combined. Some of the layers are essential, such as the input and output layers, but some of them are optional in our scheme. The general structure is the following:

\begin{enumerate}[topsep=0pt,itemsep=-1ex,partopsep=1ex,parsep=0ex] 

    \item \textbf{Input layer} (predictors)
    
    \item \textbf{Convolutional set} [optional]
    \begin{enumerate}[topsep=0pt,itemsep=-1ex,partopsep=1ex,parsep=0ex]
        \item 1D-Convolutional layer
        \item Non-linear Activation layer
        \item Pooling layer [optional]
    \end{enumerate}

    \item \textbf{1D-Conversion layer}

    \item \textbf{Dense set}
    \begin{enumerate}[topsep=0pt,itemsep=-1ex,partopsep=1ex,parsep=0ex]
        \item Dense layer
        \item Non-linear Activation layer
        \item Dropout layer [Optional]
    \end{enumerate}
    
    \item \textbf{Dense layer with single neuron} + Linear Activation layer

    \item \textbf{Output layer} (target)
    
\end{enumerate}

The Input layer's role~\cite{Keras2024Input} is to allocate the input data, and it has a shape equal to a single time window predictor: $N_{\rm channels} \times N_{\rm points}$. This information is first processed by the convolutional set, although in some architectures this step is skipped. Each convolutional layer~\cite{Keras2024Conv1Dlayer} is described by the number of filters~$N_{\rm f}$ and the kernel size~$k$, and its function is to extract patterns by correlating neighboring points within the input data~\cite{Li2022AProspects}. Each filter is a one-dimensional matrix of size~$k$ with randomly assigned values, which convolves the information from the different channels and returns a reduced series with length $(N_{\rm points}-k+1)$. As there are many filters in a single convolutional layer, the original input information is expanded to $N_{\rm f}$ dimensions (one for each filter), with a final output shape~$N_{\rm f} \times (N_{\rm points}-k+1)$. Immediately after, the convolutional output is transformed by a non-linear activation layer~\cite{Keras2024LayerActivations} which applies a mapping function, such as the hyperbolic tangent (tanh) or rectified linear unit (ReLU), keeping the original dimensions. Additionally, the resulting series may be convolved by a Pooling layer~\cite{Keras2024PoolingLayers}, which applies local transformations such as taking the average or maximum value, effectively reducing the size of the series.

The 1D-Conversion is a simple but crucial step to transition from the Convolutional set to the Dense set. It converts the multi-dimensional information from the last Convolutional layer (or directly from the Input layer if the convolutional set is skipped) into a one-dimensional (1D) array, either by rearranging the elements without altering the input values (``flattening")~\cite{Keras2024ReshapingLayers}, or by first reducing each dimension to its average value (``global'', similar to a Pooling layer~\cite{Keras2024PoolingLayers}) and then rearranging the elements.

Within the Dense set, a Dense layer~\cite{Keras2024CoreLayers} is defined by the number of neurons~$n$, each of them having as many internal parameters as the size of the input series (weights) plus one (bias). The output of the layer is a 1D array with length~$n$, which is immediately transformed by a non-activation function. Additionally, a Dropout layer~\cite{Keras2024DropoutLayer} can randomly ``turn off" some of the outputs, meaning that they do not contribute to the next layer.

In addition to the layers' structure, the ML algorithm has global hyperparameters that define the learning process. The loss function~\cite{Keras2024Losses}, such as the Mean Absolute Error, computes the difference between the predictions and the target in each training cycle, assigning a score that the algorithm seeks to minimize with each training cycle. In order to improve future predictions, the internal parameters are tuned with a back-propagation algorithm~\cite{Lecun2015DeepLearning}, and the magnitude of these adjustments are regulated by the global optimizer and learning rate~\cite{Keras2024Optimizers}.

Within the local and global hyperparameters in the Machine Learning architecture, we explore those which we consider critical, as listed in \textcolor{blue}{Table~\ref{SI_tab_ML_arch}}, along with the options that we will analyze in the optimization stages.

\begin{table}[h!]\footnotesize
    \caption{\textbf{Hyperparameter options for the  Machine Learning architecture.}
    Meanings: ReLU, ELU and tanh mean Rectified Linear Unit, Exponential Linear Unit and Hyperbolic Tangent, respectively; Adam is the Adaptive Moment Estimation optimizer, while Adadelta and Adamax are variations of the former.}
    \label{SI_tab_ML_arch}
    \begin{center} \begin{tabular}{ccc}
        \hline \hline
        \textbf{Layer} & \textbf{Hyperparameter} & \textbf{Options} \\
        \\
        \multirow{2}{*}{Input} & Channels & 1, 2, 3 \\
        & Time window's points & From 1 to 124 \\
        \cmidrule{1-1}
        \multirow{2}{*}{1D-Convolutional} & Kernel size & From 4 to 16 \\
        & Filters & 16, 32 \\
        \cmidrule{1-1}
        Non-linear activation & Activation function & ReLU, ELU, tanh \\
        \cmidrule{1-1}
        Pooling & Kernel size & 0, 2 \\
        \cmidrule{1-1}
        1D-Conversion & Averaging function & Global, Flattening \\
        \cmidrule{1-1}
        Dense & Number of neurons & From 128 to 2048 \\
        \cmidrule{1-1}
        Dropout & Dropping fraction & 0, 0.2, 0.4 \\
        \hline \hline
        \textbf{Global} & \textbf{Hyperparameter} & \textbf{Options} \\
        \\
        Optimizer & Function & Adam, Adadelta, Adamax \\
        \cmidrule{1-1}
        Learning rate & Rate & From $10^{-12}$ to $10^{1}$ \\
        \hline \hline
    \end{tabular} \end{center}
\end{table}

 
\subsection{Training protocol}
\label{SI_Training_Protocol}

We use the same training protocol for any ML algorithm, with small variations for the best performing architecture (final training), see \textcolor{blue}{Table~\ref{SI_tab_Training_Protocol}}. As the main characteristics, we include: initialization method (for internal parameters)~\cite{Keras2024Initializers}, loss function; validation fraction; batch size (number of samples that are propagated through the network at a time); maximum number of epochs (training cycles); and the early stop criterion~\cite{Keras2024EarlyStopping}, which includes a triggering condition within a patience period, and a minimum number of epochs for training. 

\begin{table}[h!]\footnotesize
    \caption{\textbf{Main characteristics of the training protocol.}
    Training parameters for every training process (``General''), with a few modification for the ``Final'', best performing ML architecture.
    Meanings: MAE is Mean Absolute Error, GlorotUniform is a Tensorflow method for uniform random initialization.}
    \label{SI_tab_Training_Protocol}
    \begin{center} \begin{tabular}{ccc}
        \hline \hline
        \textbf{Training parameter} & \textbf{General} & \textbf{Final} \\
        \\
        Initialization & GlorotUniform & GlorotUniform \\
        \cmidrule{1-1}
        Loss function & MAE & MAE \\
        \cmidrule{1-1}
        Validation fraction & 25\% & 25\% \\
        \cmidrule{1-1}
        Batch size & 512 & 256 \\
        \cmidrule{1-1}
        Max. Epochs & 200 & 400 \\
        \cmidrule{1-1}
        Early Stop trigger & No improvement in MAE & No improvement in MAE \\
        \cmidrule{1-1}
        Early Stop patience & 15 epochs & 30 epochs \\
        \cmidrule{1-1}
        Early Stop minimum & 30 epochs & 60 epochs \\
        \hline \hline
    \end{tabular} \end{center}
\end{table}

\begin{figure}
    \centering
    \includegraphics[width=0.8\linewidth]{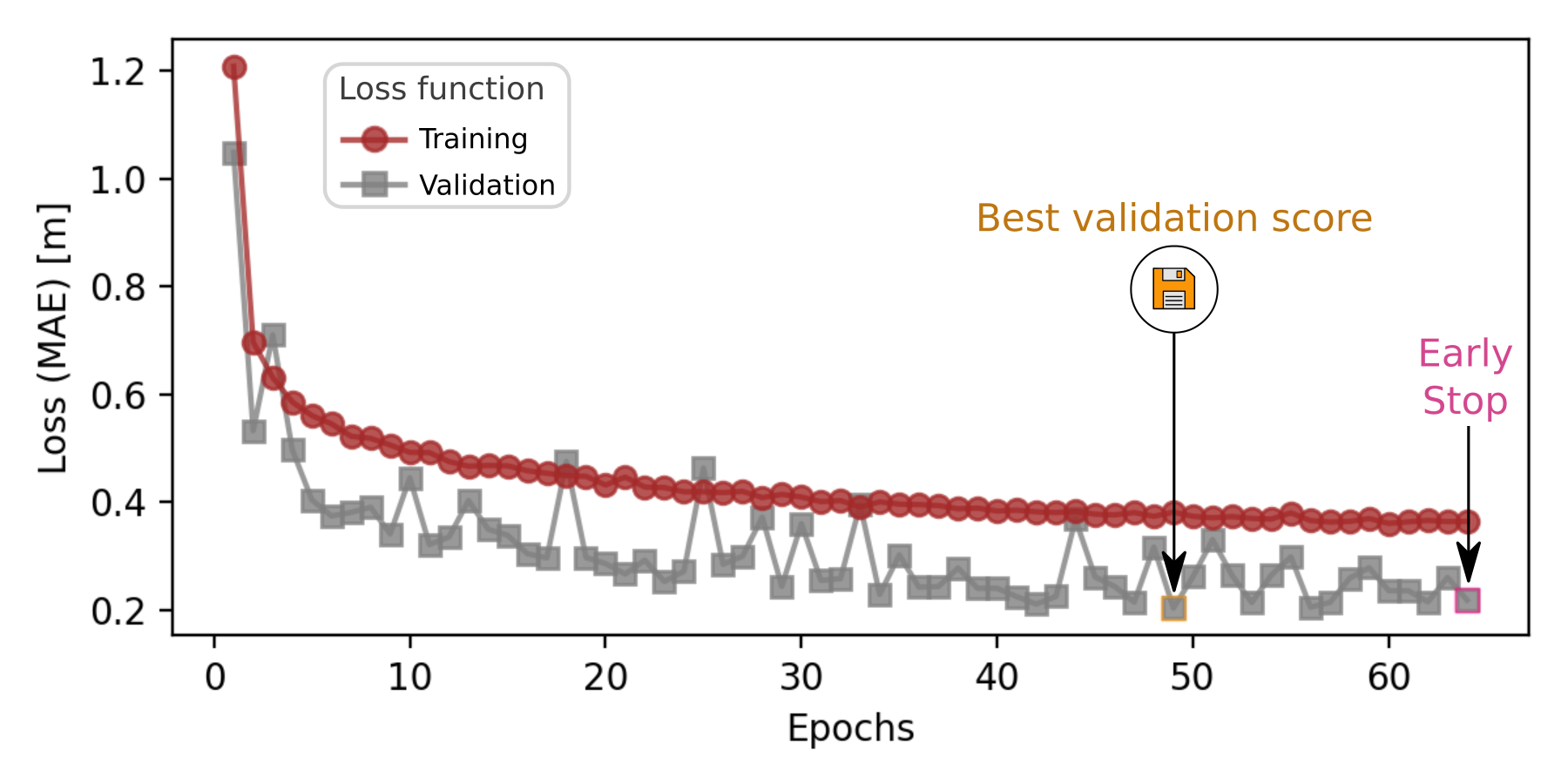}
    \caption{\textbf{Training example: Early stopping.}
    The Machine Learning algorithm is trained on the training dataset along a number of cycles (epochs), up to a maximum number of~200. At the end of each epoch, the predictions are compared with the ground truth and evaluated using the Mean Absolute Error loss function, in both the training (red circles) and validation (gray squares) datasets. If there is no improvement in the validation score (meaning lower values) during the last 15~epochs, the training process is stopped and the algorithm's internal parameters are restored to the best configuration (epoch~49).
    }
    \label{fig_SI_Training_Example}
\end{figure}

The Early Stopping condition is fundamental to prevent overfitting. In principle, the ML algorithm can be trained during the maximum number of epochs, but an early stop can be triggered at any epoch if the Loss function, evaluated on the validation dataset, has not improved (meaning the value has not decreased) in the last $N$~epochs (``Early Stop patience''), as exemplified in \textcolor{blue}{Figure~\ref{fig_SI_Training_Example}}. In this case, the algorithm's internal parameters are restored to the best configuration. There is a special rule at the beginning of the training, the ``Early Stop minimum'' condition, in which the Early Stopping can only occur after the first $M$~epochs.

\section{Optimization stages}
\label{SI_Optimization_Stages}

The Machine Learning architecture proposed in the main text is the result of a sequence of optimization stages, which analyze one or a few hyperparameters at a time. While in the main text we describe three of these stages, the complete process includes 6~stages, as summarized in \textcolor{blue}{Table~\ref{SI_tab_Training_Stages}}.

\begin{table}[h!]\footnotesize
    \caption{\textbf{Training stages for Machine Learning algorithms.}
    Meanings: ``1D-Conv" is one-dimensional Convolutional layer; ``$k$" is kernel size; ``$+$" means a connection between two consecutive layers; ``$n$" is number of neurons; ``AF" is Activation Function; ``tanh" is Hyperbolic tangent; ``Adam" is Adaptive Moment Estimation optimizer; ``$f$" is filters.}
    \label{SI_tab_Training_Stages}
    \begin{center} \begin{tabular}{ccc}
        \hline \hline
        \textbf{Name} & \textbf{Focus} & \textbf{Best configuration} \\
        \\
        Stage 1 & Input magnetic components & $(B_{\rm X},B_{\rm Y},B_{\rm Z})$ \\
        \cmidrule{1-1}
        Stage 2 & Time window's length & 40 points (4~s) \\
        \cmidrule{1-1}
        \multirow{3}{*}{Stage 3} & {1D-Conv kernels} & $(k=8) + (k=4)$ \\
        & Pooling layers & No Pooling layers \\
        & Dense neurons & $(n=1024) + (n=512)$ \\
        \cmidrule{1-1}
        \multirow{3}{*}{Stage 4} & Non-linear AF & tanh \\
        & Optimizer & Adam \\
        & Learning rate & 0.0001 \\
        \cmidrule{1-1}
        \multirow{3}{*}{Stage 5} & 1D-Conv filters & $(f=16) + (f=16)$ \\
        & 1D-Conversion Averaging & Flatten \\
        & Dropout fraction & 0 (no dropout) \\
        \hline \hline
    \end{tabular} \end{center}
\end{table}

In every stage, we have evaluated several options for ML architectures, using the frames $F(0^{\circ},0^{\circ},0^{\circ})$, $F(67^{\circ},59^{\circ},120^{\circ})$ and $F(58^{\circ},61^{\circ},90^{\circ})$, and 3 different random initialization seeds (0, 1 and 2). Therefore, for each ML architecture there are 9 independent tracking accuracy results, computed with a 1-meter tolerance.
In the following, we provide the full description of all stages, including the analyzed hyperparameter options and the architectures considered in each case.


\subsection{Stage 1: input magnetic components}

In this stage, we focus on the combinations of input magnetic components, using our original full-vector magnetic field~$\vec{B}$ measurements. We study the following options:

\begin{itemize}[topsep=0pt,itemsep=-1ex,partopsep=1ex,parsep=0ex]
    \item Scalar field: $|\vec{B}|$.
    \item Single magnetic component: $B_{\rm X}$, $B_{\rm Y}$ and $B_{\rm Z}$.
    \item Two magnetic components: $(B_{\rm X},B_{\rm Y})$, $(B_{\rm X},B_{\rm Z})$ and $(B_{\rm Y},B_{\rm Z})$.
    \item Full-vector: $(B_{\rm X},B_{\rm Y},B_{\rm Z})$.
\end{itemize}

The following hyperparameters are fixed, common to all ML architectures:

\begin{itemize}[topsep=0pt,itemsep=-1ex,partopsep=1ex,parsep=0ex]
    \item Time window points: 20.
    \item Non-linear activation functions: ReLU.
    \item 1D-conversion layer: flattening type.
    \item Optimizer: Adam.
    \item Learning rate: 0.005.
    \item No Pooling nor Dropout layers.
\end{itemize}

We use two different ML architectures, with the following structure:

\begin{itemize}[topsep=0pt,itemsep=-1ex,partopsep=1ex,parsep=0ex]
    \item \textbf{Architecture 1}: The Convolutional set includes a single 1D-Convolutional layer with 32~filters and kernel size~5; the Dense set includes two Dense layers with 512 neurons each.
    \item \textbf{Architecture 2}: No convolutional set; the Dense set includes three Dense layers with 1024, 512 and 512 neurons, in that order.
\end{itemize}

Refer to \textcolor{blue}{Figure~4A} in the main text to see the results. From this stage, we choose to select the full-vector magnetic field $(B_{\rm X},B_{\rm Y},B_{\rm Z})$ as the optimal combination of input components.


\subsection{Stage 2: time window's length}

In this stage, we focus on the time window's length~$\Delta t$. As our measurement's time resolution is $dt=$~0.1~s, the number of points contained in a time window is $N_{\rm points} = \Delta t / dt$. We study the following $N_{\rm points}$ options:

\begin{itemize}[topsep=0pt,itemsep=-1ex,partopsep=1ex,parsep=0ex]
    \item From 1 to 9 (inclusive), in steps of 1 point.
    \item From 10 to 18 (inclusive, in steps of 2 points.
    \item From 20 to 56 (inclusive), in steps of 4 points.
    \item From 60 to 124 (inclusive), in steps of 8 points.
\end{itemize}

The following hyperparameters are fixed, common to all ML architectures:

\begin{itemize}[topsep=0pt,itemsep=-1ex,partopsep=1ex,parsep=0ex]
    \item Input magnetic components: $(B_{\rm X},B_{\rm Y},B_{\rm Z})$.
    \item Non-linear activation functions: ReLU.
    \item 1D-conversion layer: flattening type.
    \item Optimizer: Adam.
    \item Learning rate: 0.005.
    \item No Pooling nor Dropout layers.
\end{itemize}

We use three different ML architectures, with the following structure:

\begin{itemize}[topsep=0pt,itemsep=-1ex,partopsep=1ex,parsep=0ex]
    \item \textbf{Architecture 1}: The Convolutional set includes a single 1D-Convolutional layer with 32~filters and kernel size~5; the Dense set includes two Dense layers with 512 neurons each. This architecture is only applicable to $N_{\rm points}\geq5$.
    \item \textbf{Architecture 2}: The Convolutional set includes two 1D-Convolutional layers with 32~filters each and kernel sizes~8 and 4, in that order; the Dense set includes two Dense layers with 512 neurons each. This architecture is only applicable to $N_{\rm points}\geq12$.
    \item \textbf{Architecture 3}: No Convolutional set; the Dense set includes three Dense layers with 1024, 512 and 512 neurons, in that order. This architecture applies for all $N_{\rm points}$ options.
\end{itemize}

Refer to \textcolor{blue}{Figure~4B} in the main text to see the results. From this stage, we select $N_{\rm points}=40$ (equivalent to $\Delta t=4$~s) as the optimal time window's length.


\subsection{Stage 3: layers' structure}
\label{SI_Optim_subsection_S3}

In this stage, we focus on the layers' structure within the ML architecture, meaning the number of Convolutional and Dense layers, along with their local hyperparameters, and the type of 1D-Conversion layer. We describe the options grouped by different sets of layers, which are later combined in all possible ways.

Possibilities for the Convolutional set:

\begin{itemize}[topsep=0pt,itemsep=-1ex,partopsep=1ex,parsep=0ex]
    \item A single 1D-Convolutional layer, with 32~filters and kernel size~5, followed by a Pooling layer of size~2.
    \item A single 1D-Convolutional layer, with 32~filters and kernel size~5, no Pooling layer.
    \item Two 1D-Convolutional layers with 32~filters each and kernel sizes~8 and 4, in that order, each one followed by a Pooling layer of size~2.
    \item Two 1D-Convolutional layers with 32~filters each and kernel sizes~8 and 4, in that order, no Pooling layers.
    \item Two 1D-Convolutional layers with 32~filters each and kernel sizes~16 and 4, in that order, each one followed by a Pooling layer of size~2.
    \item Two 1D-Convolutional layers with 32~filters each and kernel sizes~16 and 4, in that order, no Pooling layers.
\end{itemize}

Possibilities for the 1D-Conversion layer:

\begin{itemize}[topsep=0pt,itemsep=-1ex,partopsep=1ex,parsep=0ex]
    \item Flattening type (just rearranges the multi-dimensional input into a single vector).
    \item Global type (first obtains the average value for each input dimension, then makes a single vector with all values).
\end{itemize}

Possibilities for the Dense set:

\begin{itemize}[topsep=0pt,itemsep=-1ex,partopsep=1ex,parsep=0ex]
    \item A single Dense layer with 256 neurons.
    \item A single Dense layer with 1024 neurons.
    \item Two Dense layers with 256 and 128 neurons, in that order.
    \item Two Dense layers with 1024 and 512 neurons, in that order.
    \item Three Dense layers with 2048, 512 and 128 neurons, in that order.
\end{itemize}

Combining all the previous options, the number of possible ML architectures is $6 \times 2 \times 5 = 60$. The following hyperparameters are fixed, common to all ML architectures:

\begin{itemize}[topsep=0pt,itemsep=-1ex,partopsep=1ex,parsep=0ex]
    \item Input magnetic components: $(B_{\rm X},B_{\rm Y},B_{\rm Z})$.
    \item Time window points: 40.
    \item Non-linear activation functions: ReLU.
    \item Optimizer: Adam.
    \item Learning rate: 0.005.
    \item No Dropout layers.
\end{itemize}

\begin{figure}[!h]
    \centering
    \includegraphics[width=0.55\linewidth]{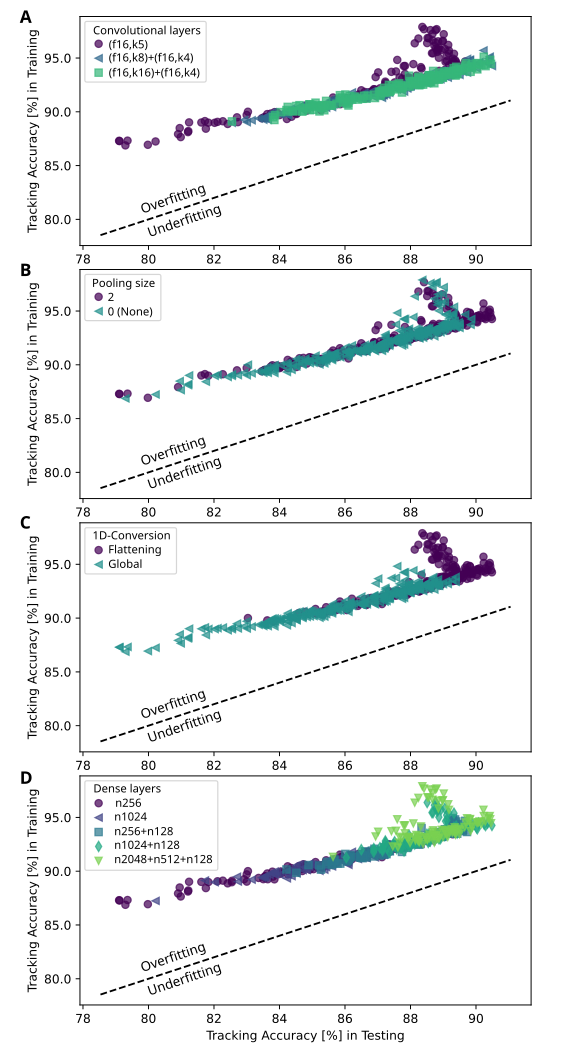}
    \caption{\textbf{Complete results for optimization stage 3: layers' structure.}
    Tracking accuracy results for each ML algorithm, comparing the scores obtained using the validation (vertical axis) and testing (horizontal axis) datasets. The overfitting and underfitting regions are separated by a dashed line. The same results are analyzed according to different groups:
    (\textbf{A}) Convolutional layers, identified by their filters~$f$ and kernel size~$k$;
    (\textbf{B}) Pooling layer size;
    (\textbf{C}) type of 1D-Conversion layer; and
    (\textbf{D}) Dense layers, identified by their number of neurons~$n$.
    }
    \label{fig_SI_Optim_S3_first}
\end{figure}

The results of this stage are shown in \textcolor{blue}{Figure~\ref{fig_SI_Optim_S3_first}}, for each ML algorithm trained with a particular initialization seed and frame. The plot compares the tracking accuracy obtained with the validation (vertical axis) and testing (horizontal axis) datasets. The results are grouped by the options for \textcolor{blue}{(A)}~Convolutional layers, \textcolor{blue}{(B)}~Pooling layers, \textcolor{blue}{(C)}~1D-Conversion layer and \textcolor{blue}{(D)}~Dense layers. The overfitting and underfitting regions are separated by a dashed line.

As we want to get the best evaluation results, the most promising architectures are those that score on the right end of the plot, preferably closer to the identity training line (i.e.\ no overfitting nor underfitting). Analyzing each group, we can generally conclude that:

\begin{itemize}[topsep=0pt,itemsep=-1ex,partopsep=1ex,parsep=0ex]
    \item Two Convolutional layers work better than a single one.
    \item The ML algorithms work better without Pooling layers.
    \item The flattening 1D-conversion layer works best.
    \item The Dense set is more efficient with higher number of Dense layers and neurons.
\end{itemize}

\begin{figure}[!h]
    \centering
    \includegraphics[width=0.75\linewidth]{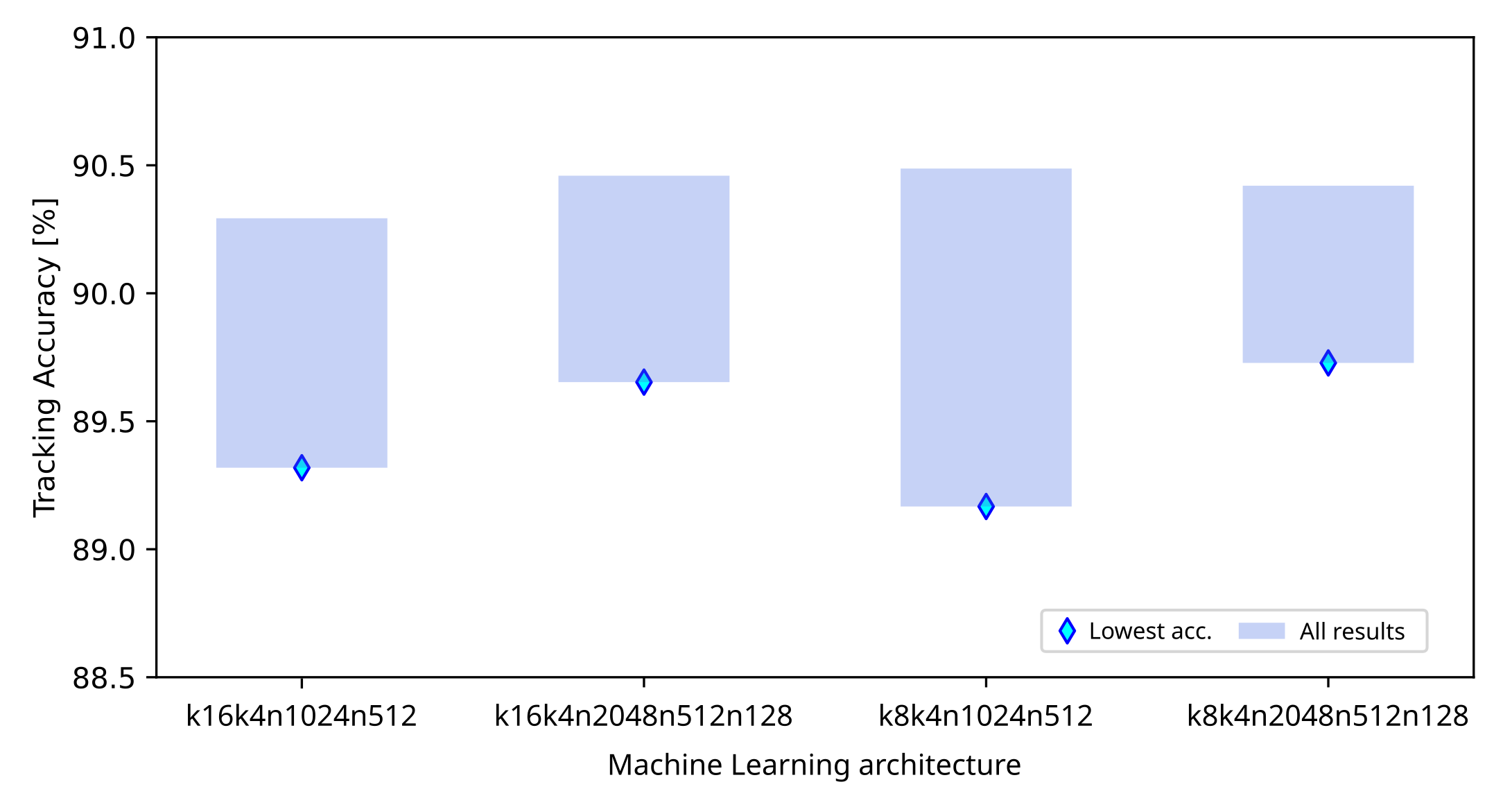}
    \caption{\textbf{Selected results for optimization stage 3: layers' structure.}
    Tracking accuracy for different ML architectures, representing the Convolutional layers with a letter ``k", followed by the kernel size and the Dense layers with a letter ``n" followed by the number of neurons. While the complete range of results for each architecture (shaded bars) includes several frames for the magnetic data and random initialization seeds, only the lowest results (blue diamonds) count as the final score.
    }
    \label{fig_SI_Optim_S3_second}
\end{figure}

Using these criteria, we select a few architectures for the final comparison, all of them with a flattening 1D-Conversion layer and no Pooling layers. We label them by representing the Convolutional layers with a letter ``k", followed by the number of filters, and the Dense layers with a letter ``n", followed by the number of neurons. As in the main text description, we summarize the results in \textcolor{blue}{Figure~\ref{fig_SI_Optim_S3_second}} by using the minimum tracking accuracy~$A_{\rm track, min}$, which combines all the results in different frames and random initialization seeds for each ML architecture.

The results indicate that the performance $A_{\rm track, min}$ is similar for all the selected architectures, between 89\% and 90\%, although architectures with a more complex Dense set are slightly more accurate. However, we also consider the computational complexity of the ML algorithm to make a choice, favoring simpler architectures, and in this case we select the simplest option ``k8k4n1024n512", with 8 and 4 kernel sizes (Convolutional set) followed by 1024 and 512 neurons (Dense set).


\subsection{Stage 4: global hyperparameters}

In this stage, we analyze the global optimizer, learning rate and the type of non-linear activation function common to all layers within the ML architecture. We describe the options for each of these hyperparameters, which are later combined in all possible ways.

Possibilities for the global optimizers:

\begin{itemize}[topsep=0pt,itemsep=-1ex,partopsep=1ex,parsep=0ex]
    \item Adaptative Moment Estimator (Adam).
    \item Adadelta.
    \item Adamax.
\end{itemize}

Possibilities for the non-linear activation functions:

\begin{itemize}[topsep=0pt,itemsep=-1ex,partopsep=1ex,parsep=0ex]
    \item Rectified linear unit (relu).
    \item Exponential linear unit (elu).
    \item Hyperbolic tangent (tanh).
\end{itemize}

Possibilities for the learning rate:

\begin{itemize}[topsep=0pt,itemsep=-1ex,partopsep=1ex,parsep=0ex]
    \item $1 \times 10^{-5}$ to $1 \times 10^{-2}$, in logarithmic decade steps, for Adadelta and Adamax optimizers.
    \item $1 \times 10^{-12}$ to $1 \times 10^{1}$, in logarithmic decade steps, for Adam optimizer (which works better, as we will explain later).
\end{itemize}

Combining all the previous options, the number of possible ML architectures is $3 \times 2 \times 4 + 3 \times 1 \times 14 = 66$. The following hyperparameters are fixed, common to all ML architectures:

\begin{itemize}[topsep=0pt,itemsep=-1ex,partopsep=1ex,parsep=0ex]
    \item Input magnetic components: $(B_{\rm X},B_{\rm Y},B_{\rm Z})$.
    \item Time window points: 40.
    \item A convolutional set including two 1D-convolutional layers with 32~filters each and kernel sizes 8 and 4, in that order.
    \item A 1D-Conversion flattening layer.
    \item A Dense set with including two Dense layers with 1024 and 512 neurons, in that order.
    \item No Dropout layers.
\end{itemize}

\begin{figure}[!h]
    \centering
    \includegraphics[width=0.55\linewidth]{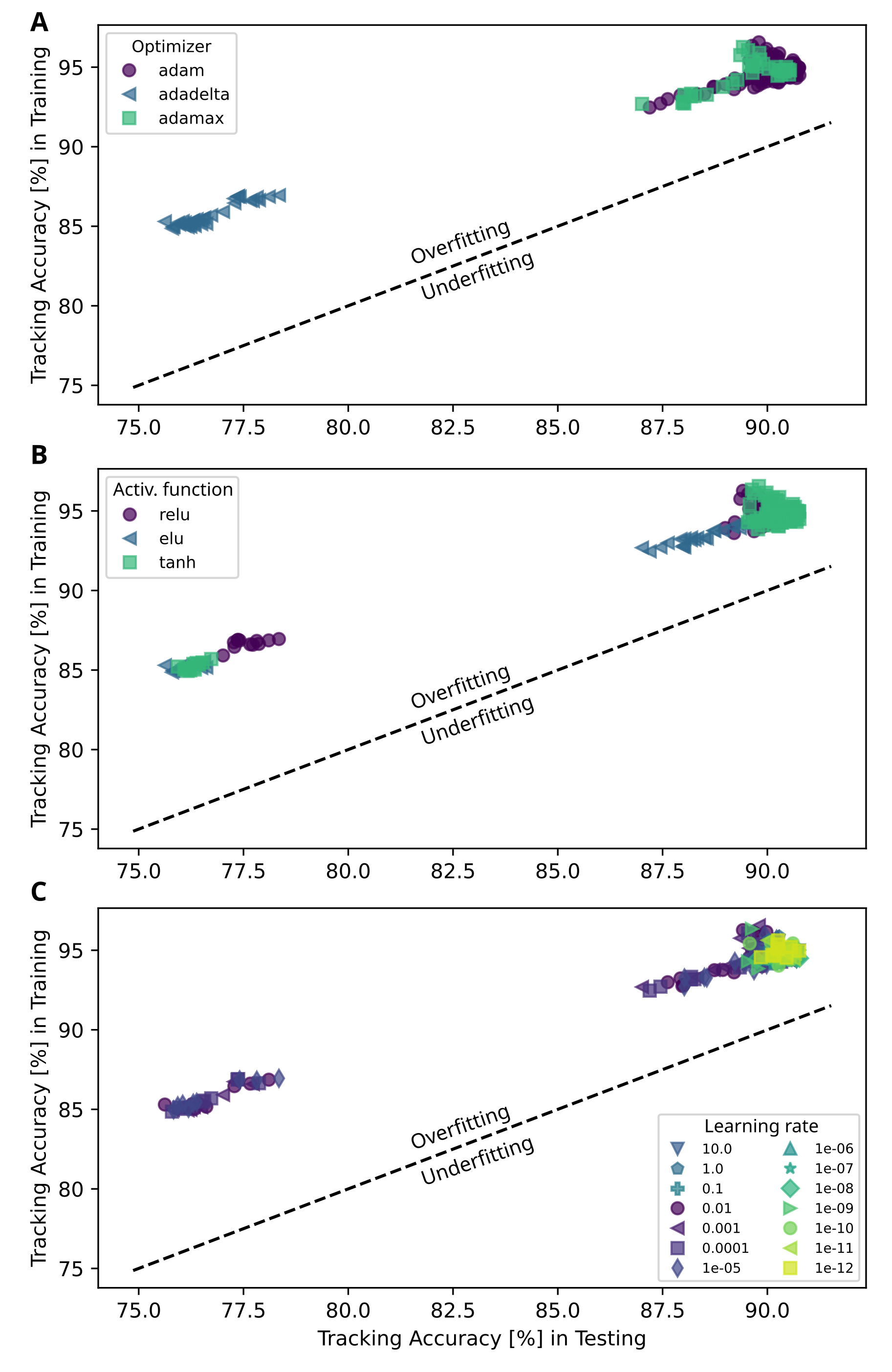}
    \caption{\textbf{Complete results for optimization stage 4: global hyperparameters.}
    Tracking accuracy results for each ML algorithm, comparing the scores obtained using the validation (vertical axis) and testing (horizontal axis) datasets. The overfitting and underfitting regions are separated by a dashed line. The same results are analyzed according to different groups:
    (\textbf{A}) optimizers;
    (\textbf{B}) activation functions; and
    (\textbf{C}) learning rates.
    }
    \label{fig_SI_Optim_S4_first}
\end{figure}

The results of this stage are shown in \textcolor{blue}{Figure~\ref{fig_SI_Optim_S4_first}}, comparing the validation and testing accuracies (explained in subsection \ref{SI_Optim_subsection_S3}). The results are grouped by the options for \textcolor{blue}{(A)}~optimizers, \textcolor{blue}{(B)}~activation functions and \textcolor{blue}{(C)}~learning rates.

The ML performances are divided into two main groups: relatively low performance (left side of the plot), associated to the Adadelta optimizer, and high performance (right side of the plot), for every other option (see \textcolor{blue}{Figure~\ref{fig_SI_Optim_S4_first}A}). If we focus our attention on the last group, we can conclude that:

\begin{itemize}[topsep=0pt,itemsep=-1ex,partopsep=1ex,parsep=0ex]
    \item The Adam optimizer is slightly better than Adamax (\textcolor{blue}{Figure~\ref{fig_SI_Optim_S4_first}A}).
    \item The tanh activation function works best (\textcolor{blue}{Figure~\ref{fig_SI_Optim_S4_first}B}).
    \item The learning rates need further analysis. (\textcolor{blue}{Figure~\ref{fig_SI_Optim_S4_first}C}).
\end{itemize}

\begin{figure}[!h]
    \centering
    \includegraphics[width=0.75\linewidth]{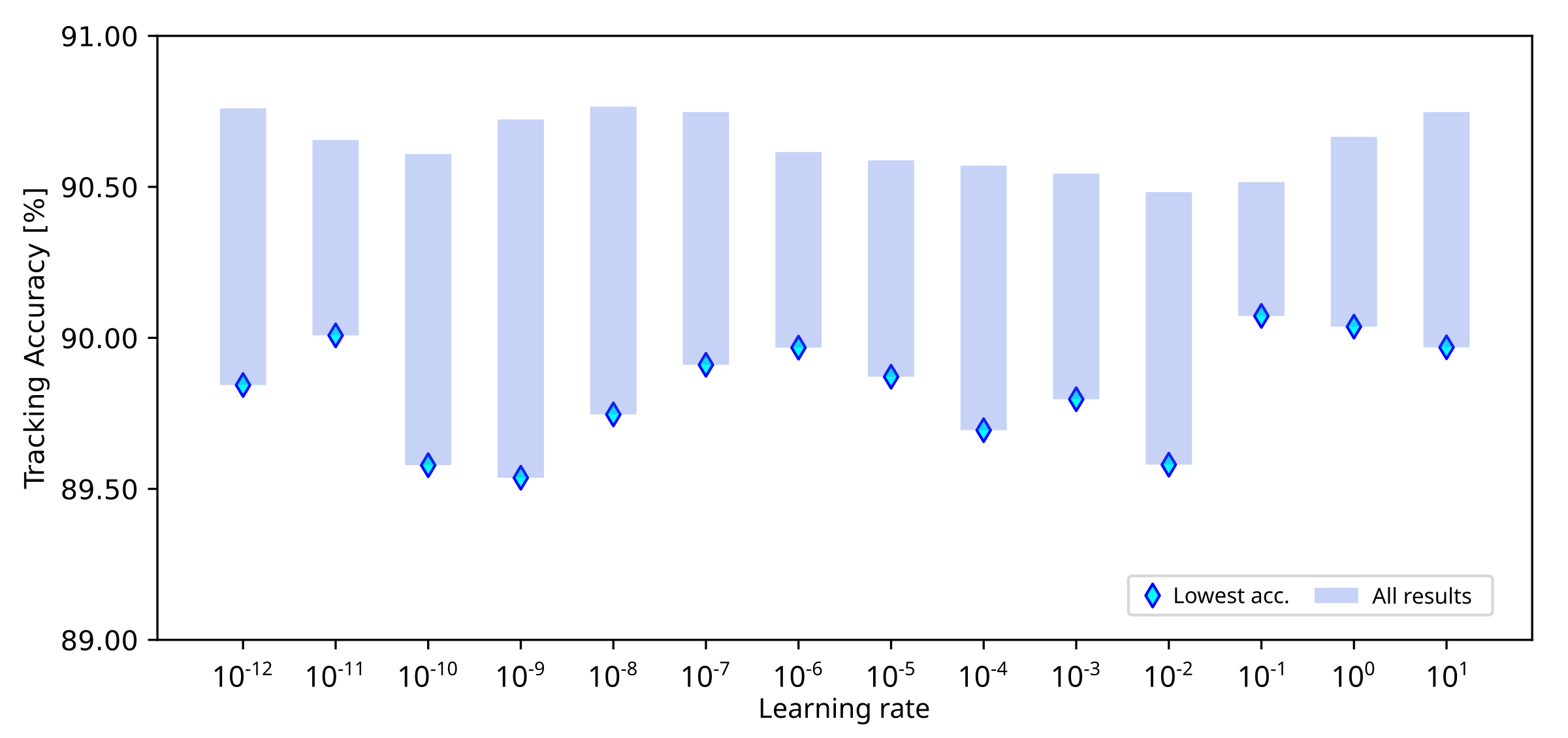}
    \caption{\textbf{Selected results for optimization stage 4: global hyperparameters.}
    Tracking accuracy for different learning rate options. While the complete range of results for each architecture (shaded bars) include several frames for the magnetic data and random initialization seeds, only the lowest results (blue diamonds) count as the final score.
    }
    \label{fig_SI_Optim_S4_second}
\end{figure}

Using these criteria, we select those architectures with the Adam optimizer and tanh activation function for the final comparison, and study the learning rate options, as shown in \textcolor{blue}{Figure~\ref{fig_SI_Optim_S4_second}}.

While the learning rate is normally a critical parameter for the training performance, in this case the algorithm appears capable of optimizing its internal parameters at any rate, with variations below 1\% for $A_{\rm track, min}$. Although the analysis suggests favoring a learning rate in the higher or lower extremes, we prefer to choose the intermediate value of 10$^{-5}$, which is closer to the standard values for the Adam optimizer ($\approx 0.001$). In this way, we avoid training the algorithm too slow or too fast, which may hinder the convergence to the global minimum of the loss function.


\subsection{Stage 5: Fine architecture (local hyperparameters)}

In this stage, we analyze the fine ML architecture: the number of filters in each 1D-Convolutional layer and the Dropout fraction after each Dense layer. We describe the options for each of these hyperparameters, which are later combined in all possible ways.

The number of filters for the first and second Convolutional layers are represented in the format $[f_1,f_2]$, respectively, while the kernel sizes (8 and 4, respectively) remain unchanged. These are the filter possibilities that we consider:

\begin{itemize}[topsep=0pt,itemsep=-1ex,partopsep=1ex,parsep=0ex]
    \item $[16,16]$.
    \item $[32,16]$.
    \item $[16,32]$.
    \item $[32,32]$.
\end{itemize}

Regarding the Dropout layers and their fractions~$f_{\rm drop}$, so far we have skipped them in the previous architectures, meaning that $f_{\rm drop}=0$. We now consider the following options, common to all Dropout layers:

\begin{itemize}[topsep=0pt,itemsep=-1ex,partopsep=1ex,parsep=0ex]
    \item $f_{\rm drop}=0$.
    \item $f_{\rm drop}=0.2$.
    \item $f_{\rm drop}=0.4$.
\end{itemize}

Combining all the previous options, the number of possible ML architectures is $4 \times 3 = 12$. The following hyperparameters are fixed, common to all ML architectures:

\begin{itemize}[topsep=0pt,itemsep=-1ex,partopsep=1ex,parsep=0ex]
    \item Input magnetic components: $(B_{\rm X},B_{\rm Y},B_{\rm Z})$.
    \item Time window points: 40.
    \item A convolutional set including two 1D-convolutional layers with kernel sizes 8 and 4, in that order.
    \item A 1D-Conversion flattening layer.
    \item A Dense set with including Dense layers with 1024 and 512 neurons, in that order.
    \item Non-linear activation functions: tanh.
    \item Optimizer: Adam.
    \item Learning rate: $1 \times 10^{-5}$.
\end{itemize}

\begin{figure}[!h]
    \centering
    \includegraphics[width=0.55\linewidth]{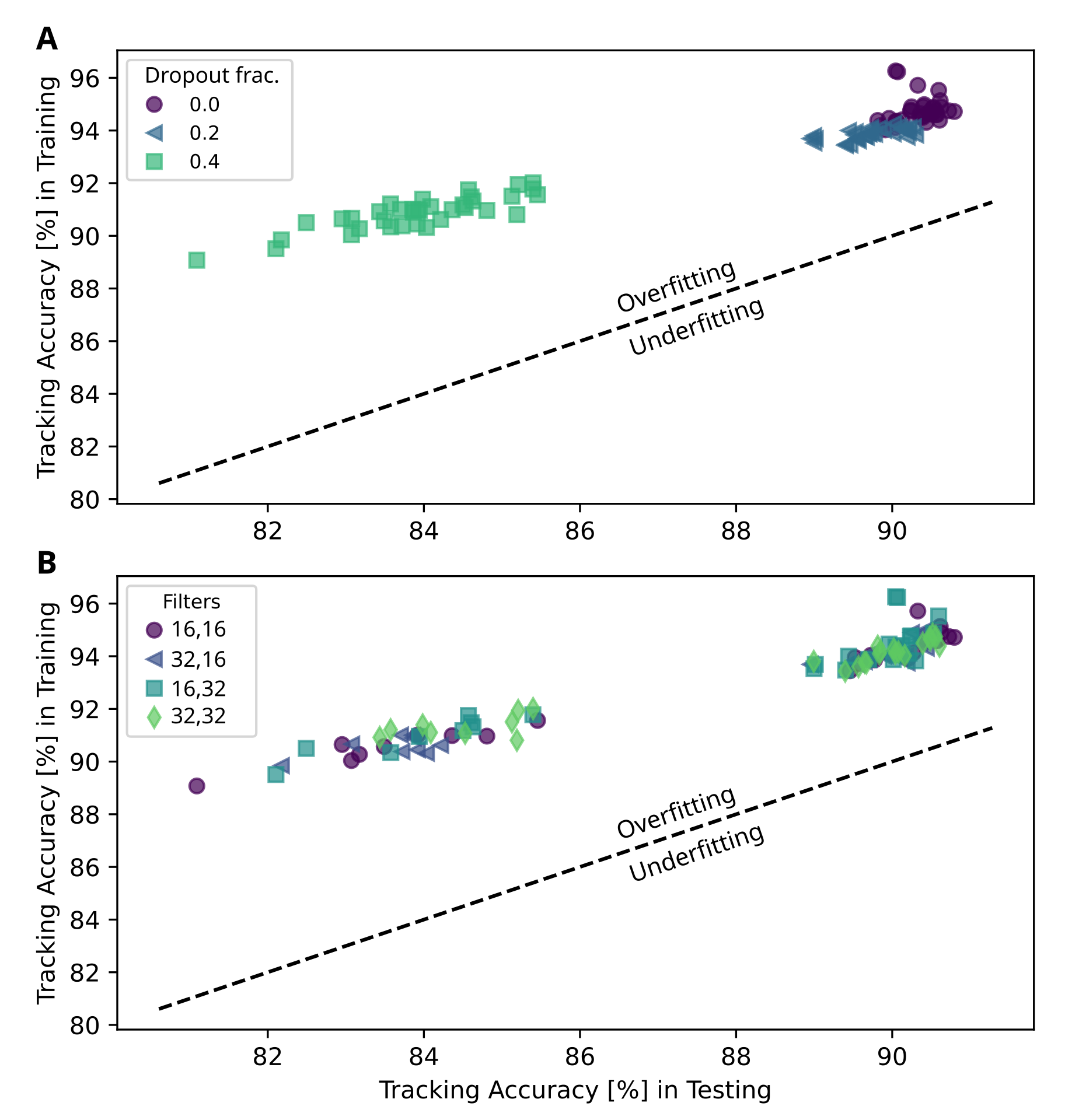}
    \caption{\textbf{Complete results for optimization stage 5: fine architecture.}
    Tracking accuracy results for each ML algorithm, comparing the scores obtained using the validation (vertical axis) and testing (horizontal axis) datasets. The overfitting and underfitting regions are separated by a dashed line. The same results are analyzed according to different groups:
    (\textbf{A}) dropout fraction after each Dense layer;
    (\textbf{B}) number of filters in the first and second 1D-Convolutional layers, respectively.
    }
    \label{fig_SI_Optim_S5_first}
\end{figure}

The results of this stage are shown in \textcolor{blue}{Figure~\ref{fig_SI_Optim_S5_first}}, comparing the validation and testing accuracies (explained in subsection \ref{SI_Optim_subsection_S3}). The results are grouped by the options for \textcolor{blue}{(A)}~dropout fraction and \textcolor{blue}{(B)}~number of filters.

\begin{figure}[!h]
    \centering
    \includegraphics[width=0.75\linewidth]{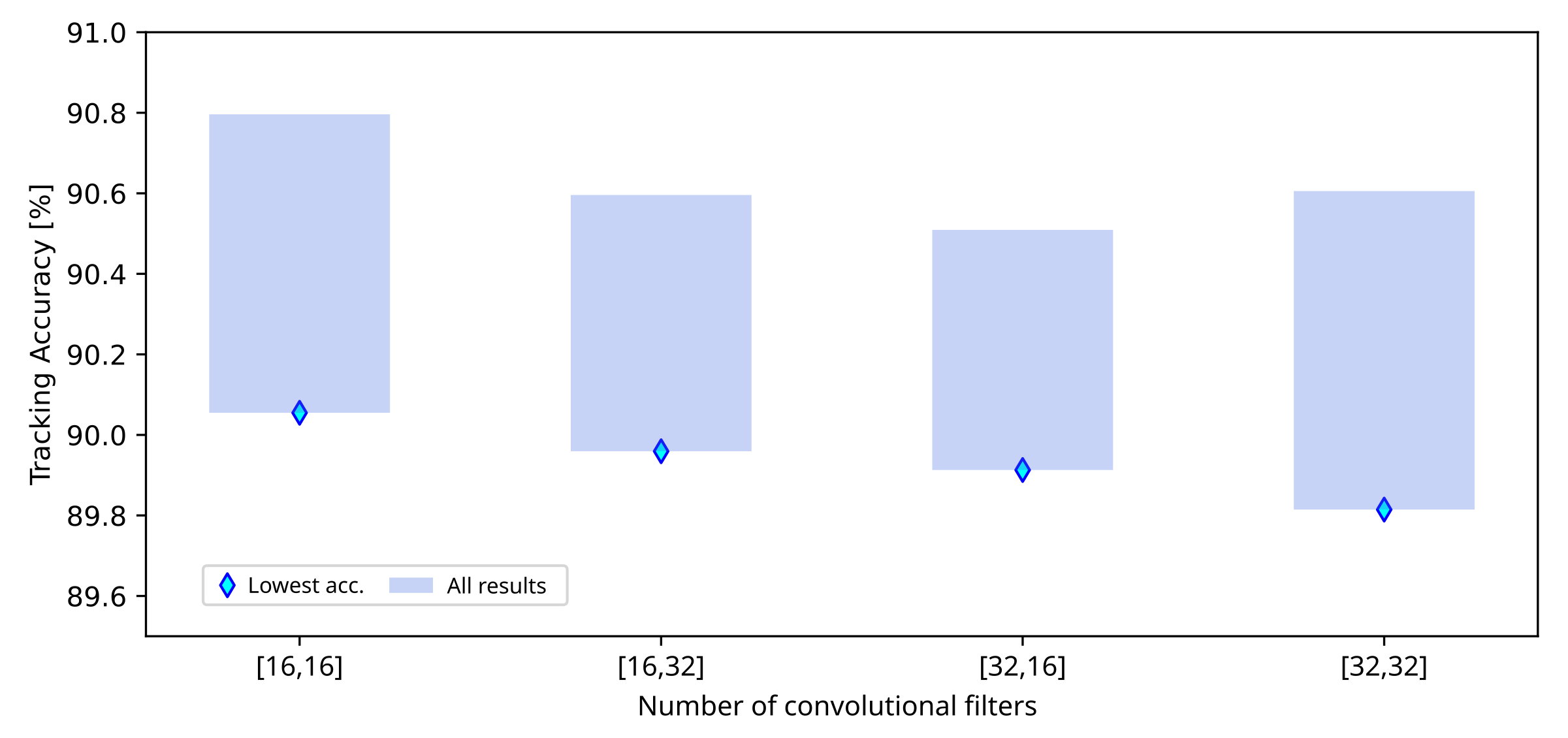}
    \caption{\textbf{Selected results for optimization stage 5: fine architecture.}
    Tracking accuracy for different number of filters $[f_1,f_2]$ for the first and second Convolutional layers, respectively. While the complete range of results for each architecture (shaded bars) include several frames for the magnetic data and random initialization seeds, only the lowest results (blue diamonds) count as the final score.
    }
    \label{fig_SI_Optim_S5_second}
\end{figure}

The Dropout fraction results (\textcolor{blue}{Figure~\ref{fig_SI_Optim_S5_first}A}) show that the ML performance is higher when the Dropout layers are not included in the Dense set. On the other hand, the number of filters (\textcolor{blue}{Figure~\ref{fig_SI_Optim_S5_first}B}) appears to have little impact. Then, we select those architectures with no Dropout layers for the final comparison, shown in \textcolor{blue}{Figure~\ref{fig_SI_Optim_S5_second}}.

The performance results indicate that the ML architecture works best with the fewest number of filters $[16,16]$, which also reduces the algorithm complexity and the risk of overfitting. As the kernel sizes are 8 and 4, respectively, using even less filters could be detrimental for the overall ML performance, because the random initialization may fail to generate the essential feature maps.


\subsection{Stage 6: training dataset size}
\label{SI_Optim_subsection_S6}

In this stage, we use all the best hyperparameters found in the previous stages, and analyze the effect of reducing the original training dataset ($\approx 6$~hours) to different fractions. In this case, the ML architecture is fixed, although we keep training different versions according to the initialization seeds and frames for the magnetic data. The hyperparameters are the following:

\begin{itemize}[topsep=0pt,itemsep=-1ex,partopsep=1ex,parsep=0ex]
    \item Input magnetic components: $(B_{\rm X},B_{\rm Y},B_{\rm Z})$.
    \item Time window points: 40.
    \item A convolutional set including two 1D-convolutional layers with 16~filters each and kernel sizes 8 and 4, in that order.
    \item A 1D-Conversion flattening layer.
    \item A Dense set with including Dense layers with 1024 and 512 neurons, in that order.
    \item Non-linear activation functions: tanh.
    \item Optimizer: Adam.
    \item Learning rate: $1 \times 10^{-5}$.
\end{itemize}

In order to reduce the original training dataset, we take the first fraction~$f$ of time windows along with their correlated targets, with $f$ ranging from 0.05 to 1 (full dataset), in steps of 0.05. Refer to \textcolor{blue}{Figure~4C} in the main text to see the results. From this stage, we conclude that larger datasets work better, but only half an hour of measurements ($f \approx 0.1$) is required to achieve a minimum tracking accuracy of 80\%.


\subsection{Best performing architecture}

\begin{figure}[!h]
    \centering
    \includegraphics[width=0.6\linewidth]{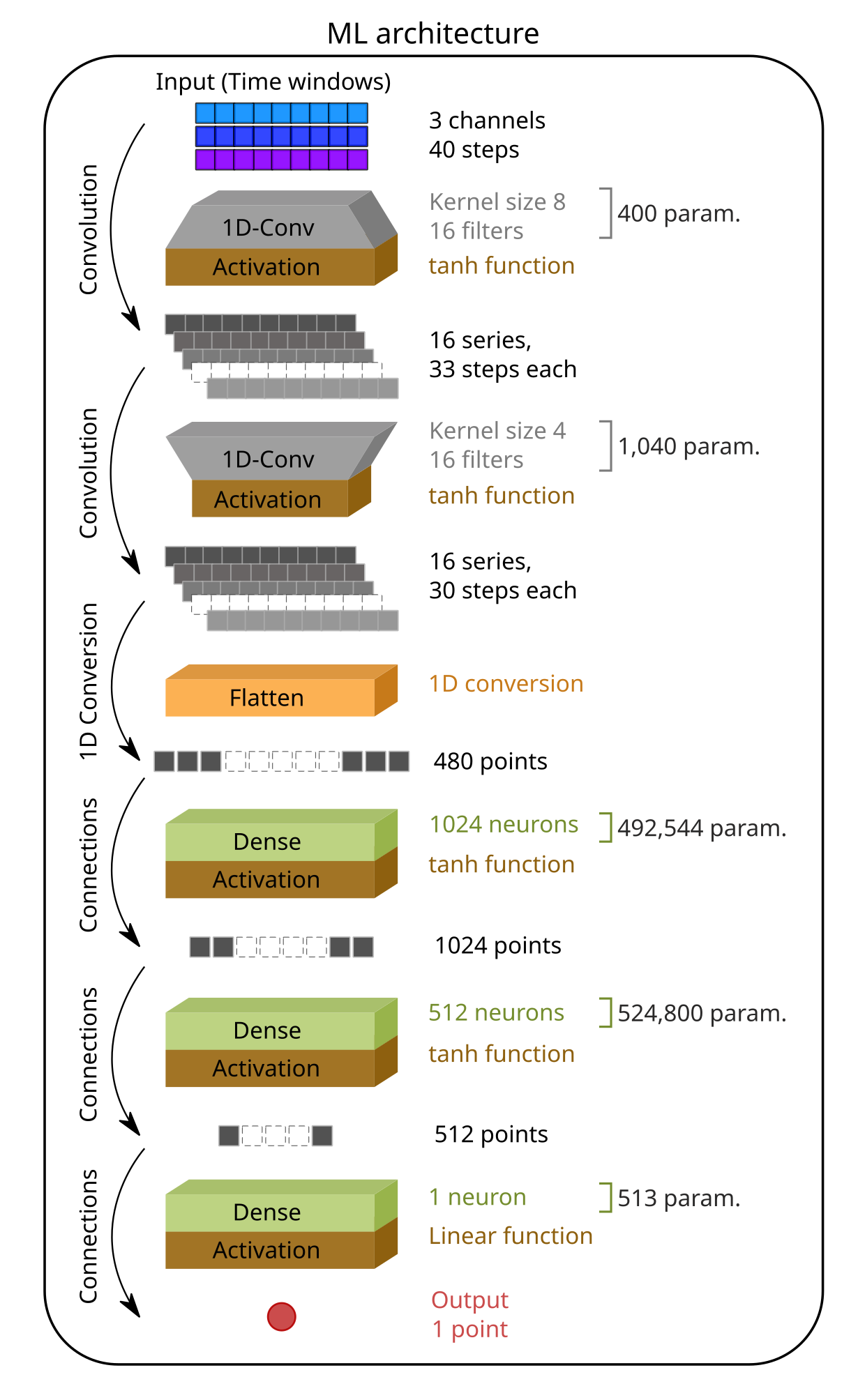}
    \caption{\textbf{Architectural description of our best performing Machine Learning algorithm.}
    The information flows from top to bottom, starting with the 3-channel, 40-point time window predictors and finally producing a single output prediction. In between, the data is processed through a Convolutional set with two layers, a 1D-Conversion layer with a flattening function, a Dense set with two layers, and finally a single-neuron Dense layer. All non-linear activation functions are Hyperbolic tangent (tanh). The arrows (left) indicate how data is transformed in each step, along with the input-output dimensions of each layer (right). The number of trainable internal parameters (if any) for each layer is described on the right.
    }
    \label{fig_SI_Example_ML_Architecture}
\end{figure}

Having optimized the ML architecture, we provide a explicit description of how the information is processed layer by layer, as shown in \textcolor{blue}{Figure~\ref{fig_SI_Example_ML_Architecture}}. There, we include the input-output dimensions in each layer and their number of trainable internal parameters. The total number of internal parameters for this architecture is 1,019,297.

\section{Data augmentation}
\label{SI_DataAugm}

In the previous section, we showed that the experimental training dataset can be reduced down to 0.5~hours while keeping the minimum tracking accuracy above 80\%. A complementary approach to reduce the measurement time even more would be to artificially augment the dataset, i.e.\ generating new synthetic data from the original measurements. We have studied this possibility by making replicas of the original dataset with the addition of random noise. Our protocol is described as follows:

\begin{enumerate}
    \item Reduce the original dataset~$\mathcal{D}$ to the first fraction~$f$, including magnetic predictors and position targets, before the normalization and time windowing processes (see Materials and Methods in the main text). The reduced dataset is identified as $\mathcal{D}_{\rm f}$.
    \item Define the augmentation number $N_{\rm augm}$ (0~means no augmentation) and the maximum noise variation $b$ for all magnetic components.
    \item Make $N_{\rm augm}$ copies of $\mathcal{D}_{\rm f}$, and add random noise to each copy of the magnetic predictors (the target positions remain unchanged). The noise is generated from a normal distribution between $-b$ and $b$ for each magnetic component. Notice that in addition to any number of copies of $\mathcal{D}_{\rm f}$, the augmented dataset will always include the original $\mathcal{D}_{\rm f}$.
    \item Normalize the magnetic predictors using the maximum values for each magnetic component among the entire augmented dataset, and process them into time windows.
\end{enumerate}

\begin{figure}[!h]
    \centering
    \includegraphics[width=0.9\linewidth]{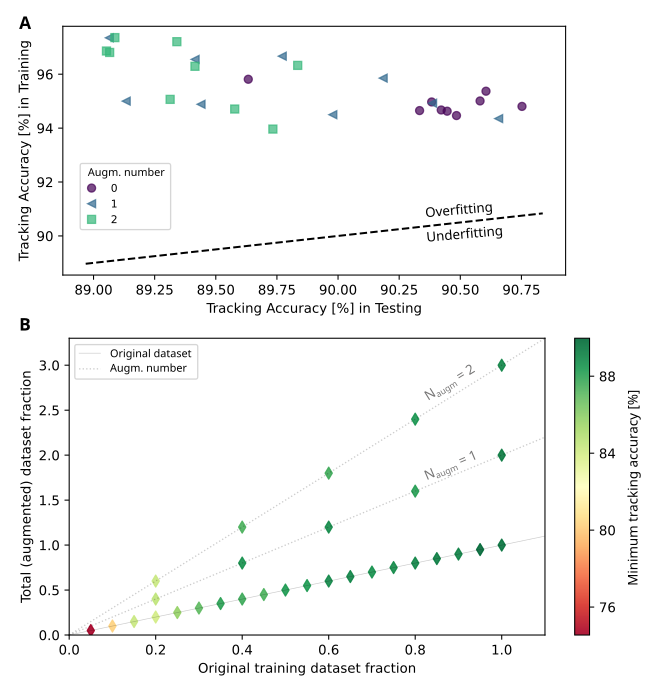}
    \caption{\textbf{Data augmentation effect on ML performance.}
    The original training dataset is artificially enlarged by replicating the original information with the addition of random noise.
    (\textbf{A}) Tracking accuracy results obtained using the validation (vertical axis) and testing (horizontal axis) datasets. The overfitting and underfitting regions are separated by a dashed line. The data is grouped according to the augmentation number (0 means no augmentation).
    (\textbf{B}) Minimum tracking accuracy (color-coded) for algorithms trained with a fraction~$f$ of the original training dataset (X-axis) augmented with additional noise (Y-axis). Values are relative to the full dataset size $f=1$. The solid line represents the results for no augmentation, while dotted lines indicate the augmentation number~$N_{\rm augm}$.
    }
    \label{fig_SI_Data_Augmentation}
\end{figure}

We investigate the validation and testing accuracies (as explained in subsection \ref{SI_Optim_subsection_S3}) using the full dataset ($f=1$) for all combinations of options $N_{\rm augm}=0,1,2$ and $b=10,20,40$~nT. The results are shown in \textcolor{blue}{Figure~\ref{fig_SI_Data_Augmentation}A}, demonstrating that augmenting the dataset ($N_{\rm augm}>0$) does not improve the ML performance, and in most cases it degrades it.

Furthermore, we study the augmentation effect for a reduced fraction of the original training dataset using a fixed noise level $b=20$~nT (which works better than the other options). The results are also compared with those obtained in subsection \ref{SI_Optim_subsection_S6} for the original dataset, as illustrated in \textcolor{blue}{Figure~\ref{fig_SI_Data_Augmentation}B}. There, the minimum tracking accuracy (color-coded) is analyzed for the fraction of the original training dataset (horizontal axis) as a function of the augmented dataset fraction (vertical axis), all values relative to the full dataset $f=1$. As we explained before for the non-augmented datasets (solid line), a larger dataset improves the ML performance. On the other hand, the augmented datasets (dotted lines, indicating $N_{\rm augm}$) have an overall neutral or negative effect on the ML accuracy.


\section{Parking accuracy per level, extended tolerance}
\label{SI_park_acc}

\begin{figure}[!h]
    \centering
    \includegraphics[width=0.85\linewidth]{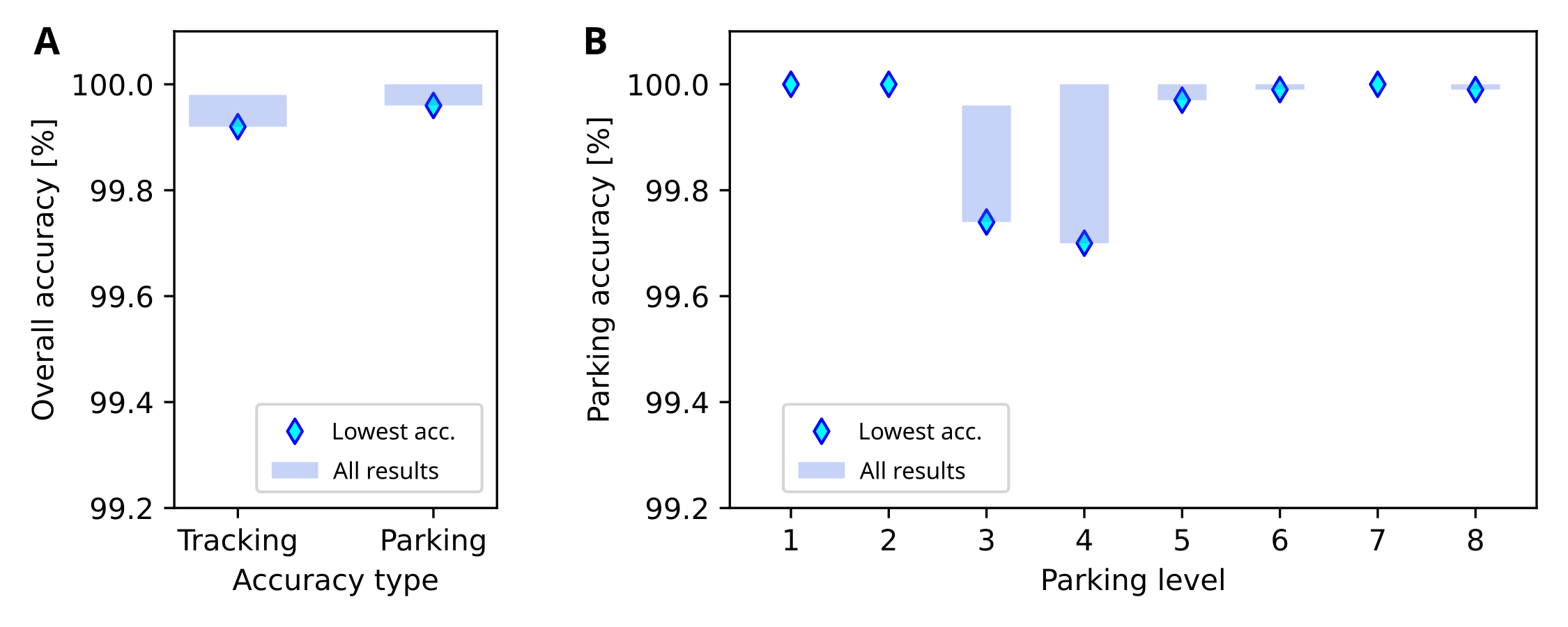}
    \caption{\textbf{ML performance for extended position tolerance}.
    (\textbf{A}) Tracking and parking accuracies evaluated in the entire testing dataset (6~hours) across 65~randomly generated frames for the magnetic data projection, using a 4-meter position tolerance. The final performance is defined as the lowest result within these frames.
    (\textbf{D}) Parking accuracy examined for each parking level.
    }
    \label{fig_SI_park_acc_lvls_tol4m}
\end{figure}

In the main text, we make a detailed analysis of the minimum parking accuracy $A_{\rm park, min}$ for each parking level, obtained using a 1-meter position tolerance. There, we demonstrated that all parking levels are identified with $A_{\rm park, min}>$90\%, except for Level~4, which shows a very poor accuracy below 50\%. We explain, however, that the wrong predictions from Level~4 are generally confused with the neighboring parking levels. This fact is demonstrated when we increase the position tolerance to 4~m and obtain $A_{\rm park, min}>$~99\% for all levels, as illustrated in \textcolor{blue}{Figure~\ref{fig_SI_park_acc_lvls_tol4m}}.

\section{Acceleration model for the elevator's position}
\label{SI_Accel_model}

The ML architecture built in the main text uses a simple linear approximation for interpolating the traveling events in-between the ground truth measurements associated to the parking intervals. This approach provides a reasonable position estimation for the general tracking task and can be applied straight forward. However, the ML algorithm can also benefit from a more sophisticated method for interpolation, increasing the precision for predicted positions at the expense of gathering more information about the system and developing a physical model.

In this section, we explain how we can incorporate an acceleration model for the elevator, based on experimental measurements, and improve the interpolation approach. Using the enhanced position targets, we repeat the training and optimization processes for the ML algorithm, confirming the same hyperparameters selection and optimal architecture. As a result, by the end of the last optimization stage, the general tracking slightly increases from 89.6\% to 91.1\%.

\subsection{Experimental measurements}

The position of the elevator as a function of time $Z(t)$ can be deduced from the acceleration profile $a_{\rm Z}(t)$, using simple equations of motion. We expect that the elevator repeats the same acceleration protocol for all flights sharing the same characteristics, mainly the number of traveled levels. Another critical factor could be the weight that is lifted, proportional to the number of passengers. However, developing such level of details for the physical model is beyond the scope of this work, and we will restrict our analysis to the number of traveled levels.

We record the acceleration profile independently of the magnetic measurements, by using the built-in accelerometer of an iPhone 14.4. We measure the complete acceleration vector~$\vec{a}$ while the elevator is operating in normal hours, for two different segments:

\begin{itemize}
    \item Segment \#1: 7 events of 1-level flights going up and 7 events of 1-level flights going down.
    \item Segment \#2: General recording, including at least one $N$-level flight in each direction, for $N$ ranging from 1 to 4, and also for $N=7$.
\end{itemize}

\begin{figure}[!h]
    \centering
    \includegraphics[width=0.75\linewidth]{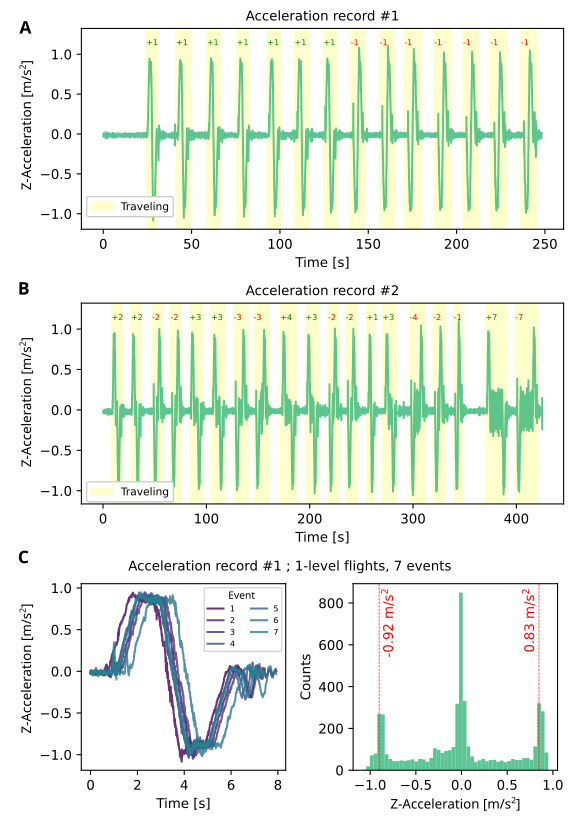}
    \caption{\textbf{Experimental measurements of the Z-acceleration profile of the elevator.}
    The recording is divided into two segments:
    (\textbf{A}) 1-level flights in any direction; and
    (\textbf{B}) general recording of $N$-level flights in any direction.
    (\textbf{C}) 
    (Left) Collection of 1-level flights going up from the first segment, compared in the same timescale.
    (Right) Histogram for Z-acceleration values, with red dashed lines indicating the most frequent values for maximum acceleration in the traveling direction (up, positive) and opposite direction (down, negative).
    }
    \label{fig_SI_Acceleration_records}
\end{figure}

The original data can be found in the GitHub repository \href{https://github.com/Fertmeneses/ML_QDM_Meneses_et_al}{Fertmeneses/ML\_QDM\_Meneses\_et\_al}. We summarize the results for $a_{\rm Z}(t)$ in \textcolor{blue}{Figure~\ref{fig_SI_Acceleration_records}}, where the \textcolor{blue}{(A)} first and \textcolor{blue}{(B)} second segments are displayed.

According to the experimental measurements, all Z-acceleration profiles follow a similar protocol, for any N-level flight:

\begin{enumerate}
    \item Accelerate towards a maximum value~$a_{\rm max,1}$ in the traveling direction, close to 1~m/s$^2$, within a time $t_1$.
    \item Reduce acceleration to 0 within a time $t_2$ and keep traveling at constant velocity $V$ for a time~$t_{\rm V}$.
    \item Accelerate towards a maximum value~$a_{\rm max,2}$ in the opposite traveling direction, close to 1~m/s$^2$, within a time $t_3$.
    \item Reduce the acceleration to~0 within a time $t_4$, also reducing velocity to 0 and stopping at the final parking level.
\end{enumerate}

For larger trips such as 7-level flights (see \textcolor{blue}{Figure~\ref{fig_SI_Acceleration_records}B}), the constant velocity step does not have the ideal null acceleration but a noisy $a_{\rm Z}\approx 0$ averaging zero. For the timescale and level of details that we need, we will disregard the acceleration noise and assume simpler movements.

On the other hand, 1-level flights do not follow the same protocol compared to the longer trips, because the traveling time is shorter and does not allow to travel at constant speed in-between the acceleration segments. A comparison of individual 1-level flights going up during the first recording segment is shown in \textcolor{blue}{Figure~\ref{fig_SI_Acceleration_records}C}, next to a histogram for the acceleration values, in which we identify the most frequent values $a_{\rm max,1}=$~0.83~m/s$^2$ and $a_{\rm max,2}=-$0.92~m/s$^2$. As we will explain later, 1-level flights are not simply equivalent to set $t_{\rm V}=0$ in our previous description, as the acceleration times $t_2$ and $t_3$ (going from maximum acceleration in the traveling direction to the opposite value) are shorter than those of longer trips. 

\subsection{Physical model}

Our physical model for the elevator's position follows the next assumptions:

\begin{itemize}
    \item All N-level flights with $N>1$ are described by the same equations: acceleration $a_{\rm N}(t)$, velocity $v_{\rm N}(t)$ and position $x_{\rm N}(t)$. The 1-level flights, instead, have separate $a_1(t)$, $v_1(t)$ and $x_1(t)$ equations.
    \item The maximum acceleration absolute values in any direction are unified into $a_0=|a_{\rm max,1}|=|a_{\rm max,2}|=$~0.87~m/s$^2$ for all flights.
    \item The acceleration times $t_1$ and $t_4$ are considered the same for all flights and unified as $\Delta t = t_1 = t_4$.
    \item For N-level flights, $N>1$, the intermediate acceleration times $t_2$ and $t_3$ are also unified as $\Delta t$. For 1-level flights, instead, they are considered equal in duration but with a different value: $t_2=t_3=\Delta t'$.
    \item For all flights, the full acceleration profile is described by a series of linear segments, without any noise.
\end{itemize}

Below, we introduce the motion equations for individual flights, each starting at $t=0$, assuming that the elevator starts parked (null acceleration and velocity) at the position~$x_0$.

\textbf{Time segments for N-level flights, $N>1$:}
\begin{align}    
    T_{\rm N,1} &= [0,\Delta t] \nonumber \\
    T_{\rm N,2} &= [\Delta t,2\Delta t] \nonumber \\
    T_{\rm N,3} &= [2\Delta t,3\Delta t] \nonumber \\
    T_{\rm N,4} &= [3\Delta t,3\Delta t+t_{\rm V}] \nonumber \\
    T_{\rm N,5} &= [3\Delta t +t_{\rm V},4\Delta t+t_{\rm V}] \nonumber \\
    T_{\rm N,6} &= [4\Delta t +t_{\rm V},5\Delta t+t_{\rm V}] \nonumber \\
    T_{\rm N,7} &= [5\Delta t +t_{\rm V},6\Delta t+t_{\rm V}] \nonumber
\end{align}

\textbf{Time segments for 1-level flights:}
\begin{align}    
    T_{1,1} &= [0,\Delta t] \nonumber \\
    T_{1,2} &= [\Delta t,2\Delta t] \nonumber \\
    T_{1,3} &= [2\Delta t,2\Delta t + 2\Delta t'] \nonumber \\
    T_{1,4} &= [2\Delta t + 2\Delta t',3\Delta t + 2\Delta t'] \nonumber \\
    T_{1,5} &= [3\Delta t + 2\Delta t',4\Delta t + 2\Delta t'] \nonumber
\end{align}

\textbf{Acceleration equations:}

\begin{equation}
    \label{Eq_a_N}
    a_{\rm N}(t)= a_0 \times \begin{cases}
    \frac{t}{\Delta t}, &
    t \in T_{\rm N,1} \\
    1, &
    t \in T_{\rm N,2} \\
    \frac{3 \Delta t-t}{\Delta t}, &
    t \in T_{\rm N,3} \\
    0, &
    t \in T_{\rm N,4} \\
    - \frac{t-(3 \Delta t+t_{\rm V})}{\Delta t}, &
    t \in T_{\rm N,5} \\
    -1, &
    t \in T_{\rm N,6} \\
    - \frac{(6 \Delta t+t_{\rm V})-t}{\Delta t}, &
    t \in T_{\rm N,7}
    \end{cases}
\end{equation}

\begin{equation}
    \label{Eq_a_1}
    a_1(t)= a_0 \times \begin{cases}
    \frac{t}{\Delta t}, &
    t \in T_{1,1} \\
    1, &
    t \in T_{1,2} \\
    \frac{(2 \Delta t+\Delta t')-t}{\Delta t'}, &
    t \in T_{1,3} \\
    -1, &
    t \in T_{1,4} \\
    -\frac{(4 \Delta t+2 \Delta t')-t}{\Delta t}, &
    t \in T_{1,5}
    \end{cases}
\end{equation}

\textbf{Velocity equations:}

\begin{equation}
    \label{Eq_v_N}
    v_{\rm N}(t) = a_0 \times \begin{cases}
    \frac{1}{2} \frac{t^2}{\Delta t}, &
    t \in T_{\rm N,1} \\
    \frac{1}{2} \Delta t+(t-\Delta t), &
    t \in T_{\rm N,2} \\
    2 \Delta t-\frac{1}{2} \frac{(3 \Delta t-t)^2}{\Delta t}, &
    t \in T_{\rm N,3} \\
    2 \Delta t, &
    t \in T_{\rm N,4} \\
    2 \Delta t-\frac{1}{2} \frac{\left[t-(3 \Delta t+t_{\rm V})\right]^2}{\Delta t}, &
    t \in T_{\rm N,5} \\
    \frac{3}{2} \Delta t-\left[t-(4 \Delta t+t_{\rm V})\right], &
    t \in T_{\rm N,6} \\
    \frac{1}{2} \frac{(6 \Delta t+t_{\rm V}-t)^2}{\Delta t}, &
    t \in T_{\rm N,7}
    \end{cases}
\end{equation}

\begin{equation}
    \label{Eq_v_1}
    v_1(t)= a_0 \times \begin{cases}
    \frac{1}{2} \frac{t^2}{\Delta t}, &
    t \in T_{1,1} \\
    \frac{1}{2} \Delta t+(t-\Delta t), &
    t \in T_{1,2} \\
    \frac{3}{2} \Delta t+\frac{1}{2} \Delta t'-\frac{1}{2} \frac{\left[(2 \Delta t+\Delta t')-t\right]^2}{\Delta t'}, &
    t \in T_{1,3} \\
    \frac{3}{2} \Delta t-\left[t-(2 \Delta t+2 \Delta t')\right], &
    t \in T_{1,4} \\
    \frac{1}{2} \frac{\left[(4 \Delta t+2 \Delta t')-t\right]^2}{\Delta t}, &
    t \in T_{1,5}
    \end{cases}
\end{equation}

\textbf{Position equations:}

\begin{equation}
    \label{Eq_x_N}
    x_{\rm N}(t)= x_0+ a_0 \times \begin{cases}
    \frac{1}{6\Delta t}t^3, &
    t \in T_{\rm N,1} \\
    \frac{\Delta t^2}{6} +\frac{\Delta t}{2} (t-\Delta t)+\frac{1}{2} (t-\Delta t)^2, &
    t \in T_{\rm N,2} \\
    \frac{8\Delta t^2}{6} +\frac{3\Delta t}{2} (t-2 \Delta t)-\frac{(3 \Delta t-t)^3}{6\Delta t}, &
    t \in T_{\rm N,3} \\
    \frac{17\Delta t^2}{6} +2 \Delta t(t-3 \Delta t), &
    t \in T_{\rm N,4} \\
    \frac{17\Delta t^2}{6} +2 \Delta t \, t_{\rm V}+2 \Delta t\left[t-(3 \Delta t+t_{\rm V})\right]- \frac{\left[t-(3 \Delta t+t_{\rm V})\right]^3}{6\Delta t}, &
    t \in T_{\rm N,5} \\
    \frac{14\Delta t^2}{3} +2 \Delta t \, t_{\rm V}+\frac{3\Delta t}{2} \left[t-(4 \Delta t+t_{\rm V})\right]-\frac{1}{2} \left[t-(4 \Delta t+t_{\rm V})\right]^2, &
    t \in T_{\rm N,6} \\
    \frac{35\Delta t^2}{6} +2 \Delta t \,t_{\rm V}-\frac{\left[(6 \Delta t+t_{\rm V})-t\right]^3}{6\Delta t}, &
    t \in T_{\rm N,7}
    \end{cases}
\end{equation}

\begin{equation}
    \label{Eq_x_1}
    x_1(t)= x_0+a_0 \times \begin{cases}
    \frac{1}{6\Delta t}t^3, &
    t \in T_{1,1} \\
    \frac{\Delta t^2}{6} +\frac{\Delta t}{2} (t-\Delta t)+\frac{1}{2} (t-\Delta t)^2, &
    t \in T_{1,2} \\
    \frac{7 \Delta t^2}{6}-\frac{\Delta t'^2}{6} +\frac{3\Delta t}{2} (t-2 \Delta t)+\frac{\Delta t'}{2} (t-2 \Delta t)+ \frac{\left[(2 \Delta t+\Delta t')-t\right]^3}{6\Delta t'}, &
    t \in T_{1,3} \\
    \frac{7\Delta t^2}{6} +\frac{2\Delta t'^2}{3} +3 \Delta t \Delta t'+\frac{3\Delta t}{2} \left[t-(2 \Delta t+2 \Delta t')\right]-\frac{\left[t-(2 \Delta t+2 \Delta t')\right]^2}{2}, &
    t \in T_{1,4} \\
    \frac{7\Delta t^2}{3} +\frac{2\Delta t'^2}{3} +3 \Delta t \Delta t'- \frac{\left[(4 \Delta t+2 \Delta t')-t\right]^3}{6\Delta t}, &
    t \in T_{1,5}
    \end{cases}
\end{equation}

\begin{figure}[!h]
    \centering
    \includegraphics[width=0.9\linewidth]{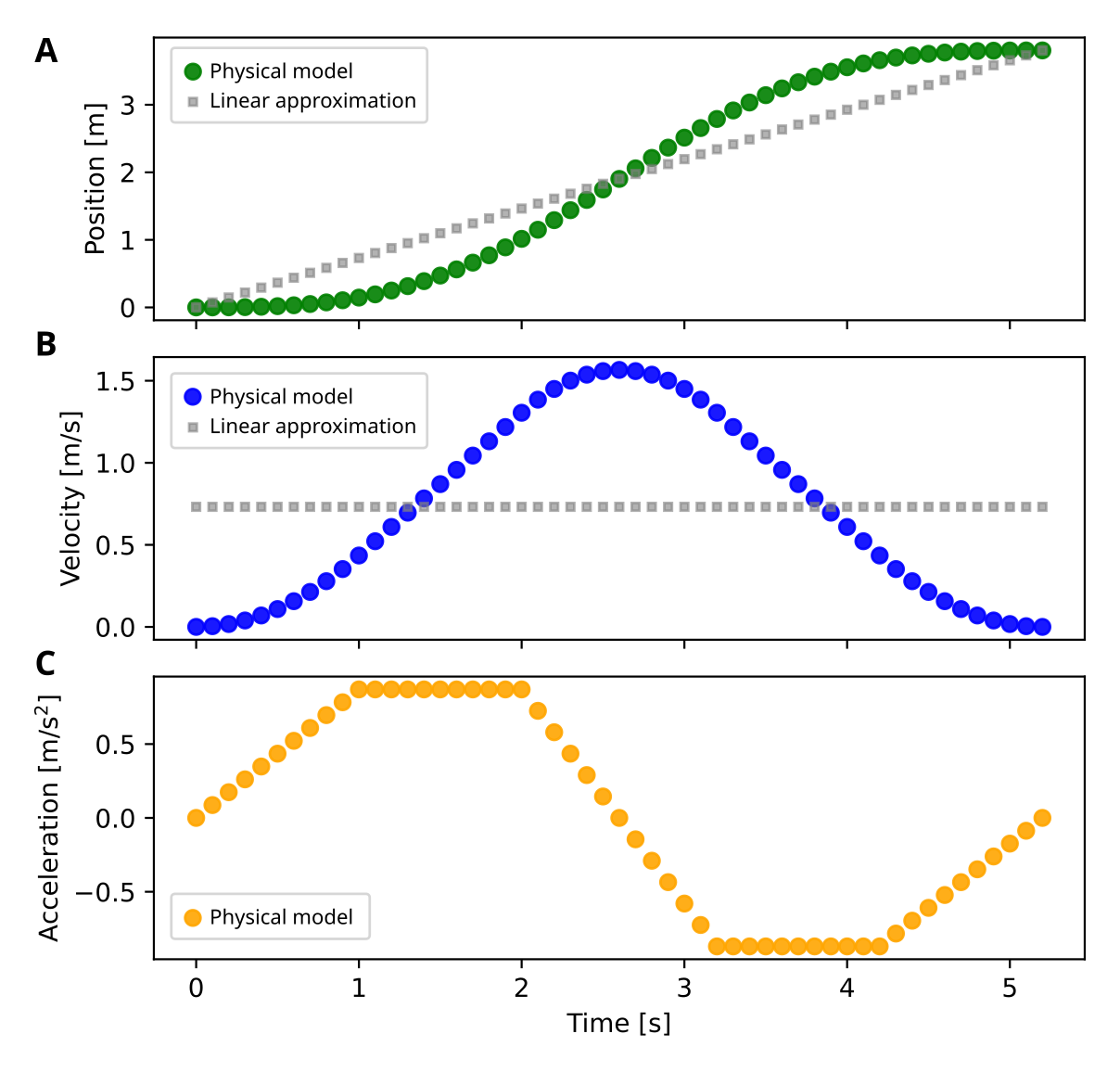}
    \caption{\textbf{Kinematics for the elevator's 1-level flights according to the physical model.}
    (\textbf{A}) Z-acceleration,
    (\textbf{B}) Z-velocity and
    (\textbf{C}) Z-position
    as a function of time for a 1-level flight going up, covering a distance of 3.8~m in 5.2~s. The linear approximation (gray squares) is shown for comparison in the velocity and position plots.
    }
    \label{fig_SI_Acceleration_Motion_1lvl}
\end{figure}

According to the experimental observations, we fix the parameters $\Delta t =$~1.0~s and $\Delta t' =$~0.6~s. As $a_0=$~0.87~m/s$^2$ is also fixed, the only free parameter is $t_{\rm V}$ for the N-level flights, $N>1$. Conversely, the 1-level flights have a fixed total traveling time of $4\Delta t + 2\Delta t'=$5.2~s, covering a total distance $\frac{7 a_0 \Delta t^2}{3} +\frac{2a_0\Delta t'^2}{3} +3 a_0 \Delta t \Delta t'=$ 3.8~m. Compared to the experimental results, the traveled distance is very close to the 3.7~m or 4.1~m spacing between adjacent levels, although our recorded traveling times are in the order of 3-4~s. The kinematics for the 1-level flights is illustrated in \textcolor{blue}{Figure~\ref{fig_SI_Acceleration_Motion_1lvl}}, also including the linear approximation for an easy comparison.

\subsection{Targets positions using the physical model}

Now that we have the physical model, we have to interpolate the ground truth parking intervals. As we have to match the experimental measurements, meaning the total traveling time~$t_{\rm total}$ and distance~$x_{\rm total}$ for each trip, it is necessary to make the model more flexible. Notice that two trips sharing the same traveled distance and direction may have different experimental values~$t_{\rm total}$.
Our criterion is to fix the following values:

\begin{itemize}
    \item $\Delta t = 2 \Delta t'$ for 1-level flights.
    \item $\Delta t \leq 1$s and as large as possible for N-level flights, $N>1$.
\end{itemize}

Under these conditions, $t_{\rm total}$ can be described as follows:
\begin{equation}
    \label{Eq_t_total}
    t_{\rm total}(\Delta t,t_{\rm V})= \begin{cases}
    5 \Delta t, &
    \text{for 1-level flights} \\
    6 \Delta t \; ; \; t_{\rm V}=0, &
    \text{for N-level flights, }N>1 \, \& \, t_{\rm total} \leq \text{6s} \\ 
    6 \Delta t + t_{\rm V} \; ; \; \Delta t=1 \text{s}, 
    & \text{for N-level flights, }N>1 \, \& \, t_{\rm total}>\text{6s}
    \end{cases}
\end{equation}

Once $\Delta t$ and $t_{\rm V}$ are fixed, the traveled distance $x_f$ must be corrected using $a_0$:
\begin{equation}
    \label{Eq_x_total}
    x_{\rm total}(a_0,t_{\rm total})= \begin{cases}
    \frac{4}{25}t_{\rm total}^{2}a_0,&
    \text{for 1-level flights} \\
    \frac{35}{216}t_{\rm total}^{2}a_0, &
    \text{for N-level flights, }N>1 \, \& \, t_{\rm total} \leq \text{6s} \\ 
    (2t_{\rm total}-\frac{37}{6} \,\text{[s]})a_0\,\text{[s]}, & 
    \text{for N-level flights, }N>1 \, \& \, t_{\rm total}>\text{6s}
    \end{cases}
\end{equation}


\subsection{Machine Learning performance with the acceleration model approach}

\begin{figure}[!h]
    \centering
    \includegraphics[width=0.75\linewidth]{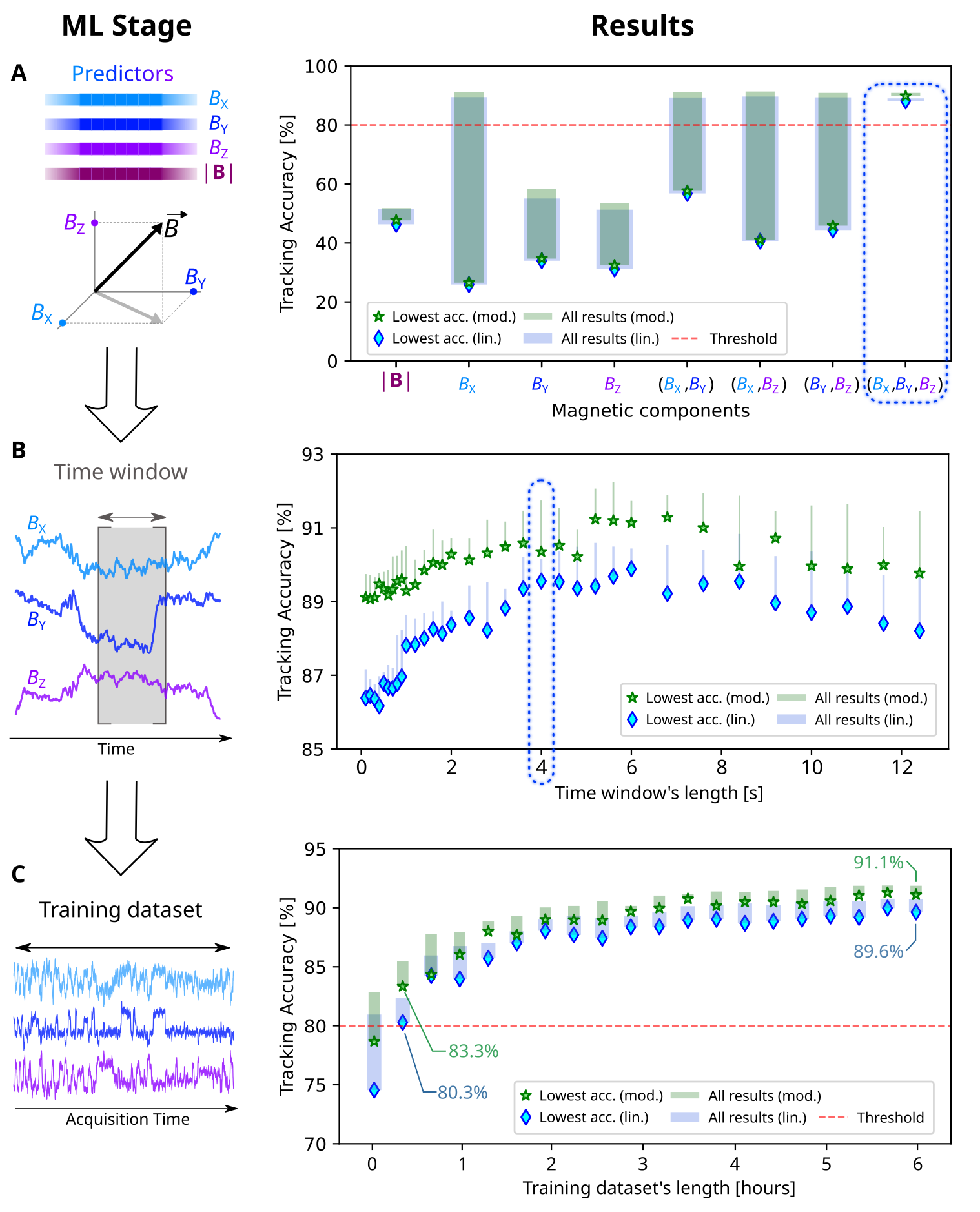}
    \caption{
    \textbf{Optimizing the Machine Learning hyperparameters for different interpolation approaches}.
    In progressive stages, multiple options for a single hyperparameter are studied, training several algorithms and computing the tracking accuracy with a 1-meter tolerance. While the complete range of results are displayed (shaded bars), only the lowest results (blue diamonds for the linear approximation ``lin.'', green stars for the physical model ``mod.'') count as the final score. We set a reference benchmark of 80\% (red dashed line). The dashed rectangles indicate the hyperparameters choice in each stage.
    (\textbf{A}) First stage: combinations of magnetic components used as input predictors.
    (\textbf{B}) Second stage: duration of time windows.
    (\textbf{C}) Final stage: impact of the training dataset size.
    }
    \label{fig_SI_Optimization_Phys_Model}
\end{figure}

Our Machine Learning architecture only requires a dataset containing magnetic predictors and target positions. Consequently, the interpolation approach that we choose, either the acceleration model or the linear approximation, only affects how we define the target positions, while the ML training protocol and optimization stages follow the same rules. Then, we have replicated the complete process of optimizing the ML hyperparameters and evaluating the best architecture using the physical model as the interpolated approach, instead of the linear approximation. The main three optimization stages shown in \textcolor{blue}{Figure~4} in the main text for the linear approximation (blue diamonds and shaded bars) are recalled in \textcolor{blue}{Figure~\ref{fig_SI_Optimization_Phys_Model}}, including the new results using the physical model (green stars and narrower shaded bars).

Remarkably, the physical model approach yields better ML performances for all hyperparameter options, proving that a better informed algorithm can provide more precise results. However, the final gain is 1.5\% in the last stage (\textcolor{blue}{Figure~\ref{fig_SI_Optimization_Phys_Model}C}), when using the optimal hyperparameters and the full training dataset, which may not justify the effort of building the physical model and collecting the experimental data for the elevator's acceleration.




\end{document}